\documentclass[a4paper,11pt]{article}
\pdfoutput=1 

\usepackage{jheppub1} 

\usepackage[T1]{fontenc} 

\usepackage{amsmath,amssymb,amsfonts,graphicx,subfigure,slashed,dsfont}

\def\bea{\begin{eqnarray}}
\def\eea{\end{eqnarray}}
\def\nn{\nonumber}
\def\ba{\begin{array}}
\def\ea{\end{array}}
\def\nn{\nonumber}
\def\Tr{\text{Tr}}

\def\sn{\text{sn}}

\title{\boldmath Holographic measurement in CFT thermofield doubles}

\author[1]{Stefano Antonini }
\author[2]{, Brianna Grado-White }
\author[3]{, Shao-Kai Jian }
\author[2]{, Brian Swingle }

\affiliation[1]{University of Maryland, College Park, MD, 20742, USA}

\affiliation[2]{Department of Physics, Brandeis University, Waltham, Massachusetts 02453, USA}

\affiliation[3]{Department of Physics and Engineering Physics, Tulane University, New Orleans, Louisiana, 70118, USA}

\emailAdd{santonin@umd.edu}
\emailAdd{bgradowhite@brandeis.edu}
\emailAdd{sjian@tulane.edu}
\emailAdd{bswingle@brandeis.edu}

\abstract{
We extend the results of arXiv:2209.12903 by studying local projective measurements performed on subregions of two copies of a CFT${}_2$ in the thermofield double state and investigating their consequences on the bulk double-sided black hole holographic dual. We focus on CFTs defined on an infinite line and consider measurements of both finite and semi-infinite subregions. In the former case, the connectivity of the bulk spacetime is preserved after the measurement. In the latter case, the measurement of two semi-infinite intervals in one CFT or of one semi-infinite interval in each CFT can destroy the Einstein-Rosen bridge and disconnect the bulk dual spacetime. In particular, we find that a transition between a connected and disconnected phase occurs depending on the relative size of the measured and unmeasured subregions and on the specific Cardy state the measured subregions are projected on. We identify this phase transition as an entangled/disentangled phase transition of the dual CFT system by computing the post-measurement holographic entanglement entropy between the two CFTs.
We also find that bulk information encoded in one CFT in the absence of measurement can sometimes be reconstructed from the other CFT when a measurement is performed, or can be erased by the measurement. 
Finally, we show that a purely CFT calculation of the Renyi entropy using the replica trick yields results compatible with those obtained in our bulk analysis.
}

\begin{document} 
\maketitle
\flushbottom

\newpage

\section{Introduction} 
\label{sec:intro}

The holographic principle---and its most concrete realization, the Anti de-Sitter/Conformal Field Theory (AdS/CFT) correspondence \cite{Maldacena:1997re,Witten:1998qj,Gubser:1998bc,Aharony:1999ti}---has suggested that gravitational spacetime is an emergent property, and in particular, that the bulk geometry is dictated by the entanglement structure of a dual, purely quantum system defined in one less dimension ~\cite{Ryu2006a,Ryu2006b,Hubeny:2007xt,Swingle:2009bg,VanRaamsdonk:2010pw,Maldacena:2013xja,Engelhardt:2014gca,Dong:2016eik,Harlow:2016vwg}.  As such, it is expected that operations in the quantum boundary theory that modify or destroy the entanglement structure will consequently modify or destroy any bulk geometric structure that existed pre-operation. Recently, such phenomena have been studied in the case of local projective measurements performed on subregions of the boundary theory. 

In particular, \cite{numasawa2016epr} initiated this work, showing that large portions of the bulk dual spacetime could indeed be destroyed (with the deleted region bounded by an end-of-the-world (ETW) brane) by postselecting a boundary subregion onto a specific class of states. This line of inquiry was furthered in \cite{Antonini:2022sfm}, which gave a detailed account of this boundary measurement in vacuum AdS$_3$/CFT$_2$ and in related tensor network models, and further showed how measurement could modify the bulk geometry and holographic dictionary via quantum teleportation. Of particular interest here, \cite{Antonini:2022sfm} showed that when measuring two disconnected boundary subregions, varying the measurement parameters could induce a phase transition corresponding to whether the remaining, unmeasured boundary regions were connected through the bulk time reflection symmetric slice or not. This bulk phase transition corresponded to an entangled/disentangled phase transition in the dual boundary theory. 

Here, we extend the results and techniques of \cite{Antonini:2022sfm} by studying post-selection in AdS$_3$/CFT$_2$ starting in the thermofield double (TFD) state of a CFT on a line, dual to an eternal, two-sided BTZ black hole in the bulk \cite{Maldacena:2001kr}. This generalization to measurements in the TFD is of particular interest for its potential implications for the reconstruction of operators behind the black hole horizon (as in \cite{kourkoulou2017pure}) and for holographic cosmologies \cite{Cooper:2018cmb,Antonini:2019qkt,Antonini:2021xar}, see also \cite{Milekhin:2022bzx} for related work. A similar endeavor was undertaken in a lower dimensional model in \cite{Antonini:2022lmg}, where postselection of the TFD state for two copies of the Sachdev-Ye-Kitaev (SYK) model dual to Jackiw-Teitelboim (JT) gravity was considered. There, the focus was on understanding the conditions under which the information contained in the entanglement wedge of one side of the TFD becomes accessible by the other side after measurement. 
For the $2+1$ dimensional case studied here, we also study this ``bulk teleportation'' between the two sides (in the sense of \cite{Antonini:2022sfm}), and additionally focus on how bulk connectivity is modified by measurement, aiming to characterize when measurements destroy the Einstein-Rosen bridge stretching between the two asymptotic regions.

We start by constructing the gravitational dual of the post-selected TFD states. To describe a local projective measurement onto a product state of a subregion, we start with the Euclidean path integral that prepares the thermofield double state, and insert a slit in the time reflection symmetric slice corresponding to the measured subregion \cite{rajabpour2015post, rajabpour2016entanglement}. Note that this local projective measurement will project the measured subregion onto a Cardy state \cite{cardy1989boundary,miyaji2014boundary}. Following  \cite{numasawa2016epr,Antonini:2022sfm}, we can build the bulk duals using the AdS/BCFT proposal \cite{takayanagi2011holographic,fujita2011aspects}. The measurement, and thus the insertion of the slit in the path integral, will correspond to the insertion of an ETW brane in the bulk anchored to the slit's boundary. Note that the CFT stress-energy tensor is divergent at the endpoints of the slit \cite{numasawa2016epr,Antonini:2022sfm}, leading to a singular dual metric. To avoid directly dealing with this complication, and again following \cite{numasawa2016epr,Antonini:2022sfm}, we perform a series of conformal transformations to map to a non-singular CFT configuration, and thus a regular bulk. 

Here, we will map our initial infinite cylinder (corresponding to the Euclidean path integral preparing the TFD state for two copies of a CFT on a line) with various slit configurations (corresponding to measurements of various subregions of the two CFTs) to a finite-length cylinder. The bulk dual of this finite cylinder can correspond to either a portion of a Euclidean BTZ black hole (cut off by a single brane connecting the two boundaries of the finite cylinder) or to a portion of Euclidean thermal AdS (cut off by two branes, each one anchored to one of the two boundaries of the finite cylinder). We then study this Hawking-Page phase transition between these two possible geometries for measurements performed on finite and semi-infinite intervals (in planar coordinates) on one or both CFTs in the TFD state. For the different measurement configurations, the phase transition will have different physical interpretations. Of particular interest are the geometrical properties of the bulk time reflection symmetric slice. To understand this, note that the time reflection symmetric slice in the Euclidean geometry will also be the time reversal symmetric slice in the corresponding Lorentzian geometry, giving the initial conditions for subsequent real time evolution. Therefore, the presence or absence of an ETW brane on the Euclidean time reflection symmetric slice will correspond to the presence or absence of an ETW brane in the Lorentzian wormhole geometry. More significantly, if the bulk Euclidean time reflection symmetric slice is connected between the two CFTs, the two asymptotic AdS boundaries where the two CFTs live will also be connected through an Einstein-Rosen bridge in the corresponding Lorentzian geometry.

In section \ref{sec:finite}, we consider projective measurements performed on a finite interval on one side of the TFD, followed by an additional Euclidean time evolution performed with the full CFT Hamiltonian. By choosing a large enough projected region, small enough euclidean time evolution, or small enough (or negative) tension of the ETW brane (corresponding to a small or negative boundary entropy \cite{affleck1991universal} of the specific Cardy state \cite{cardy1989boundary} that we project onto), we can tune the system into the BTZ phase, such that there is an ETW brane on the time reflection symmetric slice. In the thermal AdS phase, conversely, the time reflection symmetric slice does not contain any brane. In both phases, the two CFTs remain connected through the time reflection symmetric slice, suggesting that they remain significantly entangled after the measurement. We confirm this expectation by computing the holographic entanglement entropy for subregions of one CFT via the RT formula and noting that it remains non-zero for non-empty subregions in both phases.

A more drastic change to the bulk geometry occurs when we turn to measuring semi-infinite intervals in either one (Section \ref{sec:infinite-one}) or both CFTs (Section \ref{sec:infinite-two}). In these cases, we also find two different phases and show that when the measured region is sufficiently large or the tension of the ETW brane is sufficiently small or negative, the time reflection symmetric slice in the Euclidean geometry can become disconnected. This suggests that the measurement destroys enough entanglement between the two CFTs to disconnect the Lorentzian wormhole. This is again verified by computing the holographic entanglement entropy between the two sides using the RT formula: in the disconnected phase, the mutual information between the two sides vanishes.

We also find that some bulk information that would have been encoded in e.g. the left side if no measurement were performed is accessible from the right side in the presence of measurement. 
In particular, we consider insertions of heavy operators in the Euclidean past in the path integral preparing the state of interest. 
We then study, both in the absence of measurement and when measurements of semi-infinite intervals are performed, whether the bulk effects of such operator insertions can be reconstructed from the left or the right CFT. 
If these effects can be reconstructed from the left in the absence of measurement and from the right in the presence of measurement, we conclude that ``bulk teleportation'' in the sense of \cite{Antonini:2022sfm} is taking place between the two sides. 
If they can be reconstructed from the same side with and without measurement, no bulk teleportation between the two sides is happening. 
Additionally, we find that there is a third possibility, namely that information associated with operator insertions in the Euclidean past can be erased by the measurement, at least in the purely geometrical approximation we work within. 
Our results indicate that both bulk teleportation and information erasure take place in either the disconnected or connected phases, with erasure occurring predominantly in the former, and teleportation occurring predominantly in the latter.

We then confirm the phase transitions between a connected/disconnected bulk via a boundary computation, finding the corresponding entangled/disentangled phase transitions in the microscopic CFT description. Following the framework of \cite{rajabpour2016entanglement}, in Section \ref{sec:cft} we calculate the Reyni entropy between the two CFTs after the measurement is performed. By focusing on the limits where the measured region is either very large or very small\footnote{These limits allow us to answer the question at hand analytically thanks to some technical simplifications arising in the corresponding formulas.} and up to subleading corrections in $N$, we again find that, upon measuring semi-infinite intervals in either one or both CFTs, the entanglement entropy between the two sides can be made zero for a sufficiently small unmeasured region.

The rest of this paper is organized as follows. In Section \ref{sec:bulk}, we define the TFD state and construct the bulk duals of various boundary measurements. Section \ref{sec:finite} outlines the general strategy of constructing the bulk spacetime, and focuses on measuring finite boundary intervals, with an additional Euclidean time evolution performed after the measurement. We then characterize a Hawking-Page transition corresponding to whether or not the dual Lorentzian geometry has an ETW brane, and calculate the entanglement entropy in both phases via the RT formula. In section \ref{sec:infinite-one}, we focus on measurements of two semi-infinite intervals on one side, and again construct the bulk duals. We then characterize a Hawking-Page transition corresponding to whether or not the dual Lorentzian geometry is connected, calculate the entanglement entropy in both phases via the RT formula, and study bulk teleportation and information erasure due to the measurement. 
In section \ref{sec:infinite-two} we repeat the previous analysis for the case of semi-infinite intervals measured in both intervals, and find analogous results. Section \ref{sec:cft} shows the accordance between the bulk holographic entanglement entropy calculation and a Renyi entropy calculation in the dual CFT system for the case of semi-infinite intervals. Finally, we end with a summary and discussion of future directions in section \ref{sec:discussion}. Technical details, including the explicit construction of all conformal transformations employed in this paper, can be found in Appendices \ref{app:a}, \ref{app:b}, \ref{append:transition}, and \ref{append:geodesic}.

\section{Holographic description of measurement in CFT thermofield doubles}\label{sec:bulk}

\begin{figure}
    \centering
    \includegraphics[width=0.5\textwidth]{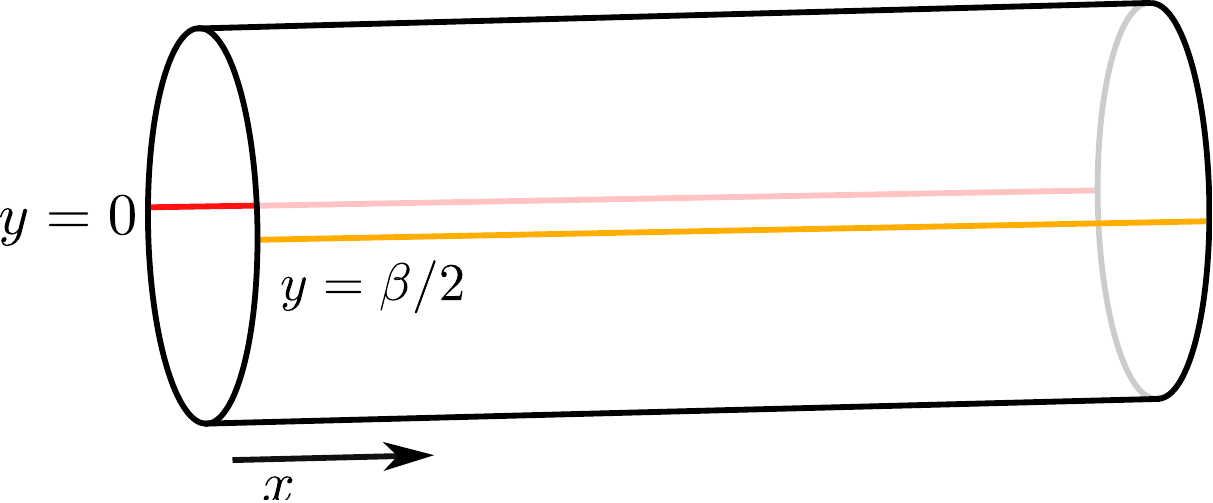}
    \caption{Eulidean state preparation for the TFD state at inverse temperature $\beta$.
    $x \in (-\infty, \infty)$ denotes the spatial coordinate of the CFT, while  the periodic $y\in [0,\beta]$ is the Euclidean time. The TFD state is obtained by slicing open the path integral along the Euclidean time reflection symmetric slice $y=0,\beta/2$. The left and right CFTs are then located at the $y=0$ and $y=\beta/2$ lines, respectively.}
    \label{fig:tfd}
\end{figure}

In this section, we consider projective measurements in a thermofield double state (TFD) of a 1+1 dimensional CFT on a line, and investigate their consequences in the dual spacetime. 
Consider a CFT with Hamiltonian $H$ and a complete set of eigenstates $|n \rangle$, 
\bea
    H | n \rangle = E_n |n \rangle.
\eea
The TFD is a purification of the thermal density matrix $e^{-\beta H}$, where $\beta$ is the inverse temperature. 
It is defined as 
\bea
    |TFD \rangle  = \frac1{\sqrt Z} \sum_{n} e^{- \beta E_n/2 } |n \rangle_L | n \rangle_R, \quad Z = \sum_n e^{-\beta E_n}
\eea
in the doubled Hilbert space $\mathcal H_L \times \mathcal H_R$, where we refer to the two subfactors as the left and right CFT (or side), respectively.
It is easy to check that tracing out either side yields the thermal density matrix, e.g. $e^{-\beta H} = \Tr_{R} \left( |TFD \rangle \langle TFD |\right)$, as expected.
The TFD state can be prepared via Euclidean path integral by slicing open the path integral over an infinite cylinder along the non-compact direction. In particular, with coordinates $x \in (-\infty, \infty)$ and $y\in [0,\beta]$ (where $y=0,\beta$ are identified), the compact $y$  is taken to be the Euclidean time, and $x$ the spatial coordinate along the line. After slicing open the path integral along the $x$-axis, the open cut at $y = 0$ will define a state for the left side, and the open cut at $y= \beta/2$ will define a state for the right side. See  Fig.~\ref{fig:tfd} for a depiction. The bulk geometry dual to the TFD state is an asymptotically AdS eternal black hole, where the left and right CFTs can be thought of as living on the left and right asymptotic boundaries, with the spacetime between them connected by an Einstein-Rosen bridge \cite{Maldacena:2001kr}.

Here, we will consider the postselected state after a partial projective measurement on a subregion $A$ of one or both the CFTs is performed. 
In particular, the postselected state will be a Cardy state~\cite{cardy2004boundary}. 
 In a lattice-discretized version of the CFT, the measurement operator consists of a tensor product of local projective measurements (LPM), $M = \left(\bigotimes_{x \in A} |\psi_x \rangle \langle \psi_x | \right)\otimes \left(\bigotimes_{x \in A^c} \mathds 1_x \right)$, where $|\psi_x \rangle$ is the state each lattice site is projected onto, $\mathds 1$ denotes the identity operator, and $A^c$ is the unmeasured region complementary to $A$. The resulting post-measurement state has zero spatial entanglement in region $A$, a property typical of Cardy states \cite{miyaji2014boundary}; see also \cite{numasawa2016epr,Antonini:2022sfm}.

On the CFT side, this measurement can be described by removing a slit in the Euclidean path integral that prepares the thermofield double state, corresponding to the measured subregion $A$ and having infinitesimal height in Euclidean time.  Similar CFT measurements were considered in~\cite{rajabpour2015post, rajabpour2016entanglement}. Analytical formulas for the entanglement entropy in setups involving one and two slits inserted in the Euclidean path integral preparing a CFT ground state have been obtained and numerically verified for simple CFTs~\cite{rajabpour2015post,rajabpour2016entanglement}. 

In this section, we will focus on the spacetime dual of the post-measurement state $M |TFD \rangle$. The holographic duals of post-selected states were explored in~\cite{numasawa2016epr,Antonini:2022sfm}.  The bulk dual of the measured state can be built using the AdS/BCFT proposal~\cite{takayanagi2011holographic,fujita2011aspects}: in particular, the boundary measurement implies the presence of an end-of-the-world (ETW) brane in the bulk. Note that, in the absence of a proper regularization procedure, the post-measurement CFT state is singular, with a divergent stress-energy tensor at the endpoints of the slit. This translates into a singular bulk spacetime metric and brane configuration. Instead of regularizing,\footnote{This could be done, for example, by giving the slits a finite height via Euclidean time evolution of the post-measurement state. Though it is not completely clear how to implement this, such an approach would yield a well-defined post-measurement state, which is a necessary pre-requisite to study post-measurement Lorentzian time evolution in the CFT, and consequently the bulk dual.} we follow \cite{numasawa2016epr,Antonini:2022sfm} and perform a series of conformal transformations (necessarily containing a branch cut), which yield a regular CFT setup. The regular bulk dual description can then be straightforwardly constructed by means of the AdS/BCFT proposal. In particular, we will map the initial infinite cylinder with slits to a cylinder of finite height, such that the slits are mapped to the two boundaries of the finite cylinder, similar to~\cite{Antonini:2022sfm}. Depending on the parameters of the measurement, the resulting Euclidean bulk dual spacetime is given by either the BTZ black hole or thermal AdS${}_3$ cut off by ETW branes anchored at the boundaries of the finite cylinder.

We will use the slit prescription described above to study measurements performed on both finite and semi-infinite intervals. For the former case, in section \ref{sec:finite} we look at projective measurements for a finite interval in one side of the TFD state as a warm-up. 
While a similar measurement has been investigated in ref.~\cite{numasawa2016epr}, we additionally consider a Euclidean time evolution using the Hamiltonian of the whole CFT following the projective measurement.\footnote{This can be regarded as another form of regularization of the post-measurement setup yielding a regular, well-defined post-measurement state \cite{numasawa2016epr}.}
We will see that this imaginary time evolution leads to a Hawking-Page phase transition: for certain measurement parameters, the bulk time reflection symmetric slice will contain an ETW brane. For the latter (semi-infinite interval) case, in Sections \ref{sec:infinite-one} and \ref{sec:infinite-two} we will be particularly interested in understanding how the Lorentzian Einstein-Rosen bridge, whose existence is associated with the entanglement between the two sides of the thermofield double, is affected by the measurement. For measurements performed on either one or both sides of the TFD, we again identify a Hawking-Page phase transition and show that in this case the Einstein-Rosen bridge is destroyed whenever we are above a phase boundary. This boundary is determined by the size of the measured region,  the temperature of the TFD state, and the brane tension. Explicitly, this destruction is signaled by the vanishing of the mutual information between the remaining, unmeasured regions on the two sides.

We further characterize how information associated with insertions of heavy operators in the Euclidean past and which would be encoded in one of the two CFTs in the absence of measurement can become accessible from the other CFT or erased by the measurement. We call the first case ``teleportation'' in the sense of the ``bulk teleportation'' studied in \cite{Antonini:2022sfm}, which in this case would take place between the two sides. In Section \ref{sec:1sidedteleportation} we analyze these phenomena when two semi-infinite intervals are measured on the left side and in Section \ref{sec:2sidedteleportation} when one semi-infinite interval is measured in each of the two CFTs.

In all our measurement configurations we will assume that the two measured intervals are projected onto the same Cardy state. This guarantees that the associated ETW branes have the same tension $T$. In the phase where a single connected brane is present (in the following we will call this the thermal AdS phase), our assumption ensures that the brane is smooth and does not have any defects. A generalization to the case where the two intervals are projected onto different Cardy states requires a treatment of non-smooth intersections among the branes along the lines of \cite{Miyaji:2022dna}; we leave such an analysis to future work.

\subsection{Finite intervals}\label{sec:finite}

\subsubsection{Slit prescription}

\begin{figure}
    \centering
    \subfigure[]{
    \includegraphics[width=0.4\textwidth]{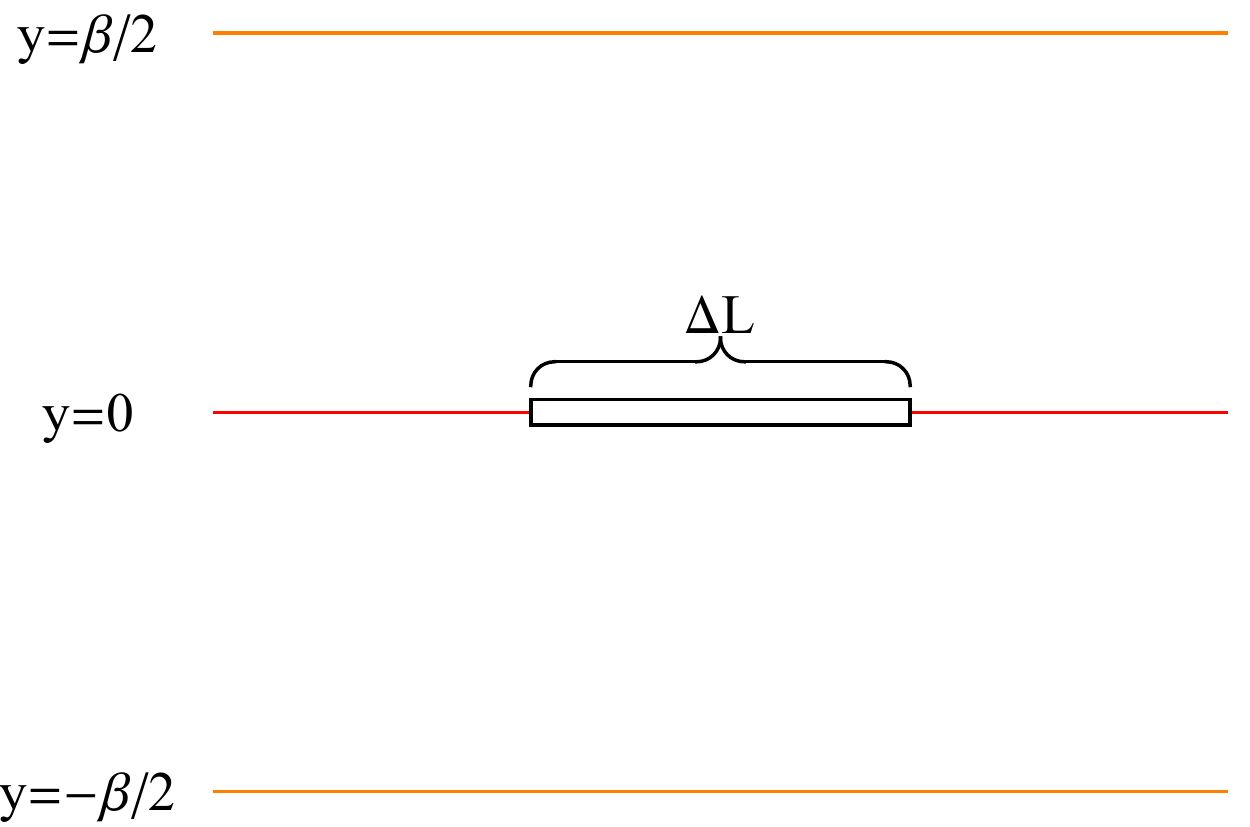}} \quad \quad \quad 
    \subfigure[]{
    \includegraphics[width=0.4\textwidth]{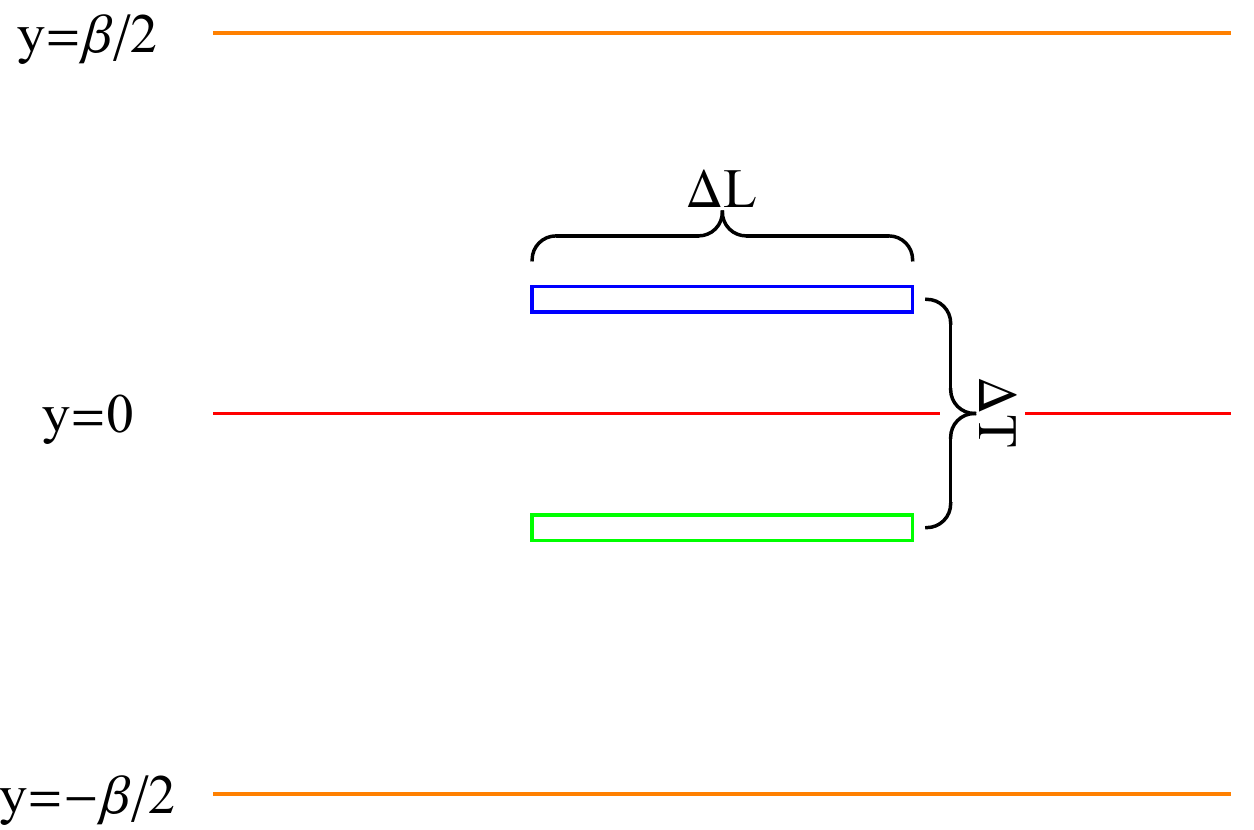}}
    \caption{The Euclidean state preparation for a TFD state with a partial projective measurement in region $A$. The red and orange lines denote the left ($y=0$) and right ($y=\beta/2$) CFTs.
    They are referred to as the time reflection symmetric lines. Note that because $y=y+\beta$, the two orange lines $y=\pm \beta/2$ are identified. (a) The removed slit region corresponding to the measured region $A$ is indicated by the black rectangle with length is $\Delta L$. (b) The setup involving projective measurement along a region $A$ of size $\Delta L$ followed by an amount $\Delta T/2$ of Euclidean time evolution. Accordingly, in the reflection symmetric Euclidean path integral there are two parallel slits separated by an imaginary time interval $\Delta T$. }
    \label{fig:measure_imagine}
\end{figure}

In this section, we focus on local projective measurements on a finite interval $A$ of length $\Delta L$ in the left CFT.

As mentioned above, the states of interest will be prepared by evolving the post-measurement state using the full CFT Hamiltonian for Euclidean time $\Delta T/2$. This regularization procedure can be thought of as starting from the single-slit path integral of Fig.~\ref{fig:measure_imagine} (a)---which describes the singular state resulting from the insertion of the projection operator---and splitting this into the two-slit Euclidean path integral depicted in Fig.~\ref{fig:measure_imagine} (b),\footnote{The additional time evolution in the finite interval case here will allow us to obtain a phase transition analogous to the one we will encounter in the semi-infinite interval case. There, however, the physical consequences will be more drastic.}. This prepares the regular state
\bea
    | \Psi \rangle \propto e^{- H_L \Delta T/2} M e^{H_L \Delta T/2}|TFD \rangle.
\eea
We take the two slits to be located at
\bea \label{eq:slit-1}
    && \text{first slit:} \quad -\frac{\Delta L}2 < x < \frac{\Delta L}2, \quad y = - \frac{\Delta T}2, \\
    && \text{second slit:} \quad -\frac{\Delta L}2 < x < \frac{\Delta L}2, \quad y = \frac{\Delta T}2.
\eea
To fully define the measurement, in addition to the two parameters $\Delta L$ and $\Delta T$, it is necessary to specify the exact Cardy state we project onto. This uniquely determines the boundary entropy of the post-measurement state \cite{affleck1991universal,cardy2004boundary}, and in turn the tension of the ETW brane \cite{takayanagi2011holographic,fujita2011aspects}. 

\begin{figure}
    \centering
\subfigure[$x$ coordinate]{
    \includegraphics[width=0.35\textwidth]{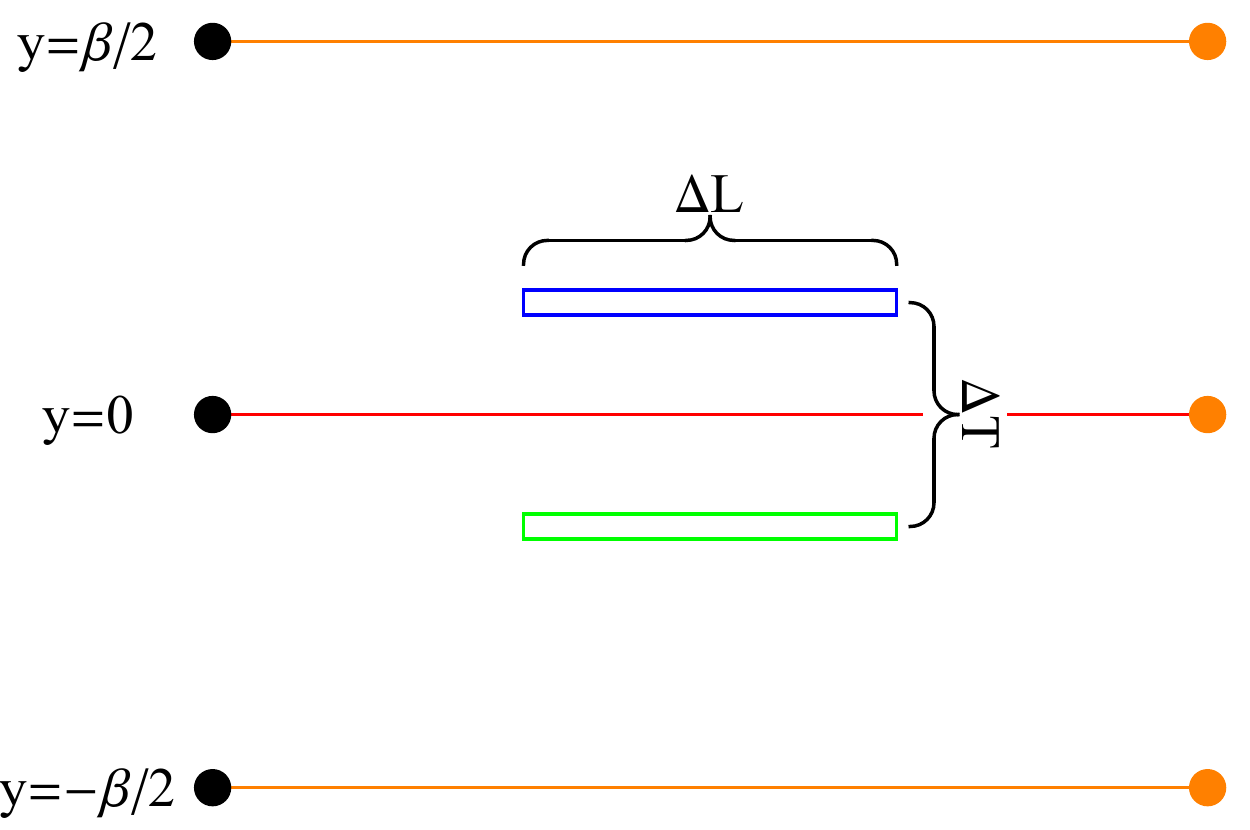}} \quad \quad \quad \quad \quad \quad
\subfigure[$X$ coordinate]{
    \includegraphics[width=0.35\textwidth]{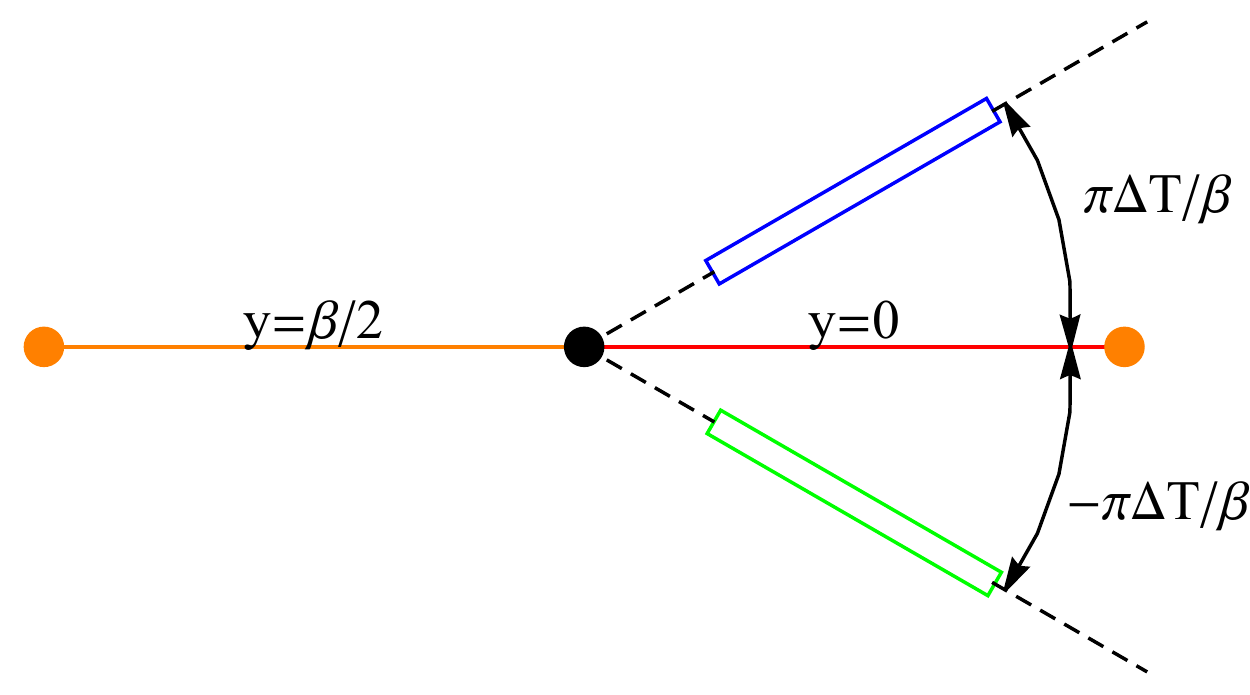}} \\\vspace{5mm}
\subfigure[$\zeta$ coordinate]{
    \includegraphics[width=0.35\textwidth]{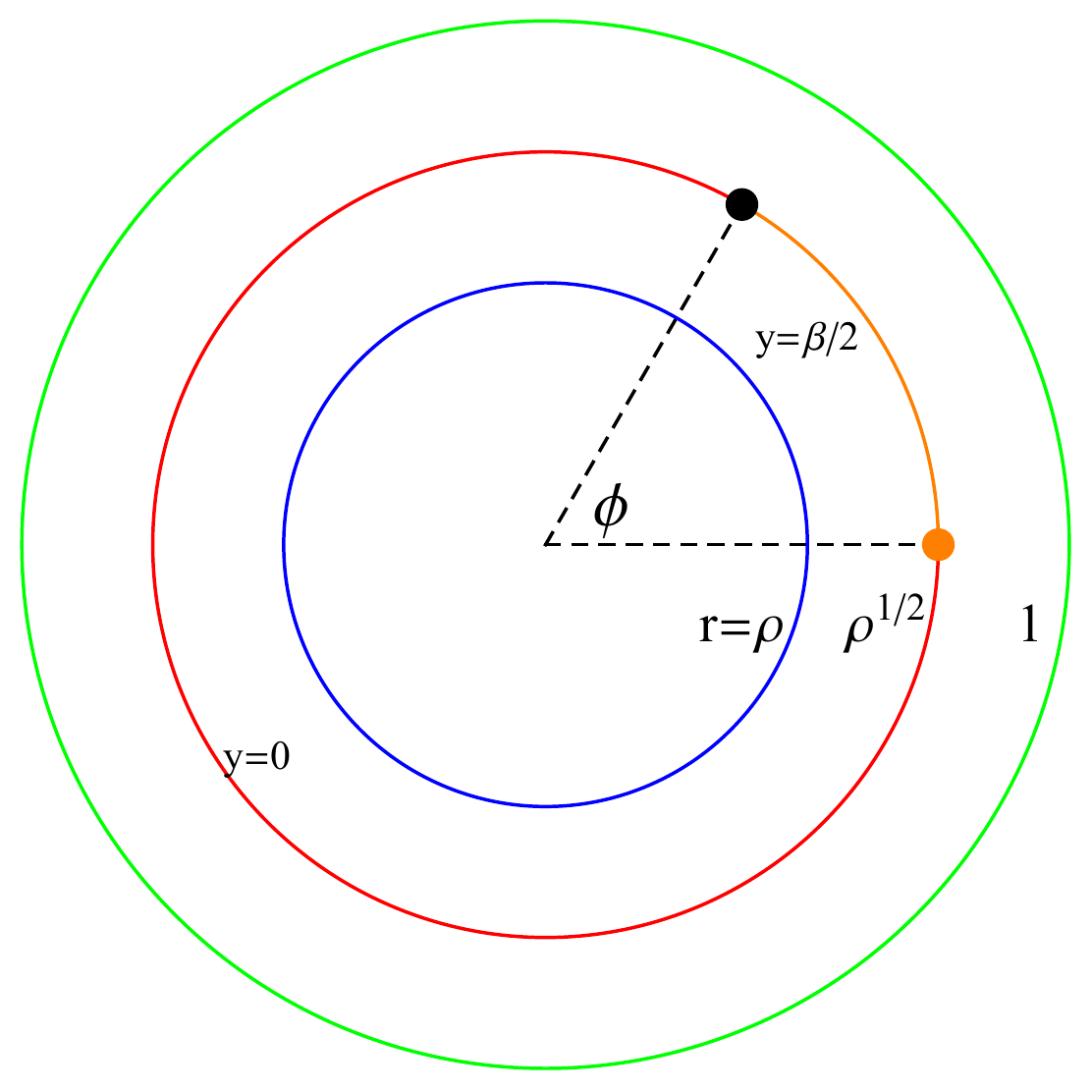}}
    \quad \quad \quad\quad \quad \quad
\subfigure[$w$ coordinate]{
    \includegraphics[width=0.35\textwidth]{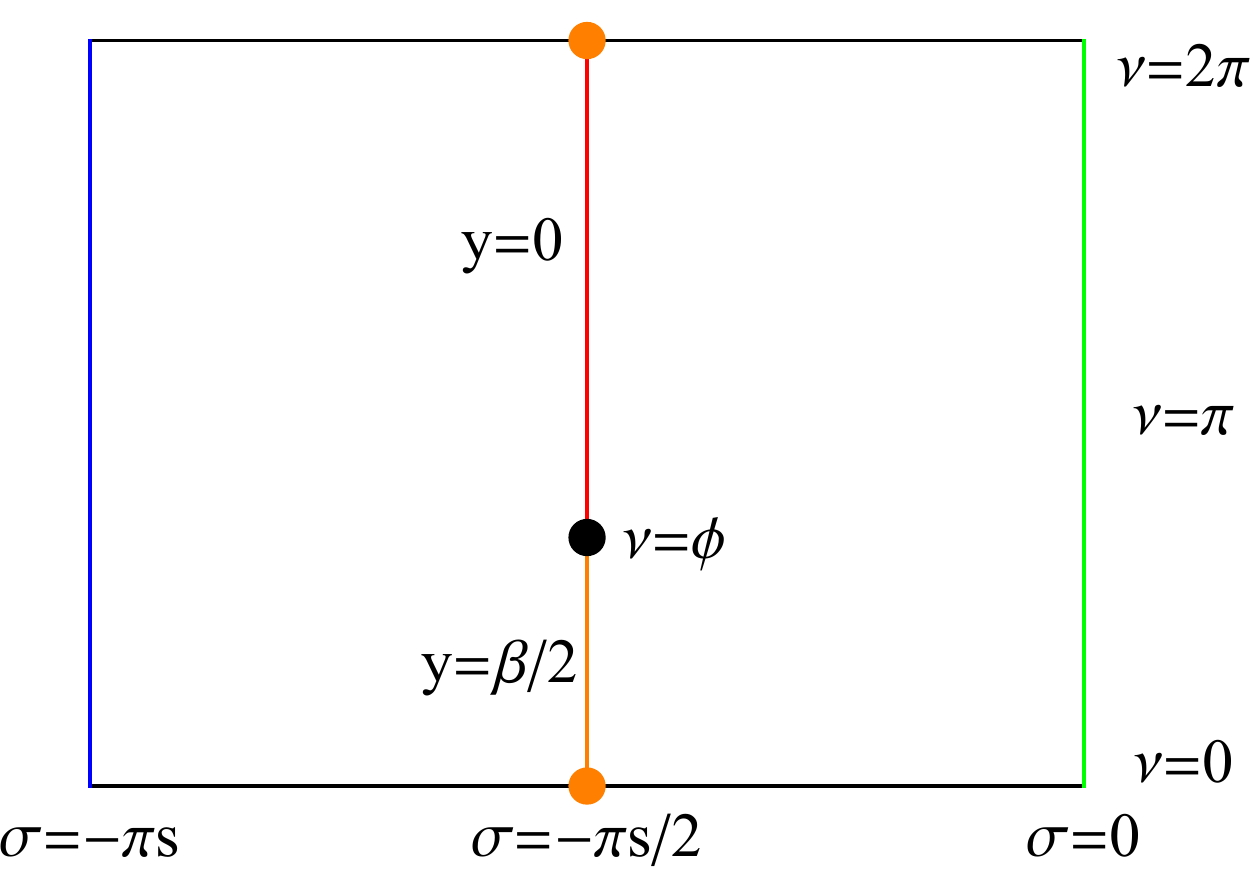}}
    \caption{The four different coordinates we use to describe measurement of finite intervals on one side of a TFD. The conformal maps between the different coordinates are given in Appendix \ref{sec:finite_conformal}. The red (orange) line denotes the left CFT at $y=0$ (right CFT at $y=\beta/2$). They are time reflection symmetric lines. The green (blue) color denotes the first (second) slit. The black (orange) dots represent spatial negative (positive) infinity $x = - \infty$ ($x=\infty$) in both left and right CFTs.
    (a) In the original coordinates, we have an infinitely long cylinder with two slits, representing a measurement  described by two parameters $\Delta L$, $\Delta T$. Note $y=y+\beta$, so the top and bottom lines are identified. 
    (b) We map this to a 2D plane with two radial slits. The angle of the slits are at $\pm\phi/2 = \pm \pi \Delta T /\beta$, with the first slit at negative $\phi$. The time reflection symmetric lines are mapped to the horizontal line, where the left CFT is mapped to the positive region, while the right is mapped to the negative region. Finally, negative spatial infinity is mapped to origin, and positive infinity is mapped to real infinity.
    (c) We next map to an annulus. The first slit maps to the outer edge $r= 1$ and the second slit to the inner edge $r= \rho$. The time reflection symmetric lines are mapped to the circle with radius $r= \sqrt \rho$, where the left CFT is mapped to the arc $\theta \in (\phi, 2\pi)$, and the right to $\theta \in (0, \phi)$. $x=-\infty$ is mapped to $r= \sqrt\rho, \theta = \phi$ (and $x=\infty$ is mapped to $r=\sqrt \rho, \theta = 0$). Here we use $\zeta = r e^{i \theta}$, and $\rho, \phi$ are two constants determined by the slits parameter $\Delta L, \Delta T$.
    (d) Finally, we map to a cylinder with $\nu = \nu + 2\pi$. The first  slit is mapped to the right edge at $\sigma =0$, and the left at $\sigma = -\pi s$). The time reflection symmetric lines are mapped to the circle $\sigma = -\pi s/2$, where the left CFT is mapped to the segment $\nu \in (\phi, 2\pi)$ and the right to $\nu \in (0,\phi)$. Finally, $x=-\infty$ is mapped to $\sigma=-\pi s/2, \nu = \phi$ and $x=\infty$ is mapped to $\sigma=-\pi s/2, \nu = 0,2\pi$).}
   \label{fig:map-finite}
\end{figure}

To study the dual spacetime, we implement a series of conformal transformations depicted in Fig.~\ref{fig:map-finite}. The details of the conformal transformations are reported in Appendix \ref{sec:finite_conformal}.
The final domain is given by a finite cylinder described by the complex coordinate $w=\sigma + i\nu$, with $\nu\sim \nu +2\pi$ and $\sigma\in [-\pi s,0]$. The composed conformal transformation leading from the original coordinates $(x,y)$ (Fig.~\ref{fig:map-finite} (a)) to the final coordinates $(\sigma,\nu)$ (Fig.~\ref{fig:map-finite} (d)) is given by
\bea
    x + i y = \frac{\beta}{2\pi} \log \left[ e^{-i \phi} \frac{\theta_4\left(\frac{-i (\sigma +i\nu) - \phi}2, e^{-\pi s/2}\right)}{\theta_4\left(\frac{-i (\sigma + i\nu)}2, e^{-\pi s/2}\right)} \right],
    \label{eq:totalconffinite}
\eea
where $\phi=2\pi\Delta T/\beta$ and $s$ is a parameter related to the ratio $\Delta L/\Delta T$, see Fig. \ref{fig:rho-finite} and equations (\ref{eq:rho}), (\ref{eq:rho2}), and (\ref{eq:s_finite}) in Appendix \ref{sec:finite_conformal}.
Under the conformal transformation (\ref{eq:totalconffinite}), the time reflection symmetric lines are mapped to the $\sigma=-\pi s/2$ circle, with the left CFT ($y=0$) given by the segment $\nu\in (\phi,2\pi)$ and the right CFT ($y=\beta/2$) by $\nu\in (0,\phi)$. Negative spatial infinity $x= -\infty$ is mapped to $\nu=\phi$, while positive spatial infinity $x=\infty$ is mapped to $\nu=0$. Finally, the first  slit is mapped to the $\sigma=0$ circle, and the second slit to $\sigma=-\pi s$. 

\begin{figure}
    \centering
    \includegraphics[width=0.5\textwidth]{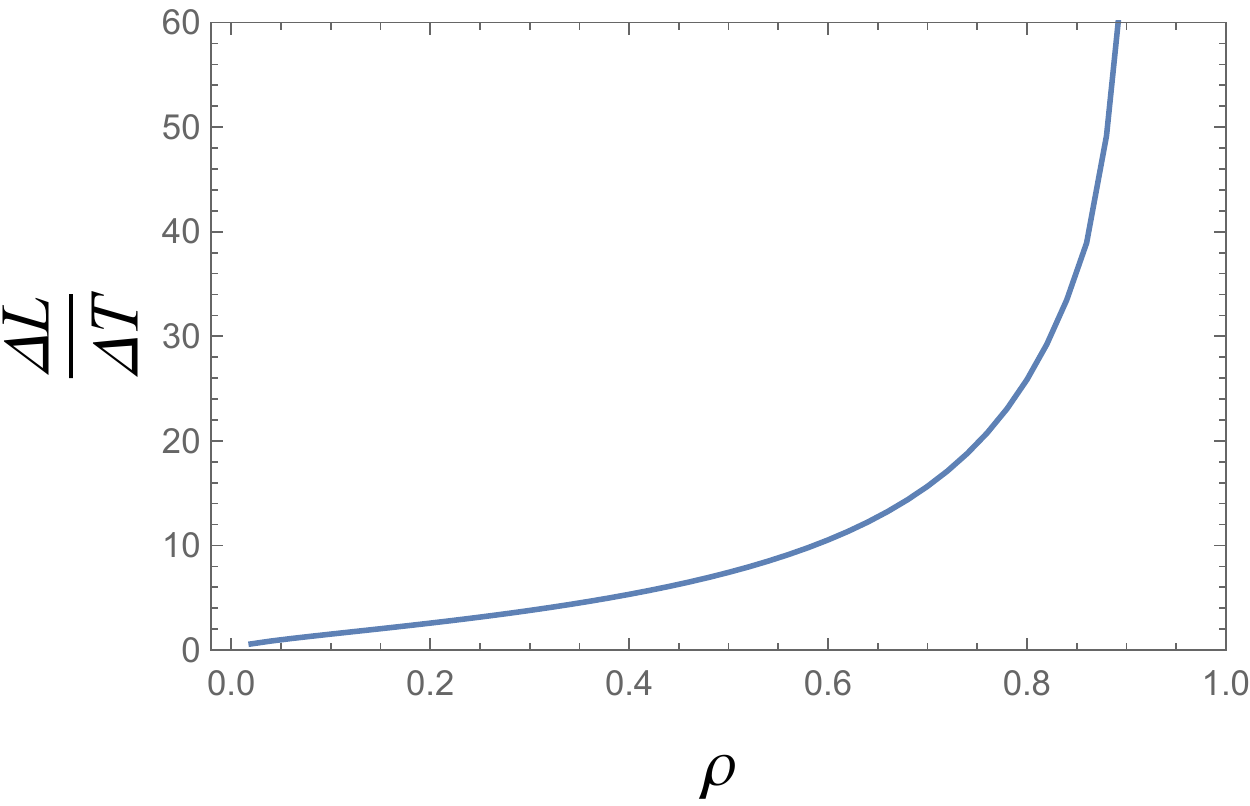}
    \caption{The ratio of measurement parameters $\frac{\Delta L}{ \Delta T}$ as a function of the parameter in our conformal transformation (equation \ref{eq:totalconffinite}) $\rho=e^{-\pi s}$ as determined by equations (\ref{eq:rho}) and (\ref{eq:rho2}) in Appendix \ref{sec:finite_conformal}. In this plot we choose $\phi=0.2$ as a representative example. }
    \label{fig:rho-finite}
\end{figure}

\begin{figure}
    \centering
    \subfigure[]{\includegraphics[width=0.4\textwidth]{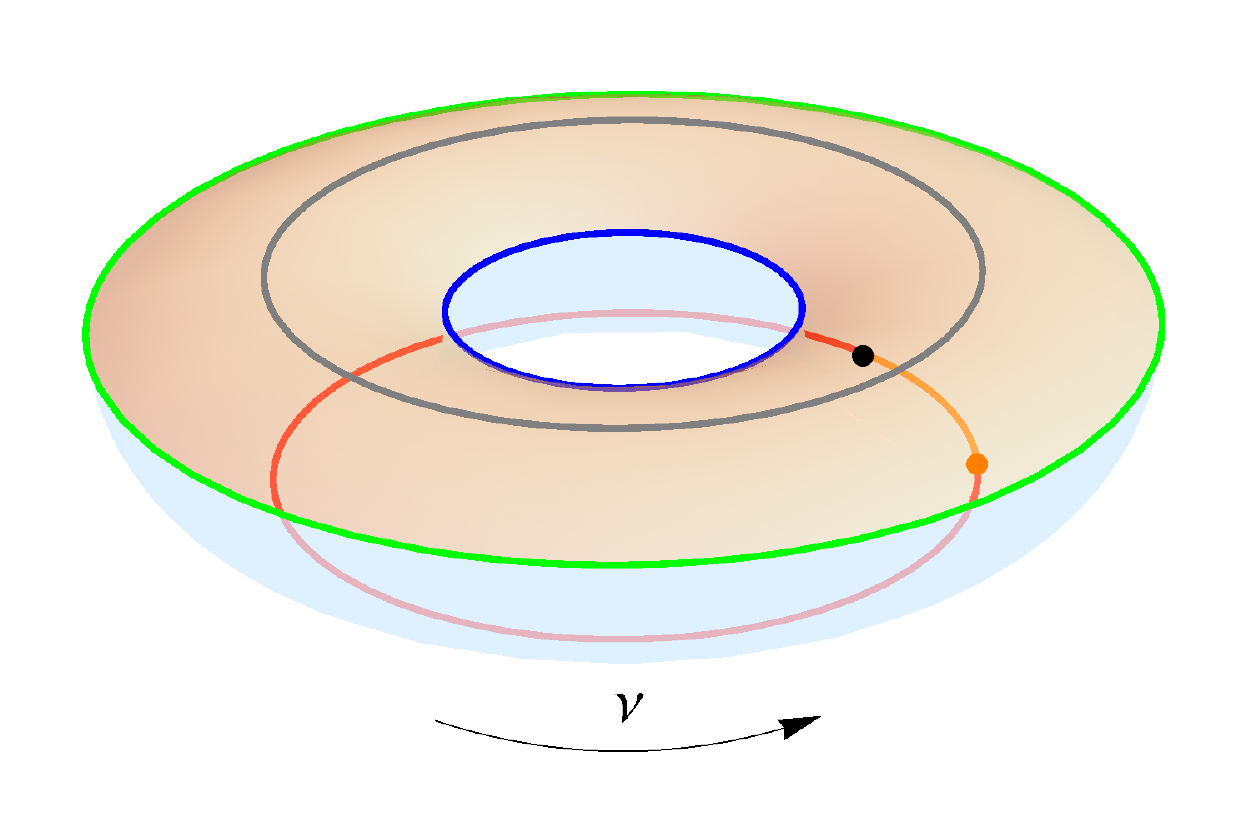}} \quad \quad
    \subfigure[]{\includegraphics[width=0.5\textwidth]{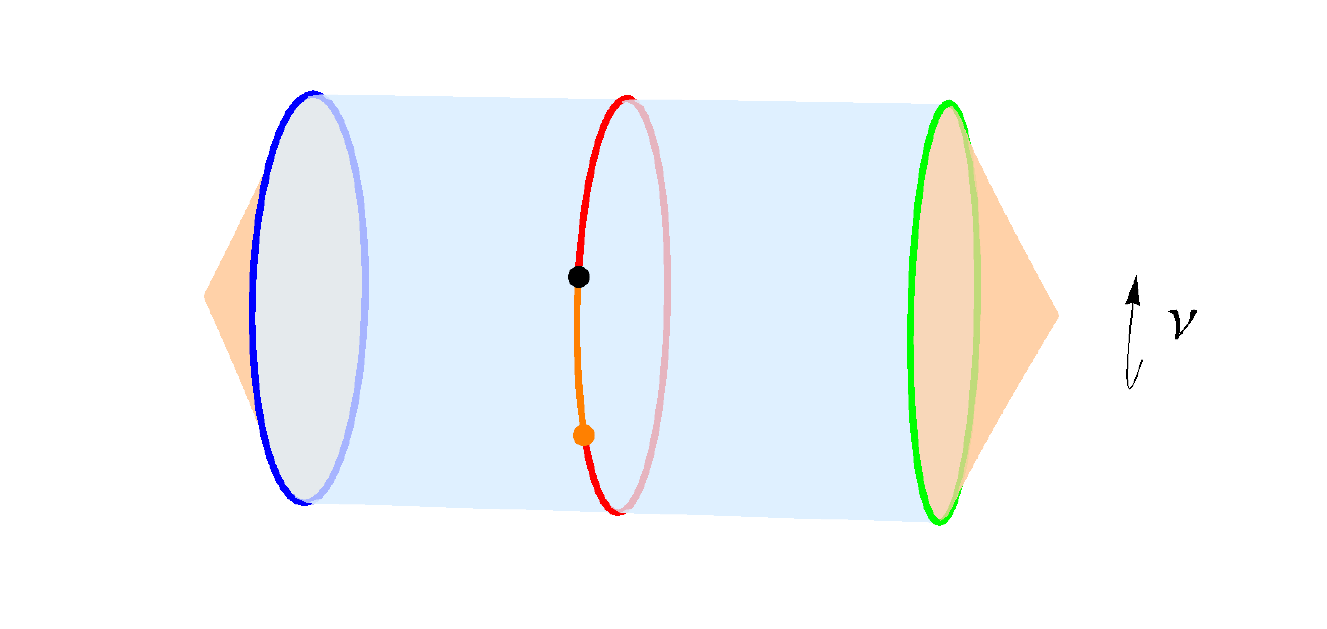}}
    \caption{Gravity dual of the measured CFT. The color convention is the same as in Fig.~\ref{fig:map-finite}. 
    The red (orange) line denotes the left CFT at the original coordinate $y=0$ (right CFT at $y=\beta/2$). They are time reflection symmetric lines. The green (blue) color denotes the first (second) slit. 
    The black (orange) dots represents spatial negative (positive) infinity in the original coordinate $x$ in both CFTs. 
    The light-blue (light-orange) surface denotes the asymptotic boundary (brane).
    $\nu \sim \nu + 2\pi$ is periodic.
    (a) BTZ black hole phase. The gray circle denotes the intersection between time reflection symmetric slice and the brane.
    (b) Thermal AdS phase. 
    There are two branes ending at $\sigma = -\pi s$ and $\sigma =0$.}
    \label{fig:gravity-dual}
\end{figure}

\subsubsection{Bulk spacetime and Hawking-Page transition}
\label{sec:transition}

Having mapped our original two slit system to the finite cylinder in $w$ coordinates, we are ready to build the holographic bulk spacetime using the AdS/BCFT duality~\cite{takayanagi2011holographic,fujita2011aspects}. 
There are two candidate phases, with the resulting euclidean spacetime described by either a portion of a BTZ black hole or of thermal AdS.  The bulk analysis performed in this section is analogous to the one carried out in \cite{Antonini:2022sfm}, although the physical interpretation of the results will be different, as the CFT setups under consideration are different. 

We consider the usual Euclidean action
\bea \label{eq:action}
    I = - \frac1{16\pi G_N} \int_{\mathcal M} \sqrt{g} (\mathcal R-2\Lambda) - \frac1{8\pi G_N} \int_{\partial \mathcal M} \sqrt{h} (K-T),
\eea
where $\Lambda = - \frac1{R^2}$ is the cosmological constant.
The second term is a boundary term on the branes, where $K$ is the trace of extrinsic curvature.
We impose Neumann boundary conditions for the brane, where $T$ is the tension.
The conventional Gibbons-Hawking-York term for the asymptotic conformal boundary is not reported in the action.


We first consider the BTZ black hole phase\footnote{\label{foot:nomenclature} Note that this nomenclature is convenient but arbitrary: it depends on which coordinate ($\sigma$ or $\nu$) we identify with the Euclidean time. Here we choose $\sigma$, implying that when the $\sigma$ circle is contractible we call the corresponding spacetime a BTZ black hole. When the $\nu$ circle is contractible we call the corresponding spacetime thermal AdS. The opposite choice leads to the opposite nomenclature. Neither choice is based on physical grounds: if we were to analytically continue the bulk geometry to study the Lorentzian version of our setup, neither $\sigma$ nor $\nu$ would be identified as Euclidean time. In fact, to study the evolution of the post-measurement state in our CFT we should instead analytically continue the original $y$ coordinate. Notice that in this paper we adopted the opposite nomenclature with respect to \cite{Antonini:2022sfm}.} with metric
\bea\label{eq:metric_BTZ}
    ds^2 = \frac{R^2}{z^2} \left( \frac{dz^2}{h(z)} + h(z) d\sigma^2  +  d\nu^2 \right), \quad h(z) = 1- \frac{z^2}{z_H^2},
\eea
where $z=0$ corresponds to the asymptotic boundary, $z_H$ is the length of the horizon (related to the temperature of the BTZ black hole by $\beta_H = 2\pi z_H$), and the periodicity of $\sigma$ is given by $2\pi z_H$ (further recall the periodicity of $\nu$ is $2\pi$). See Fig.~\ref{fig:gravity-dual} (a) for a depiction. 
Note that in the BCFT, $\sigma$ has a finite range $\sigma\in [-\pi s,0]$, where the two boundaries at $\sigma=-\pi s,0$ correspond to the two slits in our original coordinates $(x,y)$.

\begin{figure}
    \centering
    \subfigure[]{\includegraphics[width=0.4\textwidth]{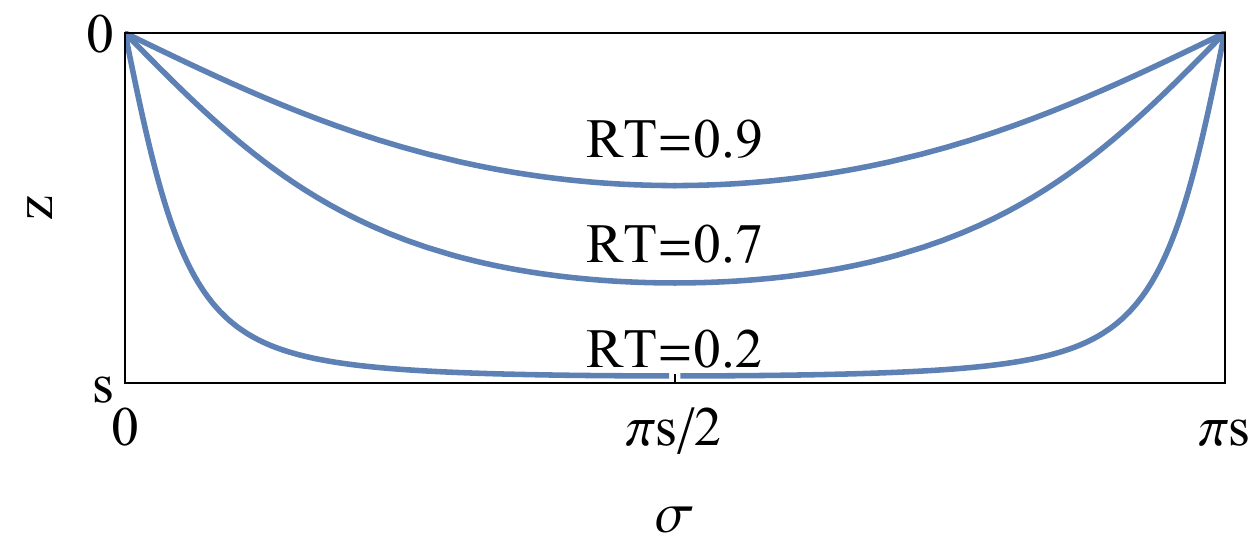}} \quad \quad \quad
    \subfigure[]{\includegraphics[width=0.4\textwidth]{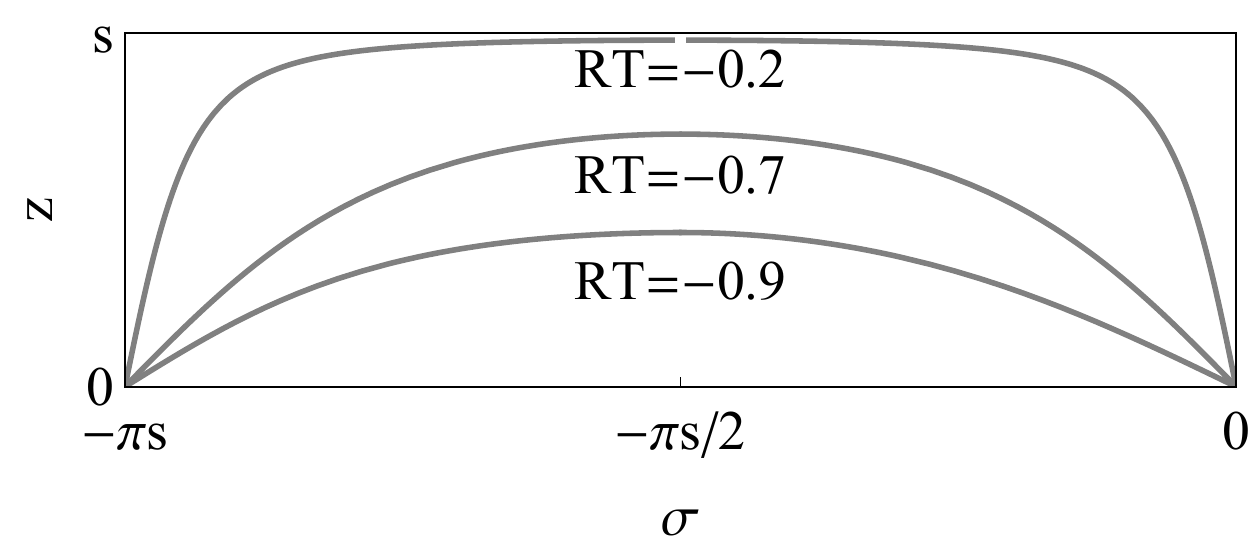}} \\ 
    \subfigure[]{\includegraphics[width=0.4\textwidth]{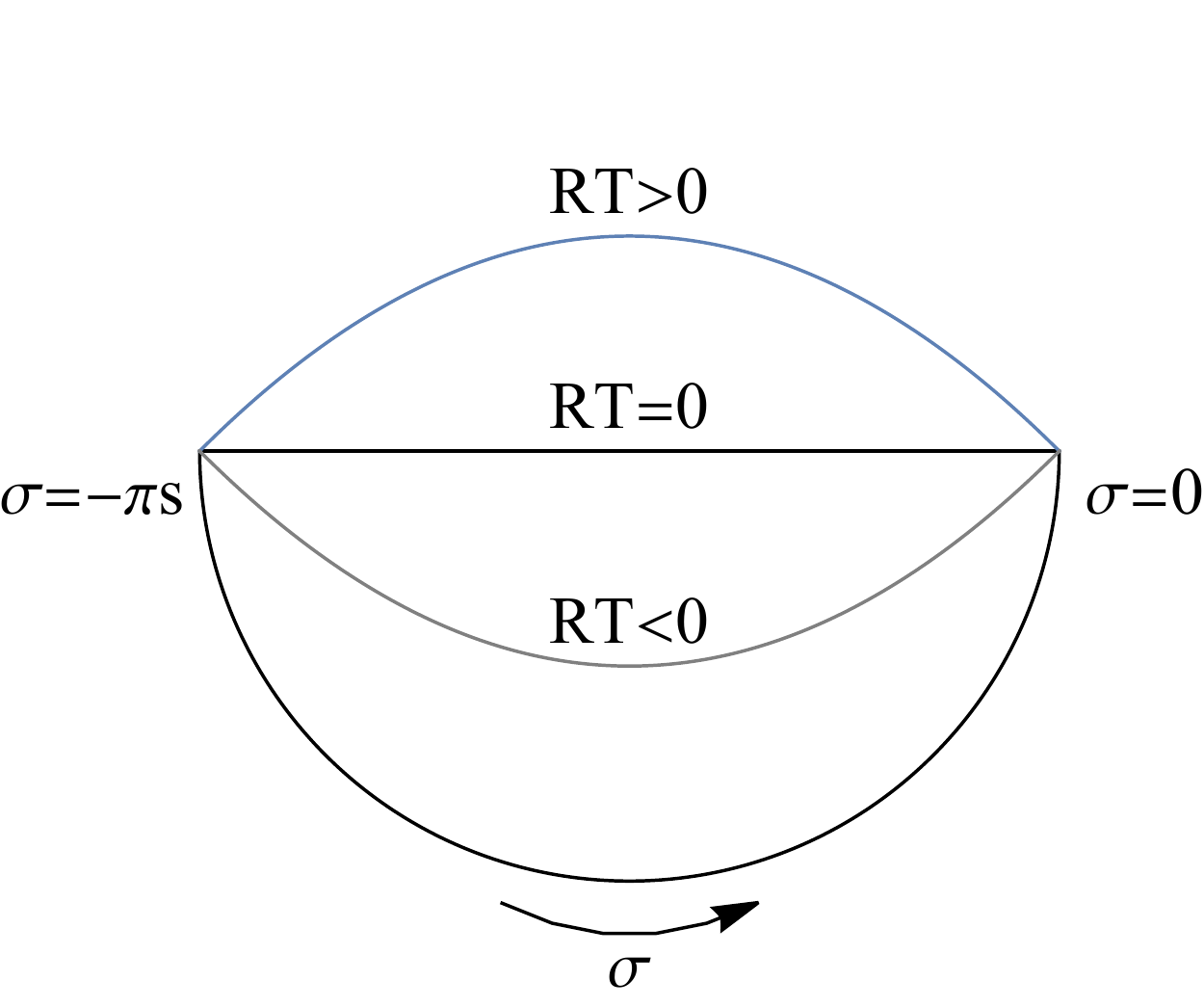}}
    \caption{Brane trajectories projected on the $(\sigma,z)$ plane. Note that the brane trajectories do not depend on $\nu$. 
    (a) Brane trajectories from (\ref{eq:brane_BTZ}) for $RT = 0.9, 0.7,0.2$. 
    (b) Brane trajectories from (\ref{eq:brane_BTZ2}) for $RT = -0.9, -0.7, -0.2$. 
    (c) Schematic trajectories of branes with different tensions. The bottom semi-circle $\sigma \in (-\pi s, 0)$ is the asymptotic boundary.
    }
    \label{fig:brane_BTZ}
\end{figure}

According to the AdS/BCFT prescription, the ETW brane must anchor at the asymptotic boundary $z=0$ at $\sigma = 0$ and $\sigma = - \pi s$.
The equation of motion for the brane yields a trajectory $(\sigma(z),z,\nu)$, which for or $T\in (0, 1/R)$ gives
\bea \label{eq:brane_BTZ}
    \sigma(z) = \begin{cases} s \tan^{-1} \left[ \frac{RTz}{s \sqrt{h - R^2 T^2}} \right], & \quad  0 < \sigma < \pi s/2 \\ \\
     s \left( \pi - \tan^{-1} \left[ \frac{RTz}{s \sqrt{h - R^2 T^2}} \right] \right), & \quad  \pi s/2 < \sigma < \pi s.
    \end{cases}
\eea
and for $T\in (-1/R, 0)$\footnote{Although mathematically sound from a bulk point of view in a bottom-up approach to holography, negative tension branes violate standard energy conditions \cite{Barcelo:2000ta}
and it is unclear whether they correspond to physically reasonable setups in the context of AdS/BCFT. Note, however, there are various top-down, explicit constructions of negative tension objects within string theory (and that thus evade any potential pathologies); see e.g. \cite{Marolf:2002np,Burgess:2002vu} for related discussions.},
\bea \label{eq:brane_BTZ2}
    \sigma(z) = \begin{cases}  s \tan^{-1} \left[ \frac{RTz}{s \sqrt{h - R^2 T^2}} \right], & \quad  -\pi s/2 < \sigma < 0 \\ \\
     s \left( -\pi - \tan^{-1} \left[ \frac{RTz}{s \sqrt{h - R^2 T^2}} \right] \right), & \quad  -\pi s < \sigma < -\pi s/2,
    \end{cases}
\eea
where we used the relationship $z_H = s$, which is determined by the requirement that the brane anchors at $\sigma=-\pi s$ and $\sigma=0$ at the boundary. See Appendix~\ref{append:brane-BTZ} for details, and in particular a derivation of equations (\ref{eq:brane_BTZ}) and (\ref{eq:brane_BTZ2}). 
Note that the periodicity of $\sigma$ at the boundary is then given by $\sigma \sim \sigma + 2\pi s$, i.e. twice the range of the coordinate. At $z_* = s \sqrt{1 - R^2 T^2} $ and $\sigma(z_*) = \frac{\pi}2 s$, $\sigma'(z_*) \rightarrow \infty $, so that there is a turning point for the brane trajectory. 
When $RT = 0$, the brane cuts off exactly half of the bulk BTZ geometry and intersects the horizon $z = s$, while for $RT \rightarrow \pm 1$, the brane locates near the boundary $z = 0$. For a positive tension brane, the brane will bulge outwards such that the retained part of the geometry is larger than half of the BTZ geometry, and it contains the horizon $z=s$. For negative tension, however, the remaining part of the spacetime is less than half, and in particular does not contain the horizon. More concretely, for a positive tension brane (i.e. when $RT>0$), the bulk region contains $0<z< s$ for $\sigma \in (-\pi s,0)$ and $z(\sigma)<z<s$ for $\sigma \in (0, \pi s)$, where $z(\sigma)$ is the inverse function of (\ref{eq:brane_BTZ}).
For negative tension ($RT<0$), the bulk region contains $0<z<z(\sigma)$ for $\sigma \in (-\pi s,  0)$, where $z(\sigma)$ is the inverse function of (\ref{eq:brane_BTZ2}).
See Fig.~\ref{fig:brane_BTZ}.

\begin{figure}
    \centering
    \includegraphics[width=0.8 \textwidth]{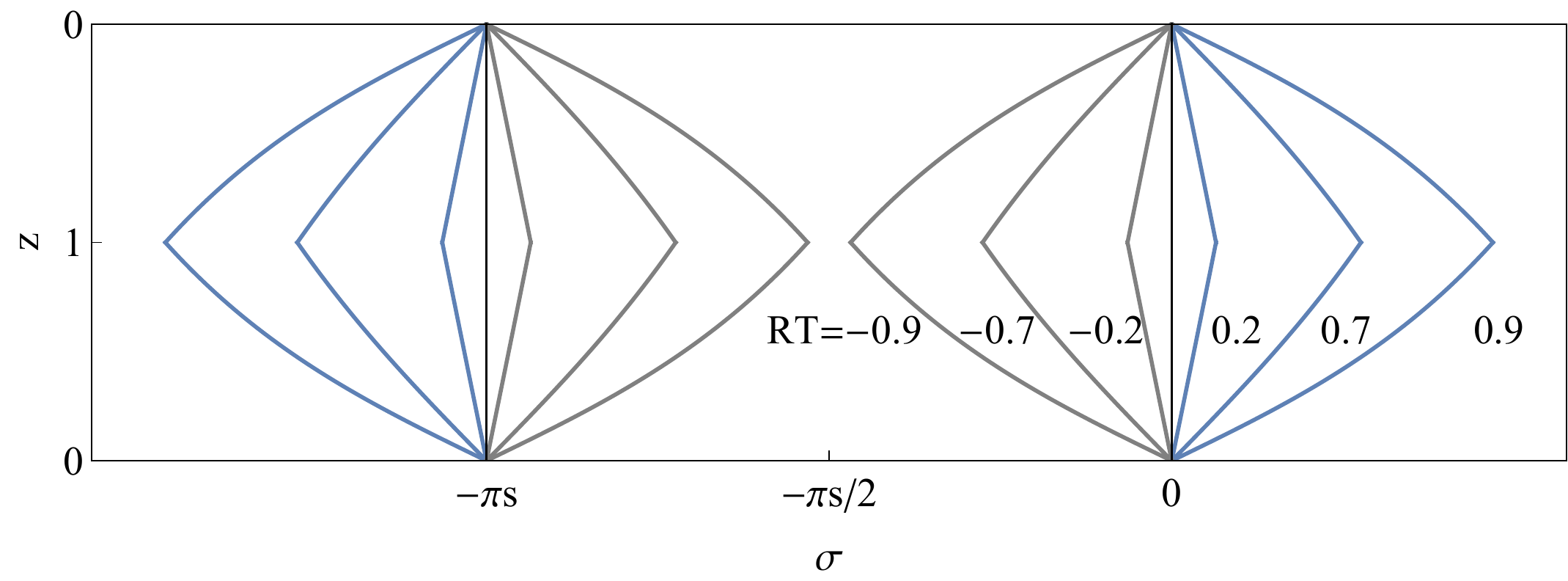}
    \caption{Brane trajectories projected on the $(\sigma,z)$ plane with fixed $\nu$ and $\nu + \pi$. Note that the brane trajectory does not depend on $\nu$. 
    The trajectories are given by (\ref{eq:brane_AdS}) for $RT = \pm 0.9, \pm 0.7,\pm 0.2$. }
    \label{fig:brane_AdS}
\end{figure}

Next, we consider the thermal AdS\footnote{Again, we emphasize that this nomenclature is arbitrary and opposite with respect to the one used in \cite{Antonini:2022sfm}, see footnote \ref{foot:nomenclature}.} phase, see Fig.~\ref{fig:gravity-dual} (b). 
In this case, the roles of $\sigma$ and $\nu$ switch, such that the metric becomes 
\bea
    ds^2 = \frac{R^2}{z^2} \left( \frac{dz^2}{f(z)} + f(z) d\nu^2  +  d\sigma^2 \right), \quad f(z) = 1- z^2. 
\eea
where the periodicity of $\nu \sim \nu + 2\pi$ fixes the maximum value of $z$ to be $z=1$. Again, the brane must anchor at the boundary $z=0$ at $\sigma = 0$ and $\sigma = - \pi s$. 
In the thermal AdS phase, there are two disconnected ETW branes whose trajectories are now given by
\bea \label{eq:brane_AdS}
    \sigma(z) = \begin{cases}
        \sinh^{-1}  \left( \frac{RTz}{\sqrt{1-R^2T^2}} \right), & \quad \sigma > -\pi s/2,  \\ \\
        -\pi s - \sinh^{-1}  \left( \frac{RTz}{\sqrt{1-R^2T^2}}  \right), & \quad \quad \sigma < -\pi s/2,
    \end{cases}
\eea
where the first line corresponds to the brane anchored at $\sigma = 0$, and the second corresponds to the brane anchored at $\sigma = -\pi s$. See Appendix~\ref{append:brane_AdS} for a derivation.
The retained bulk geometry is given by $\sigma \in \left(-\pi s - \sinh^{-1}  \left( \frac{RTz}{\sqrt{1-R^2T^2}},  \right) ,\sinh^{-1}  \left( \frac{RTz}{\sqrt{1-R^2T^2}} \right) \right) $, $z\in (0,1]$, and $\nu\in [0,2\pi]$---see Fig.~\ref{fig:gravity-dual} (b). 
We plot several brane trajectories in Fig.~\ref{fig:brane_AdS} for different tensions, both positive and negative.  
As above in the BTZ phase, here in the thermal AdS phase the remaining bulk region is larger for greater tension. 
Moreover, if the tension is sufficiently negative, the two branes anchored at the boundary at $\sigma = -\pi s$ and $\sigma = 0$ will intersect. 
To avoid the complication of this possibility, we restrict the tension to be $T\in (-T_*, 1/R)$, where the minimal tension $T_*$ is implicitly defined through
\bea
    \sinh \frac{\pi s}2 = \frac{RT_*}{\sqrt{1 - R^2 T^2_*}}.
    \label{eq:limitension}
\eea

\begin{figure}
    \centering
    \includegraphics[width=0.5 \textwidth]{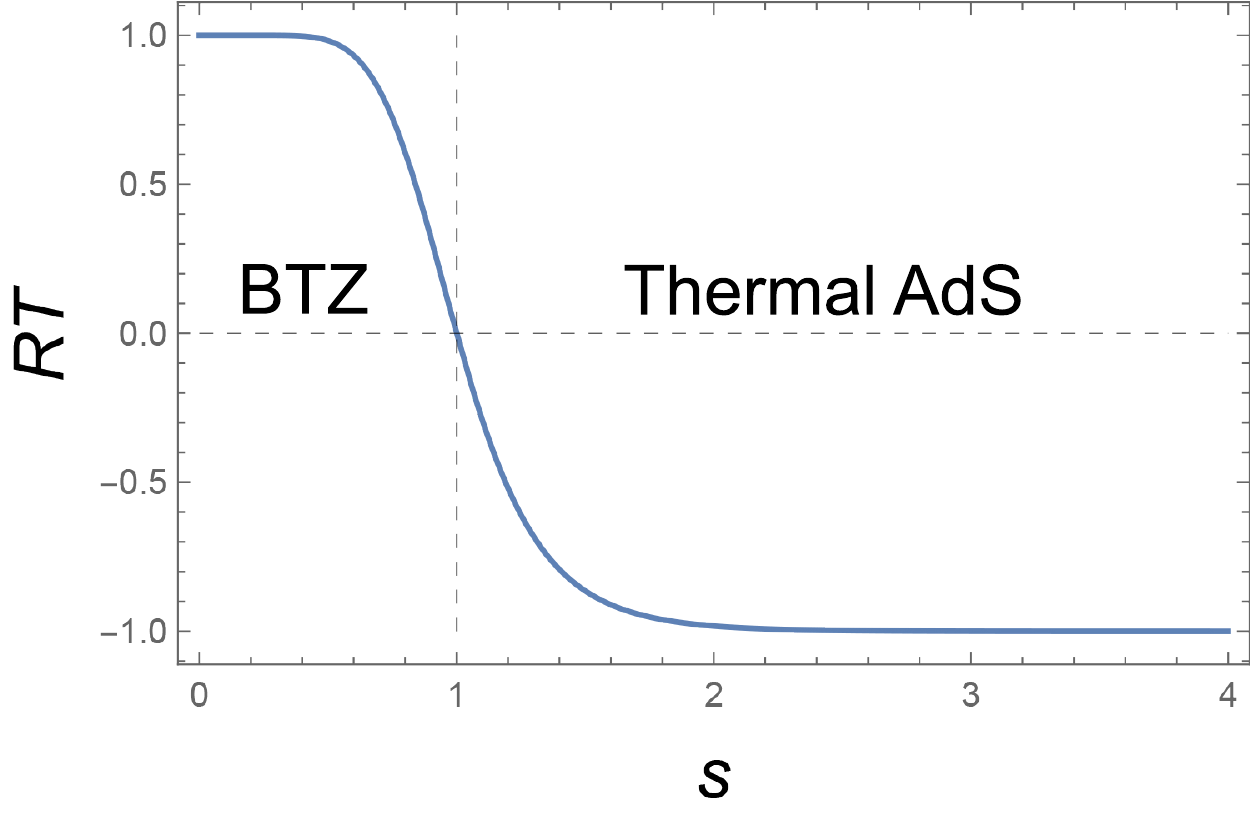}
    \caption{The phase diagram in the $s$-$RT$ plane.}
    \label{fig:transition}
\end{figure}

In order to find whether the BTZ or thermal AdS phase is the dominant one in the Euclidean gravitational path integral, we can evaluate the Euclidean action (\ref{eq:action}) on-shell to find the phase with the least action \cite{fujita2011aspects}. The details of the action evaluation are left to Appendix~\ref{append:transition}, though we summarize the results here. The phase boundary between the BTZ black hole and thermal AdS is given by a curve in the $s$-$RT$ plane given by
\bea \label{eq:transition}
     RT = \tanh \left[\frac{\pi}2 \left(\frac{1}{s_c}- s_c \right) \right].
\eea
where $s_c = s_c(T)$ denotes the critical value of $s$ for any given brane tension $T$.
As shown in Fig.~\ref{fig:transition}, for a fixed value of the tension $T$ (corresponding to the specific choice of Cardy state we are projecting on), the BTZ black hole phase is dominant for small values of $s$, and thus to measurement parameters satisfying $\Delta L\gg \Delta T$ (see Fig. \ref{fig:rho-finite} and Appendix \ref{sec:finite_conformal}). On the other hand, the thermal AdS phase is dominant for large values of $s$ (corresponding to the measurement parameters satisfying $\Delta L\ll \Delta T$). Alternatively, if we first fix a value of $s$ (i.e. for given values of the measurement parameters $\Delta L$ and $\Delta T$), the BTZ black hole phase is dominant for small values of the tension $T$, while the thermal AdS phase is dominant for large values of $T$. It is thus possible to tune across the phase boundary by appropriately choosing either the measurement parameters $\Delta L$ and $\Delta T$ or the specific Cardy state we postselect on.

The two different phases described will correspond to two qualitatively different dual spacetimes. To better understand the properties of the Lorentzian spacetimes of interest\footnote{Which are those obtained by analytic continuation after identifying the original $y$ coordinate to be the Euclidean time.} dual to these two phases, we focus on the time reflection symmetric slice, which is invariant under analytic continuation and is therefore a spatial slice of the Lorentzian geometry as well. In our final coordinate system, this slice is given by $\sigma=-\pi s/2$ on the boundary and by the $\sigma=\pm \pi s/2$ in the bulk (see Figs. \ref{fig:map-finite} (d) and \ref{fig:gravity-dual}). This corresponds to the slice $y=0,\pm \beta/2$ in the original boundary coordinates, i.e. the slice where the measured TFD state for the two CFTs is prepared by the boundary Euclidean path integral.

In the BTZ black hole phase, the time reflection symmetric slice spans from the asymptotic boundary \big(at $\sigma = -\pi s/2, z= 0, \nu \in (0,2\pi)$\big) to the brane \big(at $\sigma = \pm \pi s/2, z= z(\pi/2), \nu \in (0,2\pi)$\big), where the $+$ sign corresponds to positive tension branes, and $z(\sigma)$ is given by inverting the first line of \ref{eq:brane_BTZ}. See fig.~\ref{fig:gravity-dual} (a). 
Therefore, in the BTZ phase, the bulk time reflection symmetric slice is cut off by the ETW brane.

On the other hand, in the thermal AdS phase, the time reflection symmetric slice is bounded only by the asymptotic boundary $\sigma = -\pi s/2, z= 0, \nu \in (0,2\pi)$ and it does not intersect the brane. 
See Fig.~\ref{fig:gravity-dual} (b). 

In both cases, note that the bulk time reflection symmetric slice is connected between the left and right CFTs. This signals the fact that if the measured region is finite-sized, the Einstein-Rosen bridge (in the original $x$ coordinates) remains intact after measurement. Note that this will not be true for the infinite intervals case considered in the next subsections. Before moving on to this case, however, we verify bulk  connectivity by explicitly computing the holographic entanglement entropy of boundary subregions on the time reflection symmetric slice, making use of the bulk spacetimes we have just constructed.

\subsubsection{Holographic entanglement entropy in the post-measurement geometry}
\label{sec:hologeodesic}

With the dual spacetime in hand, we can now turn to computing the holographic entanglement entropy of intervals $[x_1, x_2]$ at $y=0$ on the boundary. We will restrict our attention to the leading order approximation given by the Ryu-Takayanagi (RT) formula \cite{Ryu2006a,Ryu2006b}. Recall that in the presence of ETW branes, the RT surface is allowed to end anywhere along the branes while respecting the homology constraint (i.e. homology is computed relative to the brane).

We first consider the BTZ black hole phase, focusing on the  $\Delta L \gg \Delta T$ limit for simplicity (which implies $s\ll 1$). 
In this case, the two parameters $\phi,s$ of the conformal map are determined by the measurement parameters $\Delta L$, $\Delta T$, and $\beta$ via
\bea
    \phi = \frac{2\pi}\beta \Delta T, \quad 
    s = \frac{2\Delta T}{\Delta L}\left(1-\frac{\Delta T}\beta \right). 
\eea
See Appendix \ref{sec:finite_conformal} for a derivation. For the time reflection symmetric lines we focus on here, the coordinate transformation between the original CFT and the final BTZ black hole coordinates (\ref{eq:totalconffinite}) reduces to 
\bea
    x_1 = \frac\beta{2\pi} \log [\pm X(e^{-\frac\pi2 s + i \nu_1})] , \quad  x_2 = \frac{\beta}{2\pi} \log [\pm X(e^{-\frac\pi2 s + i \nu_2})],
    \label{eq:xbtz}
\eea
where the $+$ sign corresponds to the left system at $y=0$ (and the $-$ sign corresponds to the right system at $y=\beta/2$), and $X$ is given by
\bea
    X(e^{-\frac\pi2 s + i \nu}) = \frac{\theta_1(\frac{\nu-\phi}2, e^{-\frac\pi2 s})}{\theta_1(\frac\nu2, e^{-\frac\pi2 s})}
    \label{eq:Xtimesym}
\eea
(see Appendix \ref{sec:finite_conformal} for details and the definition of the theta functions).

We also need to map the cutoff from the original CFT to the BTZ/thermal AdS coordinates. According to the general coordinate transformation constructed in~\cite{roberts2012time}, we have
\bea \label{eq:cft-cutoff}
    \epsilon = a \left|\frac{\partial X}{\partial \nu} \right|^{-1}, 
\eea
where $a$ is the UV cutoff in the CFT.

\begin{figure}
    \centering
    \includegraphics[width=0.5\textwidth]{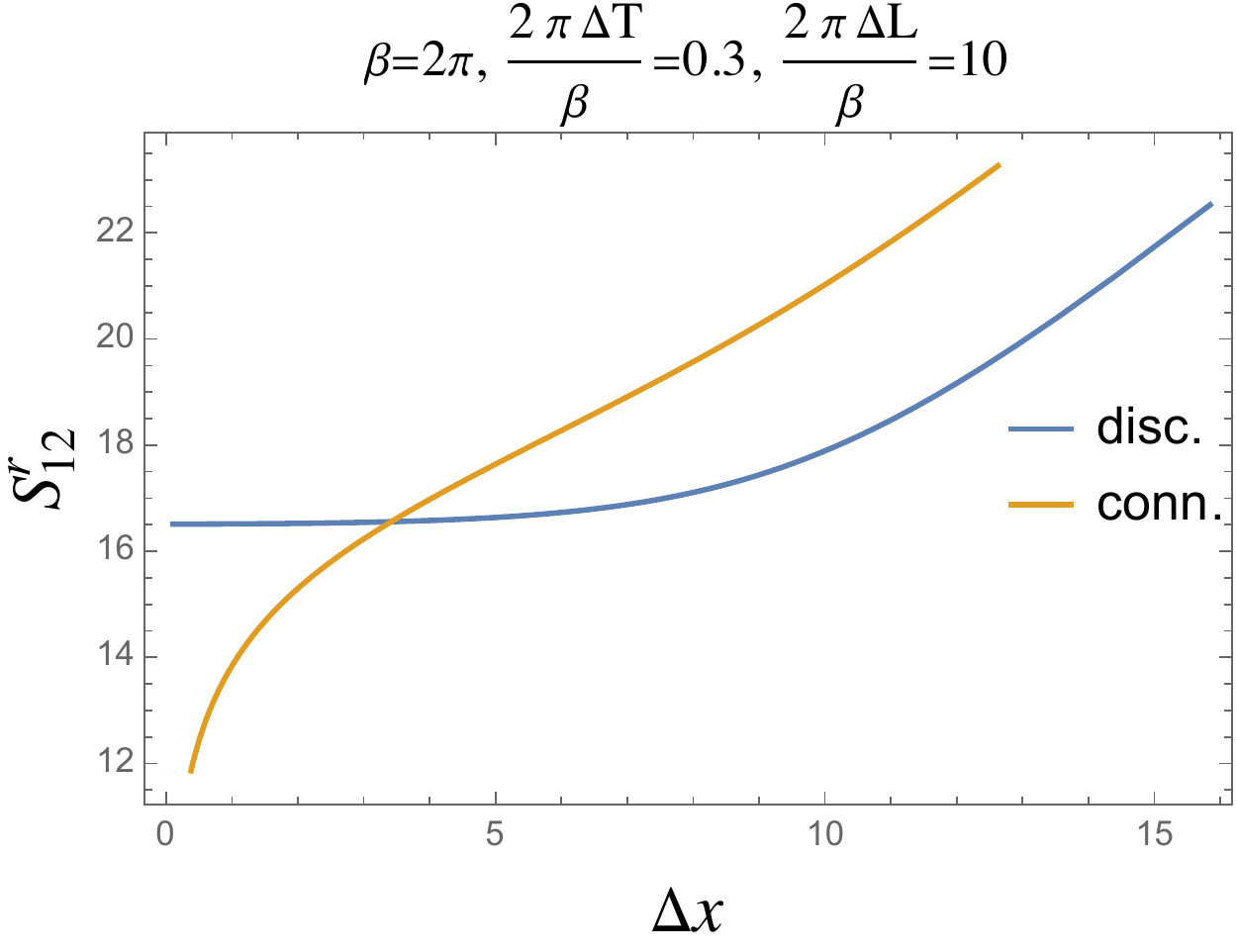}
    \caption{Entanglement entropy of interval $[- \frac{\Delta x}2, \frac{\Delta x}2]$ in the right system in the BTZ phase. The tension is taken to be zero and we set $R=1$.}
    \label{fig:ee_finite}
\end{figure}

Now, let us compute the holographic entanglement entropy of the interval $[x_1, x_2]$ in the right system (we therefore pick the $-$ sign in (\ref{eq:xbtz})) using the RT formula. In the BTZ black hole coordinates, these two points are given by 
\bea
    (\sigma_1, \nu_1, z_1) = (- \frac\pi2 s, \nu_1, \epsilon(\nu_1)),  \\
    (\sigma_2, \nu_2, z_2) = (- \frac\pi2 s, \nu_2, \epsilon(\nu_2)).
\eea
There are two candidate RT surfaces, one connected and one disconnected that reaches from the $y=\beta/2$ line to the time reflection symmetric slice on the brane. In the connected phase, the length of the RT surface is given by (see Appendix~\ref{append:geodesic})
\bea\label{eq:rt_finite_distance}
    D_{12} = R \cosh^{-1}\left\{\frac{z_H^2}{z_1 z_2}\left[ \cosh \frac{\nu_1-\nu_2}{z_H} - \sqrt{\left(1- \frac{z_1^2}{z_H^2}\right)\left(1- \frac{z_2^2}{z_H^2}\right)}\cos\frac{\sigma_1 - \sigma_2}{z_H} \right] \right\},
\eea
where we have used the fact that $z_H = s$. In the disconnected phase, the two segments of the RT surface end on the brane, with endpoints fixed by symmetry to be
\bea
\begin{aligned}
     (\sigma_3, \nu_3, z_3) &= (\frac\pi2 s, \nu_1, z_H \sqrt{1-(RT)^2}),  \\
    (\sigma_4, \nu_4, z_4) &= (\frac\pi2 s, \nu_2, z_H \sqrt{1-(RT)^2})
\end{aligned}
\eea
 for the segments anchored at $(\sigma_1, \nu_1, z_1)$ and $(\sigma_2, \nu_2, z_2)$ respectively. The length of the geodesic segments are again given by equation (\ref{eq:rt_finite_distance}), such that the holographic entanglement entropy is given by
\bea
    S_{12}^r = \frac{1}{4G_N} \min\{D_{12}, D_{13} + D_{24}\}. 
\eea
The entanglement entropy for a specific choice of $\Delta L \gg \Delta T$ and $\beta$ is depicted in Fig.~\ref{fig:ee_finite}. We plot the entanglement entropy of the interval $[- \frac{\Delta x}2, \frac{\Delta x}2]$ in the right system as a function of its size $\Delta x$. 
As expected, the connected phase dominates for small $\Delta x$, while the disconnected phase dominates for large $\Delta x$, i.e. when the interval is large enough that having two disconnected segments ending on the brane reduces the length of the RT surface.

The thermal AdS phase corresponds (for fixed tension $T$) to large values of $s$. We can then study the limit $\frac{\Delta L}{\beta}\ll\sin(\frac{\pi}{\beta}\Delta T)$ (see Appendix \ref{sec:finite_conformal} and Fig. \ref{fig:rho-finite}), for which
\bea
    e^{-\pi s} = \left( \frac{\pi \Delta L}{4\beta \sin \frac{\phi}2} \right)^2, \quad \phi = \frac{2\pi}\beta \Delta T. 
\eea

\begin{figure}
    \centering
    \includegraphics[width=0.5\textwidth]{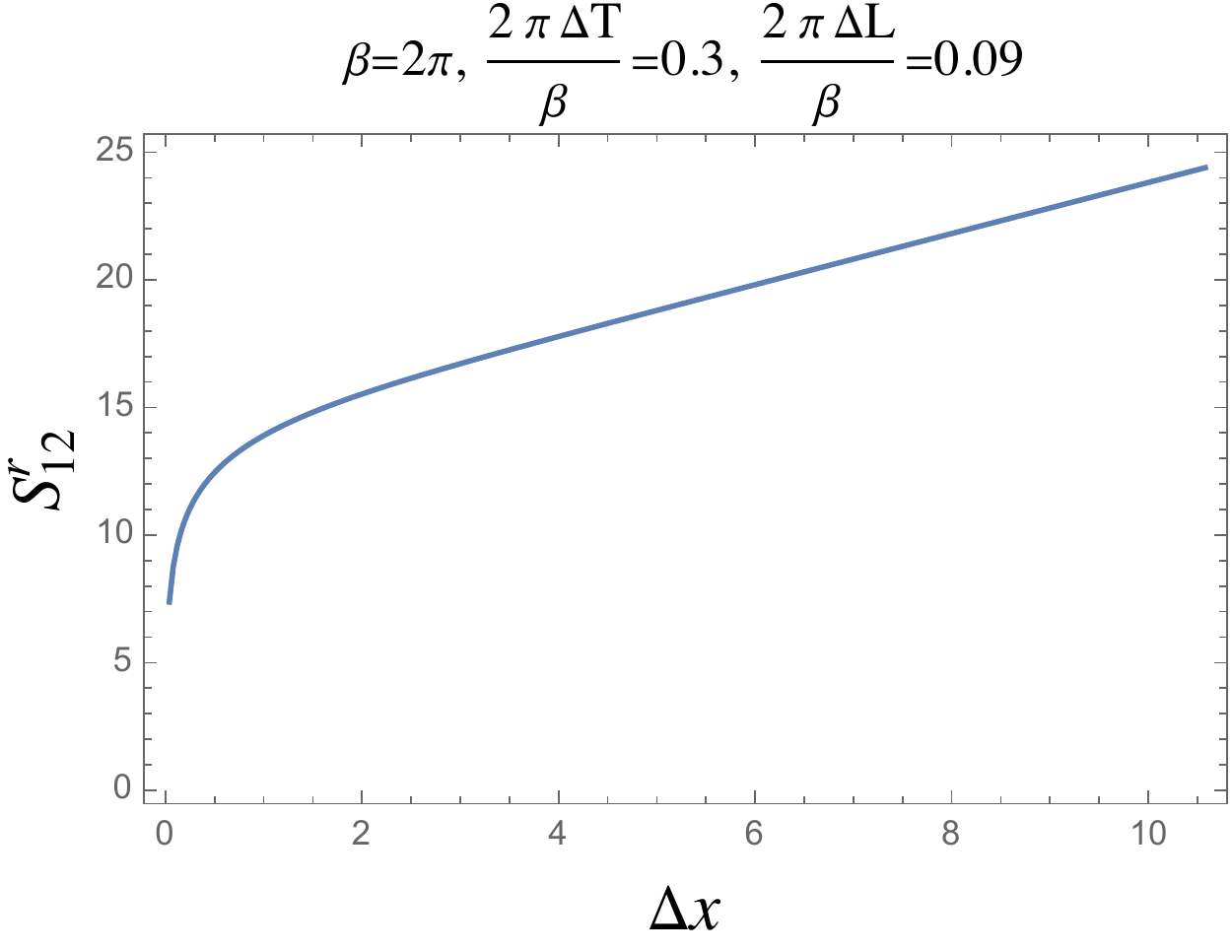}
    \caption{Entanglement entropy of interval $[- \frac{\Delta x}2, \frac{\Delta x}2]$ in the right system int the thermal AdS phase. 
    The tension is taken to be zero and we set $R=1$.}
    \label{fig:ee_finite_therm}
\end{figure}

The conformal maps (\ref{eq:xbtz})-(\ref{eq:Xtimesym}) for the time symmetric slice and the relationship (\ref{eq:cft-cutoff}) between cutoffs in the two coordinate systems still hold in this limit. In the thermal AdS phase there is only one, connected candidate class of RT surfaces because, as we have pointed out at the end of Section \ref{sec:transition}, the brane does not intersect the time reflection symmetric slice.
In the thermal AdS spacetime, the geodesics between the two endpoints  $(\sigma_1,\nu_1,z_1) = (-\frac{\pi}{2}s,\nu_1,\epsilon(\nu_1) )$ and $(\sigma_2,\nu_2,z_1) =(-\frac{\pi}{2}s,\nu_2,\epsilon(\nu_2))$ 
can again be computed via equation \ref{eq:rt_finite_distance}, except with $\nu$ and $\sigma$ switched, and with $z_h$ set to one. The entropy is then given by 
\begin{equation}
    S^r_{12}  = \frac{1}{4 G_N} R \cosh^{-1} \left\{ \frac{1}{z_1 z_2} \left[ 1-\sqrt{(1-z_1^2)(1-z_2^2)}\cos[\nu_1 - \nu_2]\right]\right\}
\end{equation}
The entanglement entropy for an interval $\left[-\frac{\Delta x}{2}, \frac{\Delta x}{2}\right]$ in the left system as a function of $\Delta x$ for a specific example is depicted in Fig. \ref{fig:ee_finite_therm}.

We remark that in both phases the entanglement entropy grows as $\Delta x$ increases. In particular, this implies that the entanglement entropy of the whole right (and therefore left) CFT after measurement is large. In other words, the two CFTs are still highly entangled after measurement. This result was to be expected, given that, as we have discussed, the bulk time reflection symmetric slice is connected between the two CFTs in both phases.

\subsection{Infinite intervals: one-sided measurement} \label{sec:infinite-one}

\subsubsection{Slit prescription}
\label{sec:infinite-one-slit}

As mentioned at the beginning of this section, the Lorentzian spacetime dual to the TFD state is a double-sided black hole, with two asymptotically AdS boundaries connected via an Einstein-Rosen bridge. Each CFT copy can be thought of as living on one of the asymptotic AdS boundaries. From the analysis of the previous subsection, it is clear that when only finite intervals are measured, there is always a connected portion of Lorentzian spacetime separating the two CFTs, as evidenced by the fact that the entanglement entropy between the two sides remains non-zero. Further, recall that both CTFs  are mapped to different segments of the $\sigma=-\pi s/2$ circle in BTZ/thermal AdS coordinates, and that the time reflection symmetric slice (whose boundary contains the two CFTs) is connected both in the BTZ and in the thermal AdS phase. This slice corresponds to the time reversal invariant slice in the Lorentzian geometry where the initial conditions for the real time evolution are given. Its connectivity implies the connectivity of the Lorentzian wormhole, which is therefore preserved after the measurement. In other words, the Einstein-Rosen bridge remains intact after the measurement, and so the two CFTs are necessarily still highly entangled.

\begin{figure}
    \centering
\subfigure[$x$ coordinate]{
    \includegraphics[width=0.35\textwidth]{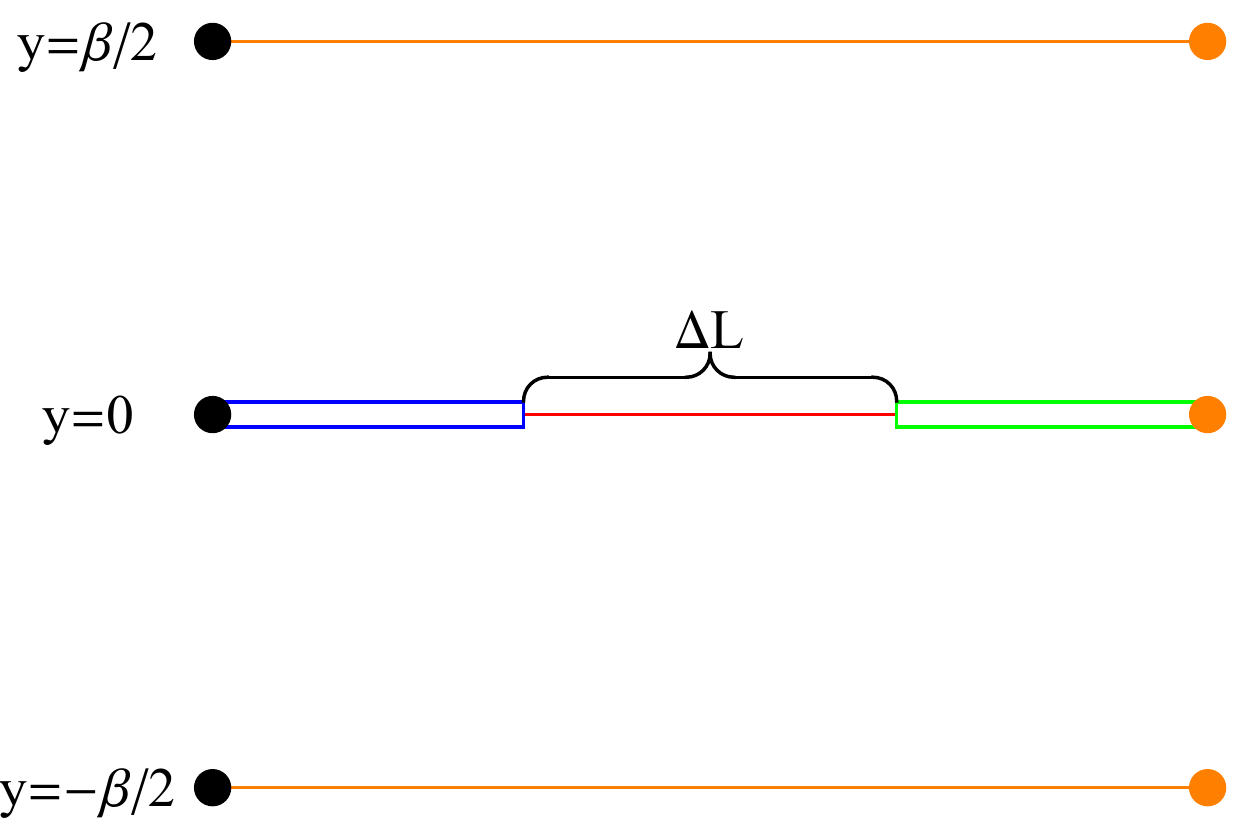}
    } \quad \quad \quad \quad \quad \quad
\subfigure[$X$ coordinate]{
    \includegraphics[width=0.35\textwidth]{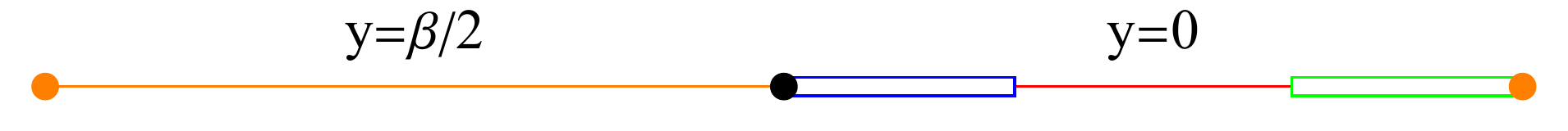}}\\\vspace{5mm}
\subfigure[$\zeta$ coordinate]{
    \includegraphics[width=0.35\textwidth]{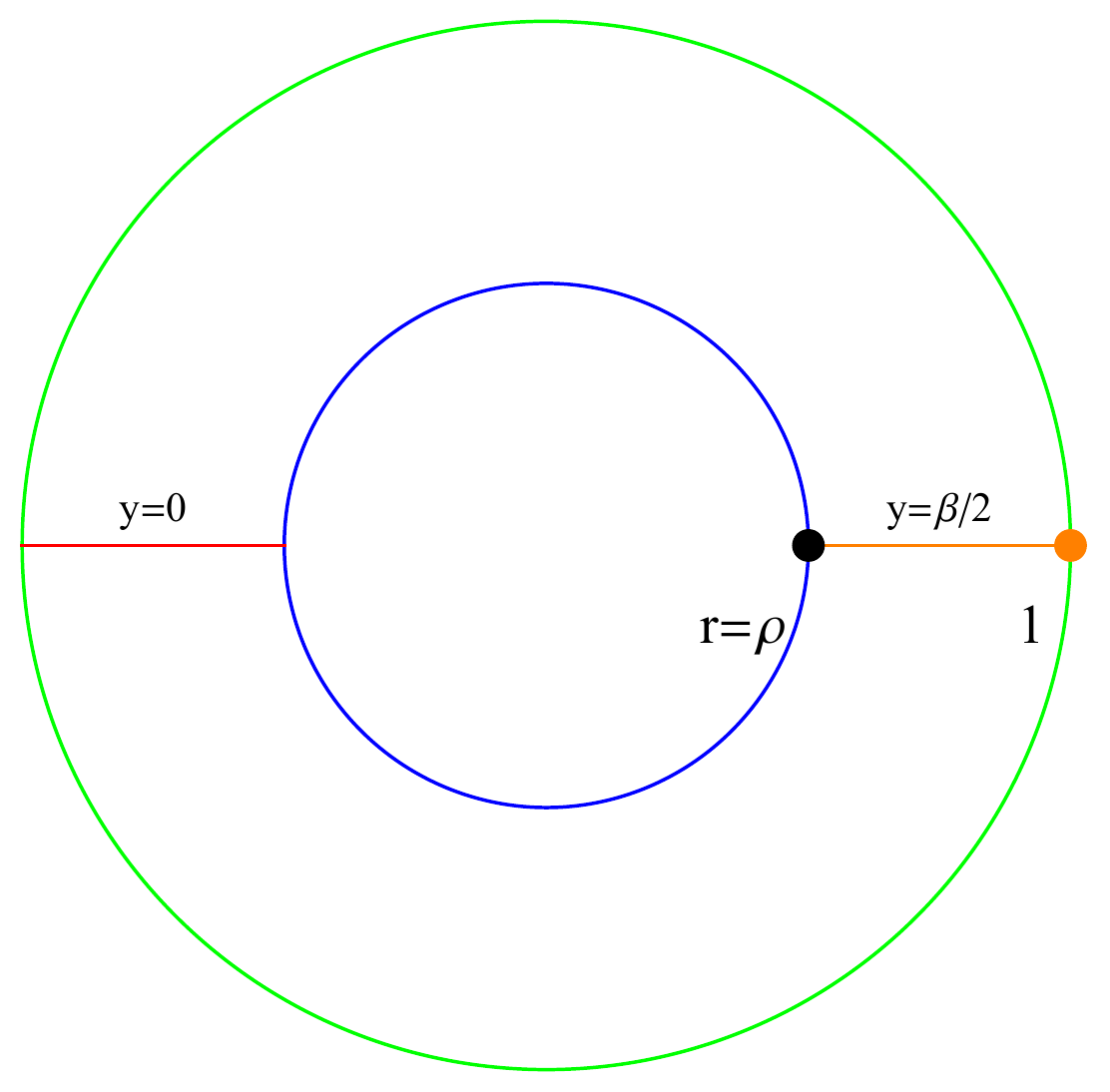}}
    \quad \quad \quad\quad \quad \quad
\subfigure[$w$ coordinate]{
    \includegraphics[width=0.35\textwidth]{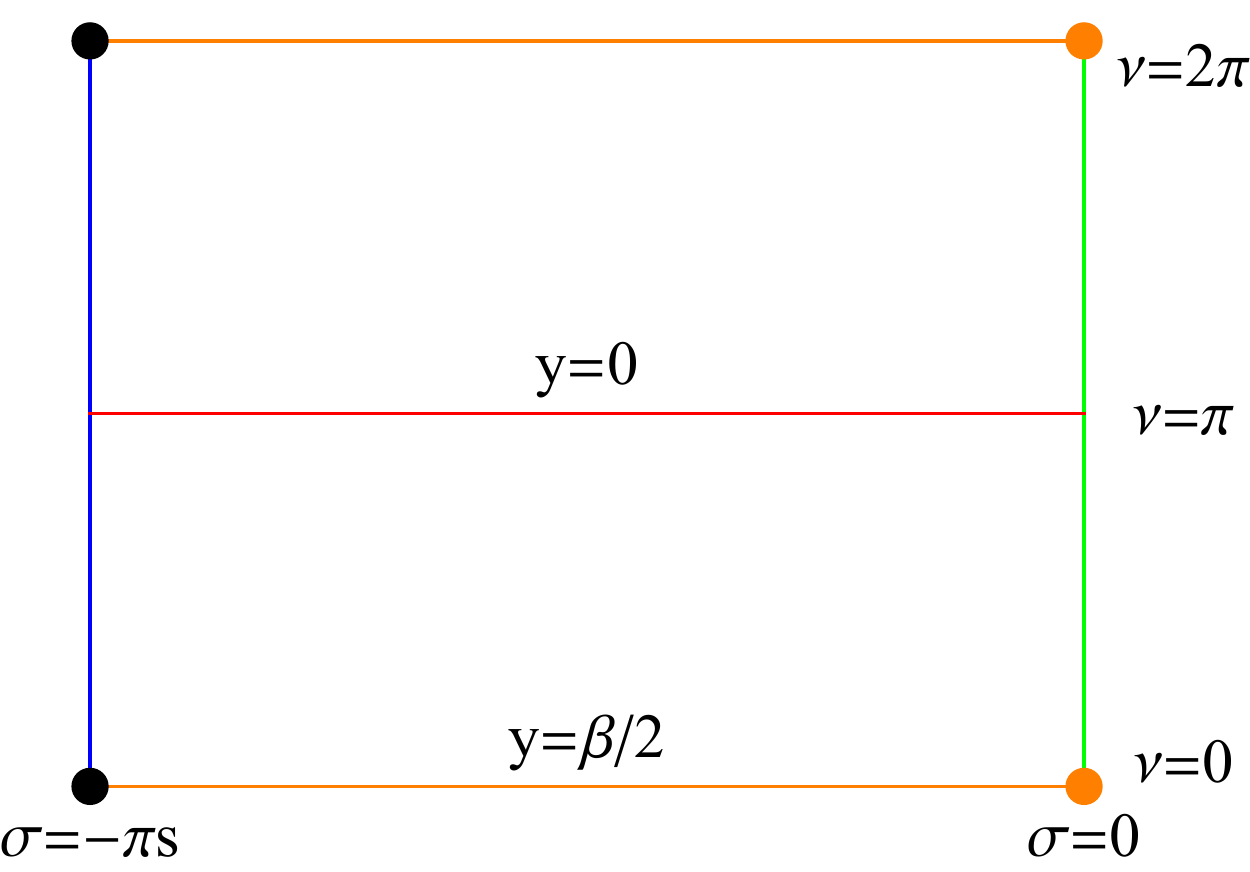}}
    \caption{The four different sets of coordinates used to describe the measurement of two semi-infinite intervals in the left CFT. The conformal maps between the different coordinates are given in Appendix \ref{append:infinite1sideconf}. The red (orange) line denotes the unmeasured part of the left side CFT at $y=0,x\in (-\Delta L/2,\Delta L/2)$ (the right side CFT at $y=\beta/2$). 
    They are time reflection symmetric lines. 
    The blue (green) color denotes the first (second) slit. The black (orange) dots represent negative (positive) infinity in $x$ in both left and right CFTs.
    (a) The initial, infinitely long cylinder with two infinite slits in the left CFT.
    $\Delta L$ is the length of the unmeasured interval. 
    Also note $y\sim y+\beta$, so the top and bottom lines should be identified. The black (orange) dots represents spatial negative (positive) infinity $x = - \infty$ ($x=\infty$) in both left and right CFTs. 
    (b) We first map to a 2D plane with two radial slits on the real line.
    The time reflection symmetric lines are mapped to the real line, where the left (right) CFT is mapped to a positive (negative) region. 
    negative (positive) infinity in $x$ is mapped to origin (infinity).
    (c) We next map to the annulus. The second slit is mapped to the outer edge $r= 1$ and the first slit to the inner edge $r= \rho$. The unmeasured part of the left CFT is mapped to the segment $\zeta \in (-\rho, -1)$, and the right CFT is mapped to $\zeta \in (\rho, 1)$.
    negative (positive) infinity in $x$ is mapped to $r= \rho, \theta = 0$ ($r=1, \theta = 0$). 
    Here we use $\zeta = r e^{i \theta}$, and $\rho$ is a constant determined by the size $\Delta L$ of the unmeasured part of the left CFT. See equation (\ref{eq:s-infinite-one}).
    (d) Finally, we map to the finite cylinder with $\nu \sim \nu + 2\pi$. 
    The first (second) slit is mapped to the left (right) edge at $\sigma =-\pi$ ($\sigma =0$). 
    The unmeasured part of the left CFT is mapped to the segment $\nu = \pi$, and the right CFT is mapped to $\nu = 0,2\pi$.
    negative (positive) infinity in $x$ is mapped to $\sigma = -\pi s, \nu = 0,2\pi$ ($\sigma = 0, \nu = 0,2\pi$), with $\rho = e^{-\pi s}$.}
   \label{fig:map-infinite}
\end{figure}

In this subsection, we will see that the measurement of infinitely long intervals in one of the two CFTs can instead ``destroy'' the Einstein-Rosen bridge and disconnect the two sides. This change in the bulk geometry reflects the fact that the two CFTs are being disentangled by the measurement. In particular, we will consider the measurement of two semi-infinite intervals in the left CFT (as depicted in Fig.~\ref{fig:map-infinite}). We find that in the BTZ black hole phase, the time reflection symmetric slice is disconnected. This implies that the connectivity of the corresponding Lorentzian wormhole is destroyed by the measurement. Therefore, the analogous Hawking-Page transition described in the previous subsections will now correspond to a connected-disconnected phase transition in the Lorentzian geometry associated with the measured TFD state, and to an entangled/disentangled phase transition in the microscopic boundary theory.

We again start with a thermofield double state with temperature $\beta$, prepared by a Euclidean path integral on a 2D infinite cylinder with coordinates $(x,y)$, and $y \sim y + \beta$. We perform a local projective measurement on two semi-infinite segments of the left CFT given by 
\bea \label{eq:slit-2-1sid}
    && \text{first slit:} \quad   x < - \frac{\Delta L}2 , \quad y = 0, \\
    && \text{second slit:} \quad \frac{\Delta L}2 < x , \quad y = 0,
\eea
 where $\Delta L > 0$ now denotes the length of the unmeasured interval. This parameter, in addition to the tension of the ETW brane (corresponding to the boundary entropy of the specific Cardy state we postselect on), fully specifies the measurement.

As in the finite interval case, in order to obtain a non-singular configuration to analyze, we implement a series of conformal transformations to map the two slits on the infinite Euclidean cylinder in $(x,y)$ coordinates to the boundaries of a finite cylinder in $(\sigma,\nu)$ coordinates. Note that here (unlike in the previous section) we do not implement any additional Euclidean time evolution, simply because we do not need to in order to obtain the phase transition of interest here.

The details of the conformal transformations are reported in Appendix \ref{append:infinite1sideconf}. This procedure allows us to build the bulk dual spacetime using the AdS/BCFT proposal in complete analogy with the previous subsection. However, note that here the two time reflection symmetric slices in the original infinite cylinder, corresponding to the locations of the two CFTs, get mapped to two disconnected segments of the final, finite cylinder.

\begin{figure}
    \centering
    \includegraphics[width=0.5\textwidth]{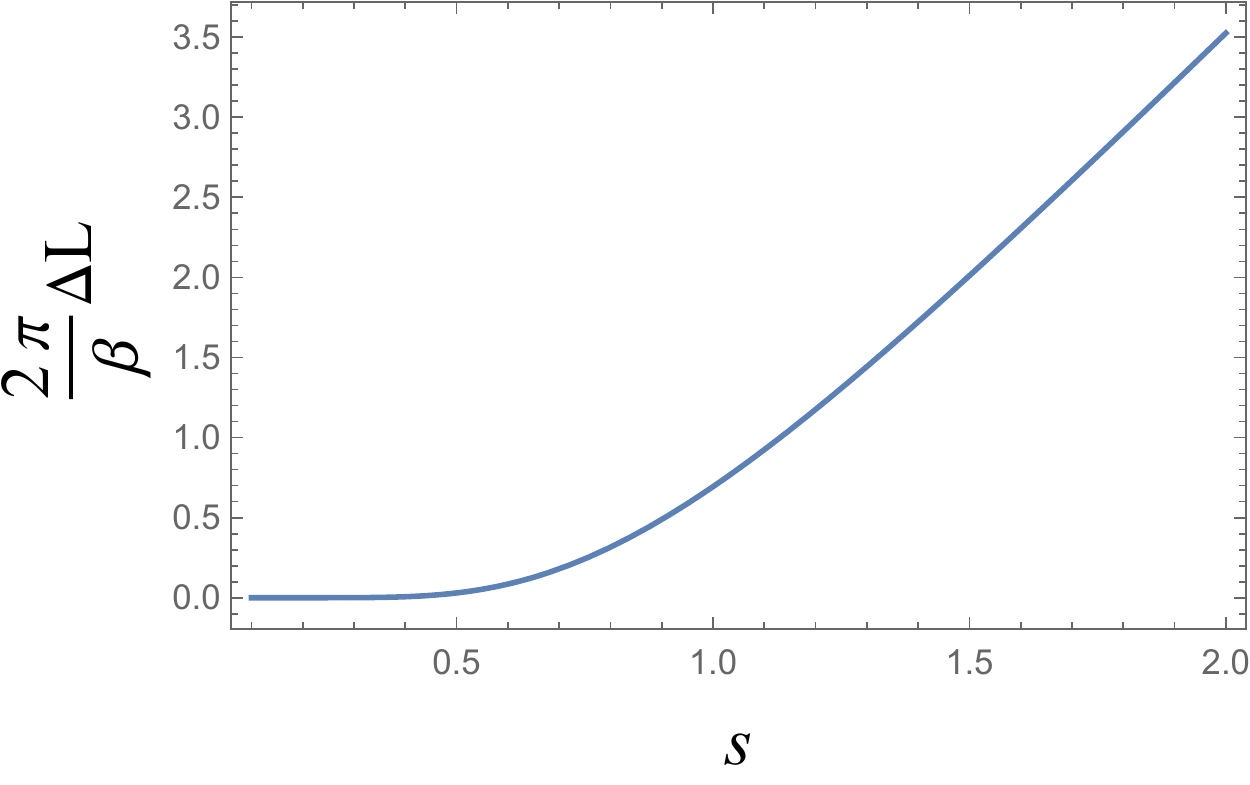}
    \caption{The relationship between the measurement parameter $\Delta L/\beta$ and the conformal transformation parameter $s$ for the case where infinite intervals are measured on one-side. Note that the parameter $s$ is a monotonic function of $\Delta L$.}
    \label{fig:s-infinite}
\end{figure}

We make use of three conformal transformations that map between four different sets of coordinates as indicated in Fig.~\ref{fig:map-infinite}. 
The composed conformal transformation from the original $(x,y)$ coordinates (Fig.~\ref{fig:map-infinite} (a)) to the final $w$ coordinate (Fig.~\ref{fig:map-infinite} (d)) is given by 
\bea
    x + i y = \frac{\beta}{2\pi} \log \left[ \frac{\theta_4^2 \left(-\frac{i(\sigma + i \nu)}2 , e^{-\pi s} \right)}{\theta_1^2 \left(- \frac{i(\sigma+ i \nu)}2 , e^{-\pi s} \right)} \right].
    \label{eq:infinite1sidedmap}
\eea
The parameter $s$ is related to the parameter $\Delta L$ through equation (\ref{eq:s-infinite-one}) (see Fig. \ref{fig:s-infinite}). Recall that here, as above, in the original $(x,y)$ coordinates the left CFT corresponds to the $y=0$ line, and the right CFT to $y=\beta/2$. In the final $(\sigma,\nu)$ coordinates, the unmeasured part of the left CFT is mapped to the $\nu=\pi$ segment and the right CFT is mapped to the $\nu=0,2\pi$ segment. The two slits are again mapped to the circles $\sigma=-\pi s$ (first slit) and $\sigma=0$ (second slit), whereas negative (positive) infinity in $x$ is mapped to $\sigma=-\pi s$, $\nu=0,2\pi$ ($\sigma=0$, $\nu=0,2\pi$).


As we will see, the fact that the right CFT and the unmeasured part of the left CFT are mapped to two disconnected segments (instead of to two segments on the same circle, as in the finite interval case of Section \ref{sec:finite}) implies that it is possible for the measurement of infinite intervals to destroy the Einstein-Rosen bridge and disconnect the Lorentzian geometry between the two CFTs. If a given measurement of infinite intervals sufficiently disentangles the two CFTs, then the remaining systems must be conformally mapped to two segments in the final, finite cylinder that are not on the boundaries of a connected bulk time reflection symmetric slice. We will see how, in the BTZ black hole phase, this is indeed the case.

\subsubsection{Bulk spacetime and Hawking-Page transition}
\label{sec:infinite-one-HP}

We have now mapped the infinite cylinder with two semi-infinite slits to the same finite cylinder we had in the finite interval case of the previous subsections.
Therefore, the Euclidean bulk dual spacetime is the same as in Sec.~\ref{sec:transition}. 
In particular, we have two phases determined by the same phase boundary~(\ref{eq:transition}), see Fig.~\ref{fig:transition}.
Note that here, $s$ is related to $\Delta L$ by equation~(\ref{eq:s-infinite-one}) in Appendix \ref{append:infinite1sideconf}. 
When the tension is zero, the transition is at $s=1$, corresponding to $\frac{2\pi}\beta \Delta L = \log 2$.

\begin{figure}
    \centering
    \subfigure[]{\includegraphics[width=0.4\textwidth]{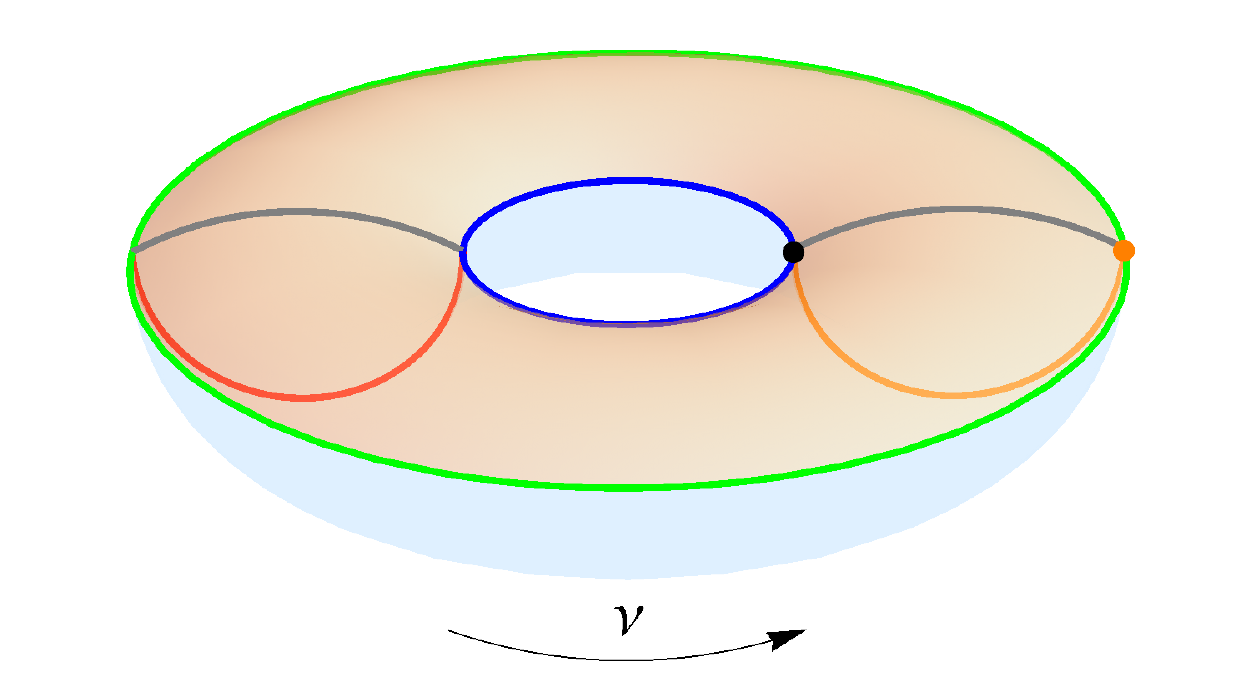}} \quad \quad
    \subfigure[]{\includegraphics[width=0.4\textwidth]{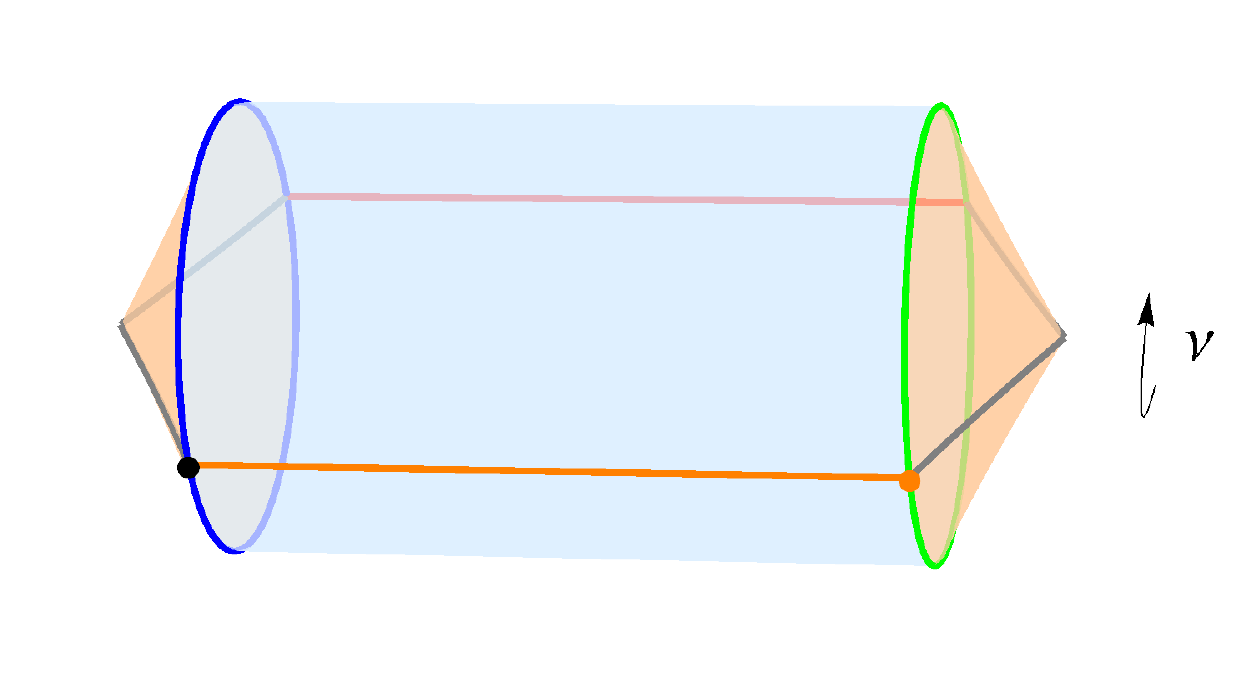}}
    \caption{Gravity dual of the infinite interval measurement setup (with two infinite intervals measured in the left CFT). The color convention is the same as in Fig.~\ref{fig:map-infinite}. 
    The red line denotes the unmeasured part of the left CFT at $y=0$, and the orange line denotes the right CFT at $y=\beta/2$. These are time reflection symmetric lines. The blue (green) color denotes the first (second) slit. 
    The black (orange) dots represents negative (positive) infinity in $x$ in both left and right CFTs. 
    The light-blue (light-orange) surface denotes the  asymptotic boundary (the ETW brane).
    The gray color denotes the intersection between the time reflection symmetric slice and the brane.
    $\nu \sim \nu + 2\pi$ is periodic.
    (a) BTZ black hole phase involving a single connected brane.
    (b) Thermal AdS phase involving two disconnected branes anchored at $\sigma = -\pi s$ and $\sigma =0$.}
    \label{fig:gravity-dual_infinite}
\end{figure}

While the discussion of the phase structure is analogous to the finite interval measurement case, the physical interpretation of the phase transition is different. In fact, in the semi-infinite interval case of this subsection we used different coordinate transformations that mapped the unmeasured part of the left CFT and the right CFT to two disconnected segments, see Fig. \ref{fig:map-infinite}. As a result, the ``measurement-induced'' Hawking-Page transition between the BTZ black hole phase and the thermal AdS phase corresponds now to a connected-disconnected phase transition of the Einstein-Rosen bridge in the associated Lorentzian spacetime, and to an entangled/disentangled phase transition between the two CFTs. To see why this is the case, consider the Euclidean spacetime associated with the two phases, depicted in Fig.~\ref{fig:gravity-dual_infinite}, and focus on the bulk time reflection symmetric slice (whose boundary contains the right CFT and the unmeasured part of the left CFT). In the $w$ coordinate (see Fig.~\ref{fig:map-infinite} (d)), this is given by the $\nu=0,\pi$ bulk slice. In the BTZ black hole phase, the time reflection symmetric slice clearly has two disconnected components: both span from the asymptotic boundary $z= 0,   \sigma \in (-\pi s,0)$ to the brane $z = z(\sigma), \sigma \in (0,\pi s)$, but one is located at $\nu=0$ while the other is located at $\nu = \pi$.
Here $z(\sigma)$ is the inverse function of $\sigma(z) = s \tan^{-1} \left[ \frac{R T z}{s \sqrt{h - R^2 T^2}} \right]$. 
Now recall that the time reflection symmetric slice in the Euclidean geometry corresponds to the time reversal symmetric slice in the associated Lorentzian geometry, on which initial data for the real time evolution is defined. Since the time reflection symmetric slice is disconnected in the BTZ black hole phase, the associated Lorentzian geometry is also disconnected. When $s$ is sufficiently small for a fixed $RT$ (so that the size $\Delta L$ of the unmeasured region in the left CFT is sufficiently small) or when the brane tension $T$ is sufficiently small or negative for a fixed value of $\Delta L$, the measurement ``destroys'' the Einstein-Rosen bridge and disconnects the Lorentzian geometry between the two CFTs, as expected. This corresponds to having zero entanglement between the unmeasured part of the left CFT and the right CFT after the measurement, as we will see shortly.

On the other hand, in the thermal AdS phase the time reflection symmetric slice is bounded by the asymptotic boundary---at $\nu = 0, z= 0, \sigma \in( -\pi s,0)$ and $\nu = \pi, z= 0, \sigma \in( -\pi s,0)$--- and the branes---at $\nu = 0 \cup \pi, \sigma = \sigma(z), z \in (0,1)$, where $\sigma(z) = \sinh^{-1} \left( \frac{RTz}{\sqrt{1-R^2T^2}} \right)$. 
In this phase, the time reflection symmetric bulk slice has only one connected component. Following the reasoning above, this implies that the Lorentzian geometry is also connected, meaning that the measurement does not destroy the Einstein-Rosen bridge. 

As expected, for fixed $RT$, the thermal AdS phase is dominant for large $s$ (i.e. when the size $\Delta L$ of the unmeasured region in the left CFT is large, see Fig. \ref{fig:s-infinite}). Note that the connected thermal phase is dominant also for large negative values of the tension for any fixed value of $s$. Having a connected post-measurement Lorentzian spacetime corresponds to having a large amount of entanglement between the unmeasured part of the left CFT and the right CFT after the measurement, as we will now describe.


\subsubsection{Holographic entanglement entropy in the post-measurement geometry}

In order to characterize the phase transition in terms of the post-measurement entanglement between the right CFT and the unmeasured part of the left CFT, we can look at the post-measurement mutual information $I^{lr}=S(l)+S(r)-S(lr)$, where $S(l)$ is the entanglement entropy of the unmeasured part of the left CFT, $S(r)$ is the entanglement entropy of the right CFT, and $S(lr)=0$ is the entanglement entropy of the union of the two, which is vanishing since the full system is in a pure state. The entanglement entropies $S(l)$ and $S(r)$ can be computed applying the Ryu-Takayanagi formula to the segments $\nu = 0, z= 0, \sigma \in( -\pi s,0)$ and $\nu = \pi, z= 0, \sigma \in( -\pi s,0)$, respectively. In the BTZ phase, the dominant RT surface is clearly the empty surface for both left and right CFTs, and therefore we get $S^l = S^r = I^{lr}=0$.
In the thermal AdS phase, the dominant RT surface sits at $z=1$, starting at the left brane and ending at the right brane. Its area is then given by
\bea
    S^l = S^r =  \frac{R}{4G_N} \int d \sigma = \frac{R}{4G_N} \left[ \pi s + 2 \sinh^{-1}\left( \frac{RT}{\sqrt{1-(RT)^2}} \right) \right].
    \label{eq:RTsurface}
\eea
Thus, the mutual information between the right CFT and the unmeasured part of the left CFT is
\bea \label{eq:mi}
    I^{lr} = 
    \begin{cases}
        0, \quad &\text{BTZ phase: } s<s_c(T) \\ \\
        \frac{R}{2G_N} \left[ \pi s + 2 \sinh^{-1}\left( \frac{RT}{\sqrt{1-(RT)^2}} \right) \right], \quad &\text{thermal AdS phase: } s> s_c(T). 
    \end{cases}
\eea
where $s_c(T)$ is given in~(\ref{eq:transition}). This result (which is analogous to the one obtained in \cite{Antonini:2022sfm}, although its physical interpretation is different) clearly shows how the Hawking-Page transition, which corresponds to a Lorentzian bulk connected/disconnected phase transition, also corresponds to an entangling/disentangling phase transition in the boundary theory. This can be readily understood: it tells us that, for any given Cardy state we project on, if we measure a region of the CFT which is too large we disentangle the right CFT from the remainder of the left CFT.

\subsubsection{Heavy operator insertions, teleportation and information erasure}
\label{sec:1sidedteleportation}

Now that we have identified our Hawking-Page phase transition as a connected/disconnected phase transition for the Lorentzian bulk (or, equivalently, an entangling/disentangling phase transition for the dual boundary theory), we can ask what portions of the bulk are encoded in each of the CFTs after the postselection is performed. In particular, we will be interested in whether bulk information encoded in one CFT in the absence of measurement becomes accessible from the other CFT in the presence of a measurement. In \cite{Antonini:2022sfm} a similar question could be sharply answered, thanks to the specific choice of setup, in terms of the connectivity of the post-measurement geometry. The results of \cite{Antonini:2022sfm} revealed that portions of the bulk that in the absence of measurement would have been encoded in some region $A$ of the boundary CFT become accessible from the complementary region $\bar{A}$ when region $A$ is measured, provided that the measurement outcome (i.e. the specific Cardy state we are postselecting on) is known.\footnote{We work here under the assumption that the measurement outcome is known to all boundary observers.} This feature was dubbed ``bulk teleportation'' \cite{Antonini:2022sfm} and showed (using tensor network models) to be in fact due to quantum teleportation of the bulk information from region $A$ to region $\bar{A}$.

In the setup considered here, there is no analogously direct way to understand if bulk teleportation between the two sides is taking place. While studying the post-measurement bulk connectivity allowed us to determine whether the two CFTs remained significantly entangled after the measurement, it does not tell us if and how much of the pre-measurement left wedge is accessible from the right CFT in the post-measurement geometry. Fortunately, there is another way we can answer this question, namely, by studying the insertions of heavy operators in the Euclidean path integral preparing the state.

\subsubsection*{Operator insertions in the absence of measurement}

Consider a CFT operator $\mathcal{O}$ dual to a bulk scalar field $\phi$ with mass $m$ in the range $1\ll mR\ll R/\ell_P$, where $\ell_P$ is the Planck length. The large mass condition $mR\gg 1$---corresponding to a large conformal dimension $\Delta_{\mathcal{O}}$ for the dual operator---guarantees that spacelike correlators of the operator $\mathcal{O}$ can be computed in the geodesic approximation \cite{Faulkner:2018faa}. The condition $mR\ll R/\ell_P$ ensures that the backreaction of the scalar field on the background geometry can be neglected. Now consider an insertion of the operator $\mathcal{O}$ at some value of the Euclidean time $y=-y^*$ for $y*\in (0,\beta/2)$ and at $x=x^*$ with $x^*\in (-\infty,\infty)$ in the boundary Euclidean path integral preparing the TFD state. Because a well-defined real Lorentzian theory can be obtained by analytic continuation only if the associated Euclidean path integral is time reflection symmetric,\footnote{Equivalently, a well-defined real Lorentzian geometry can be obtained by analytic continuation only if the associated Euclidean gravitational path integral, and in particular the associated Euclidean saddle, is time reflection symmetric.} we must consider a second insertion of the operator $\mathcal{O}$ at $y=y^*$, $x=x^*$. 

\begin{figure}
    \centering
    \includegraphics[width=0.4\textwidth]{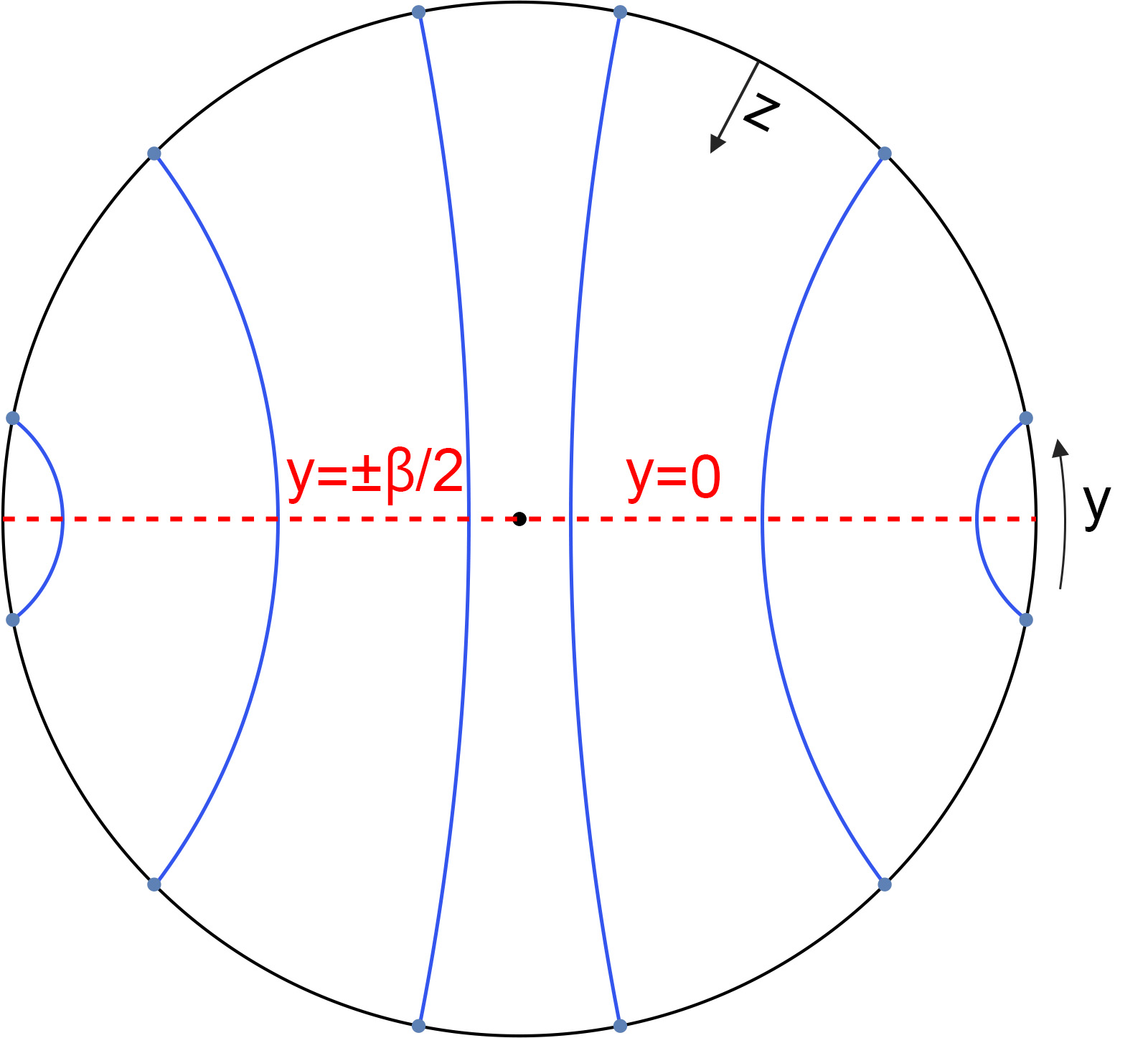}
    \caption{The time reflection symmetric geodesics in the $x=x^*$ slice of the BTZ black hole. The insertion points are at $y=\pm y^*$ on the boundary and the associated bulk Euclidean geodesics are depicted in blue. The red dashed line indicates the $y=0,\pm\beta/2$ time reflection symmetric slice. For $y^*\in (0,\beta/4)$ the geodesics cross the time reflection symmetric slice at $y=0$ and for $y^*\in (\beta/4,\beta/2)$ they cross the reflection symmetric slice at $y=\pm\beta/2$.}
    \label{fig:euclgeod}
\end{figure}

As we have discussed, the Euclidean path integral without any operator insertions prepares the TFD state on the $y=0,\pm\beta/2$ slice. 
The same path integral modified by the operator insertions at $y=\pm y^*$, $x=x^*$ prepares on the same slice a TFD state perturbed by the operator insertion. We can then ask whether the effect of the operator insertion is detectable from the left (at $y=0$) or the right (at $y=\pm\beta/2$) CFT. 
In the geodesic approximation, this question is straightforward to answer --- we can consider the time reflection symmetric bulk geodesic in the $y-z$ plane connecting $y=-y^*$ and $y=y^*$, and determine whether it crosses the time reflection symmetric slice at $y=0$ or $y=\pm\beta/2$.\footnote{As noted above, these geodesics can be simply computed using the background Euclidean BTZ metric because we are working within a range of scalar masses in which backreaction of the scalar field on the geometry is negligible.} In fact, in the Lorentzian spacetime obtained by analytic continuation, the $y=0$ slice corresponds to the portion of the time reversal symmetric slice in the left wedge and encoded in the left CFT, whereas the $y=\pm\beta/2$ slice corresponds to the portion in the right wedge and encoded in the right CFT. If, for example, the geodesic intersects the symmetric slice at $y=0$, the effect of the operator insertion is to create a particle in the left wedge, whose properties can be reconstructed from the left CFT. For $y^*\in (0,\beta/4)$ the geodesics cross the time reflection symmetric slice at $y=0$ and for $y^*\in (\beta/4,\beta/2)$ they cross the reflection symmetric slice at $y=\pm\beta/2$ (see Figure \ref{fig:euclgeod}).

\subsubsection*{Bulk teleportation and information erasure in the presence of measurement}

\begin{figure}
    \centering
    \subfigure[]{
    \includegraphics[width=0.45\textwidth]{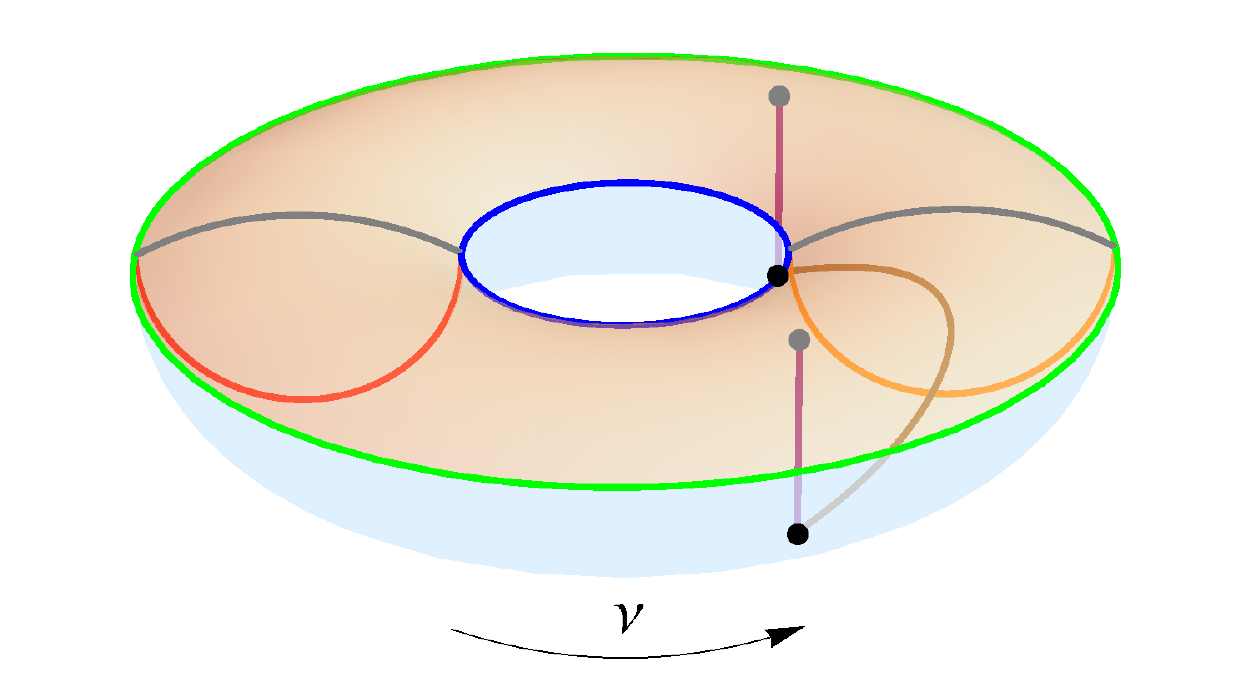}
    } \quad \quad 
\subfigure[]{
    \includegraphics[width=0.45\textwidth]{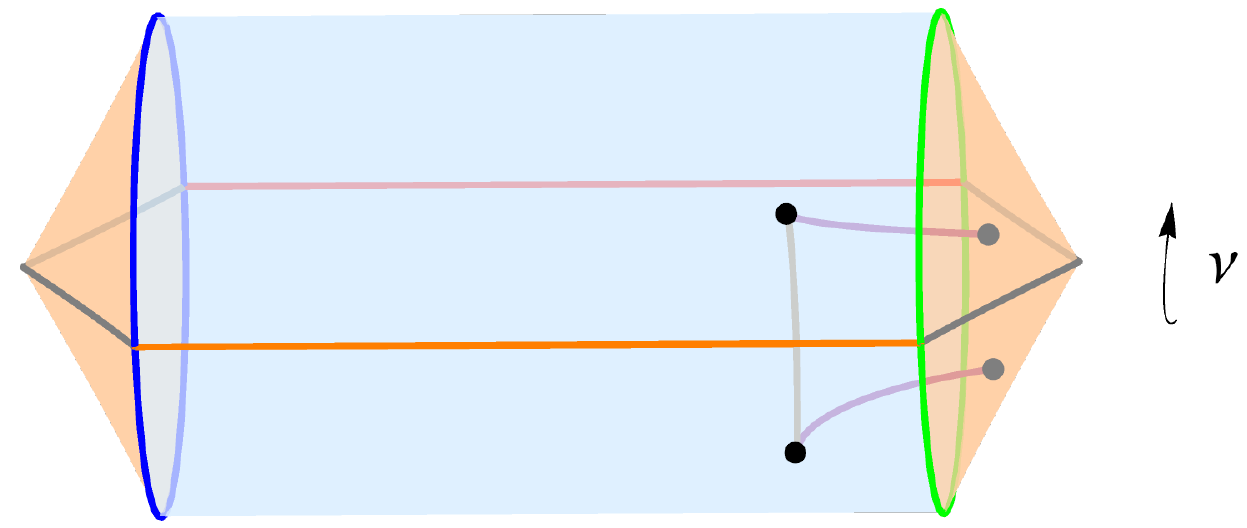}}
    \caption{Time reflection symmetric geodesics in the BTZ (left) and thermal AdS (right) phases. The symmetric insertion points on the AdS boundary are indicated by black dots. The constant-$\sigma$ connected geodesic is represented by a brown line connecting the two insertion points and passing through the $\nu=0,\pi$ reflection symmetric slice. The constant-$\nu$ disconnected geodesic is represented by two purple lines anchored at the insertion points and ending on the ETW brane, with the intersection points between the geodesic and the brane represented by gray dots. Note that the disconnected geodesic does not intersect the reflection symmetric slice.}
    \label{fig:postmeasurementgeod}
\end{figure}

In order to understand how the information associated with these operator insertions is affected by the measurement, we can carry out the following procedure:
\begin{enumerate}
    \item Consider operator insertions analogous to those described above at given points $(x^*,-y^*)$ and $(x^*,y^*)$ in our path integral with slits in the original $(x,y)$ coordinates depicted in Figure \ref{fig:map-infinite} (a).
    \item Map the two points to the points $P_1=(\sigma^*,\nu^*)$, $P_2=(\sigma^*,2\pi-\nu^*)$ in BTZ/thermal AdS coordinates using the map (\ref{eq:infinite1sidedmap}), with $\sigma^*\in (-\pi s,0)$ and $\nu^*\in (0,\pi)$.\footnote{Recall that in BTZ/thermal AdS coordinates the unmeasured part of the left CFT is mapped to $\nu=\pi$, the right CFT to $\nu=0,2\pi$ and the time reflection symmetric slice is therefore mapped to the $\nu=0,\pi$ slice, with $\nu=0$ and $\nu=2\pi$ identified.}
    \item In the bulk dual spacetime---which can be either thermal AdS or the BTZ black hole cut off by ETW branes---compute the shortest, reflection symmetric (around the $\nu=0,\pi$ slice) geodesic anchored at the points $P_1$ and $P_2$. For each choice of points $P_1$, $P_2$ there are two candidate geodesics: a constant-$\sigma$ connected geodesic which crosses the reflection symmetric slice, and a constant-$\nu$ geodesic with two disconnected segments ending on the ETW brane (see Figure \ref{fig:postmeasurementgeod}). Since both geodesic are anchored at the points $P_1$, $P_2$ and they are both reflection symmetric, the correct geodesic to consider in our analysis is the shorter of the two. 
    \item If the constant-$\sigma$ connected geodesic is the shortest one, determine whether it crosses the reflection symmetric slice at $\nu=0$ or at $\nu=0,2\pi$. Similar to what we observed in the absence of measurement, if $\nu^*\in (0,\pi/2)$ the geodesic crosses the reflection symmetric slice at $\nu=0,2\pi$, whereas if $\nu^*\in (\pi/2,\pi)$ it crosses the reflection symmetric slice at $\nu=\pi$.
\end{enumerate}

If the constant-$\sigma$ connected geodesic is the shortest one, there are two possibilities. First, notice that the RT surface calculation carried out in the previous subsection implies that the $\nu=0,2\pi$ portion of the reflection symmetric slice is part of the entanglement wedge of the right CFT, whereas the $\nu=\pi$ portion of the same slice is part of the entanglement wedge of the left CFT. Now consider an operator insertion with $y^*\in (0,\beta/4)$, whose associated geodesic in the absence of measurement crosses the time reflection symmetric slice at $y=0$ and whose effect is therefore detectable from the left CFT. If, in the presence of measurement, the point is mapped to an insertion point with $\nu^*\in (\pi/2,\pi)$, the constant-$\sigma$ geodesic crosses the reflection symmetric slice at $\nu=\pi$, implying that the effect of the operator insertion is still detectable from the left CFT. However, if the point is mapped to an insertion point with $\nu^*\in (0,\pi/2)$, then the constant-$\sigma$ geodesic crosses the reflection symmetric slice at $\nu=0,2\pi$, implying that the effect of the operator insertion is detectable from the right CFT after the measurement is performed. In the latter case, bulk teleportation in the sense of \cite{Antonini:2022sfm} is taking place between the two sides: part of the pre-measurement entanglement wedge of the left CFT is encoded in the right CFT after the measurement is performed. A completely analogous reasoning can be carried out for insertion points with $y^*\in (\beta/4,\beta/2)$.

On the other hand, if the shortest geodesic associated with an insertion with $y^*\in (0,\beta/4)$ is the constant-$\nu$ disconnected one, the geodesic ends on the brane and it does not intersect the portion of the time reflection symmetric slice accessible from either CFT. Therefore, we can conclude that the operator insertion does not affect the post-measurement Lorentzian geometry (at least in the purely geometrical approximation we are focusing on): the measurement ``erases'' the information associated with the operator insertion. This suggests that part of the pre-measurement entanglement wedge of the left side is destroyed by the measurement, as we could have suspected.

In order to complete our analysis, we compute the length of the connected and disconnected geodesics in the BTZ and thermal AdS spacetimes cut off by ETW branes. Let us first restrict our analysis to $\nu^*\in (\pi/2,\pi)$ and $\sigma^*\in (-\pi s/2,0)$. In the BTZ phase we again use equation (\ref{eq:rt_finite_distance}), obtaining
\begin{equation}
    L_{conn}^{btz}=R\cosh^{-1}\left\{\frac{s^2}{\epsilon^2}\left[\cosh\left(\frac{2(\pi-\nu^*)}{s}\right)-\left(1-\frac{\epsilon^2}{s^2}\right)\right]\right\}
    \label{eq:btzconn}
\end{equation}
for the connected geodesic and
\begin{equation}
    L_{disc}^{btz}=2\min_{z_b}\left(R\cosh^{-1}\left\{\frac{s^2}{\epsilon z_b}\left[1-\sqrt{\left(1-\frac{\epsilon^2}{s^2}\right)\left(1-\frac{z_b^2}{s^2}\right)}\cos\left(\frac{\sigma_b^{btz}(z_b)-\sigma^*}{s}\right)\right]\right\}\right)
    \label{eq:btzdisconn}
\end{equation}
for the disconnected geodesic, where $\epsilon$ is the bulk IR cutoff, we have identified $z_H=s$, and $\sigma_B^{btz}(z_b)$ is the brane position given by the first lines of equations (\ref{eq:brane_BTZ})-(\ref{eq:brane_BTZ2}) for positive and negative tension branes, respectively. Note that to obtain the length of the disconnected geodesic we minimized over the location of the endpoint of the geodesic on the brane---given by $(z_b,\sigma_b^{btz}(z_b),\nu^*)$ for the component of the geodesic at $\nu=\nu^*$---and we multiplied by 2 to account for the two disconnected components of the geodesic at $\nu=\nu^*$ and $\nu=2\pi-\nu^*$. By symmetry, the geodesic lengths for $\nu^*\in (0,\pi/2)$ and $\sigma^*\in (-\pi s,-\pi s/2)$ are given by the same formulae after replacing $\pi-\nu^*\to \nu^*$ and $\sigma^*\to -\sigma^*-\pi s$, with $\sigma_B^{btz}(z_b)$ still given by the first lines of equations (\ref{eq:brane_BTZ})-(\ref{eq:brane_BTZ2}).

In the thermal AdS phase, we can again use equation (\ref{eq:rt_finite_distance}), exchanging $\sigma \leftrightarrow \nu$ and setting $z_H=1$. For $\nu^*\in (\pi/2,\pi)$ and $\sigma^*\in (-\pi s/2,0)$ we get
\begin{equation}
    L_{conn}^{th}=R\cosh^{-1}\left\{\frac{1}{\epsilon^2}\left[1-\left(1-\epsilon^2\right)\cos (2\left(\pi-\nu^*\right))\right]\right\}
    \label{eq:thermalconn}
\end{equation}
for the connected geodesic and 
\begin{equation}
    L_{disc}^{th}=2\min_{z_b}\left(R\cosh^{-1}\left\{\frac{1}{\epsilon z_b}\left[\cosh\left(\sigma_b^{th}(z_b)-\sigma^*\right)-\sqrt{\left(1-\epsilon^2\right)\left(1-z_b^2\right)}\right]\right\}\right)
    \label{eq:thermaldisconn}
\end{equation}
for the disconnected geodesic, where $\sigma_b^{th}(z_b)$ is the brane position now given by the first line of equation (\ref{eq:brane_AdS}). As in the BTZ phase, the geodesic lengths for $\nu^*\in (0,\pi/2)$ and $\sigma^*\in (-\pi s,-\pi s/2)$ are given by the same formulae after replacing $\pi-\nu^*\to \nu^*$ and $\sigma^*\to -\sigma^*-\pi s$, with $\sigma_B^{th}(z_b)$ still given by the first lines of equation (\ref{eq:brane_AdS}).

\begin{figure}
    \centering
    \subfigure[]{
    \includegraphics[width=0.35\textwidth]{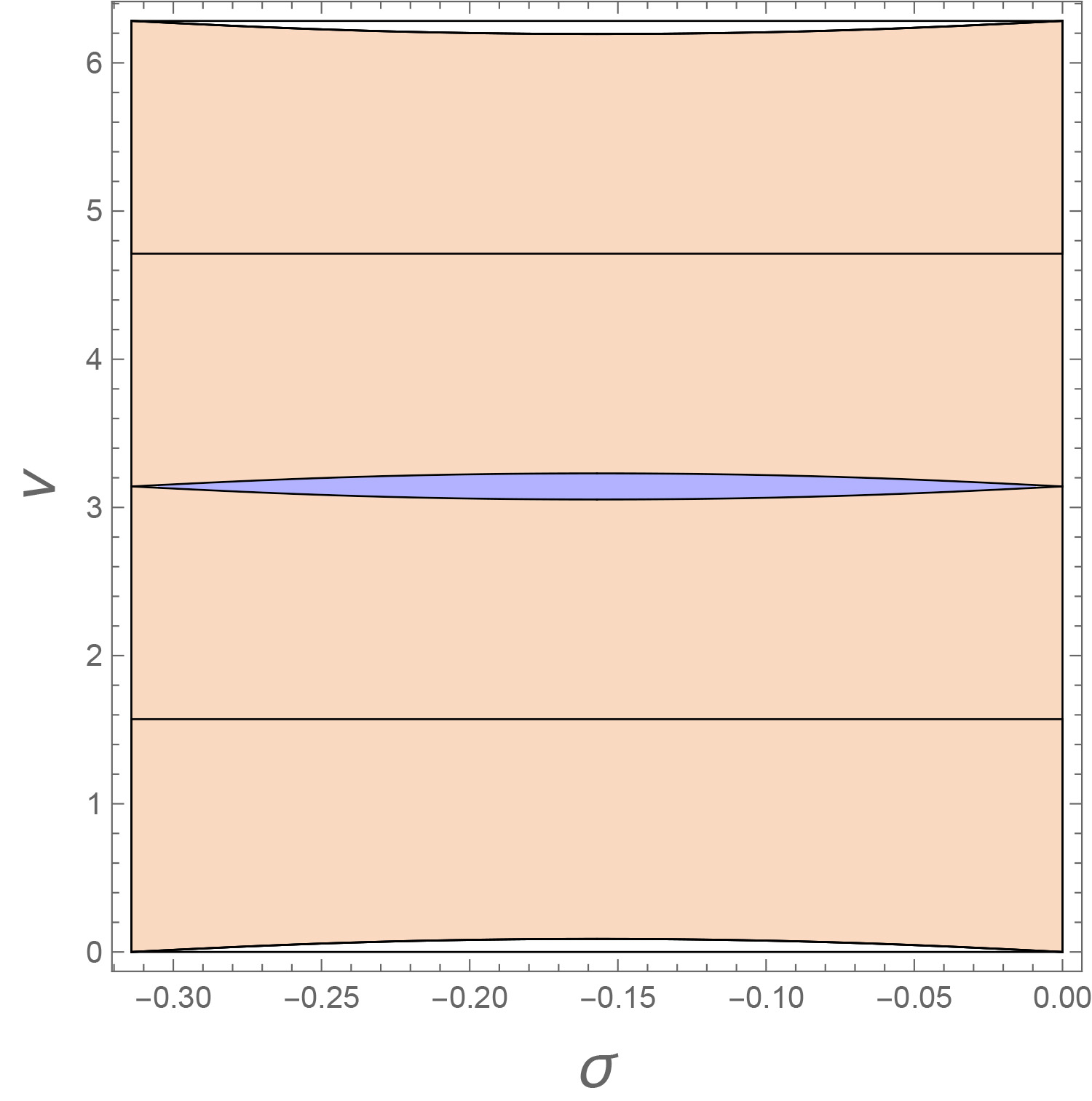}
    } \quad \quad \quad \quad \quad \quad
\subfigure[]{
    \includegraphics[width=0.35\textwidth]{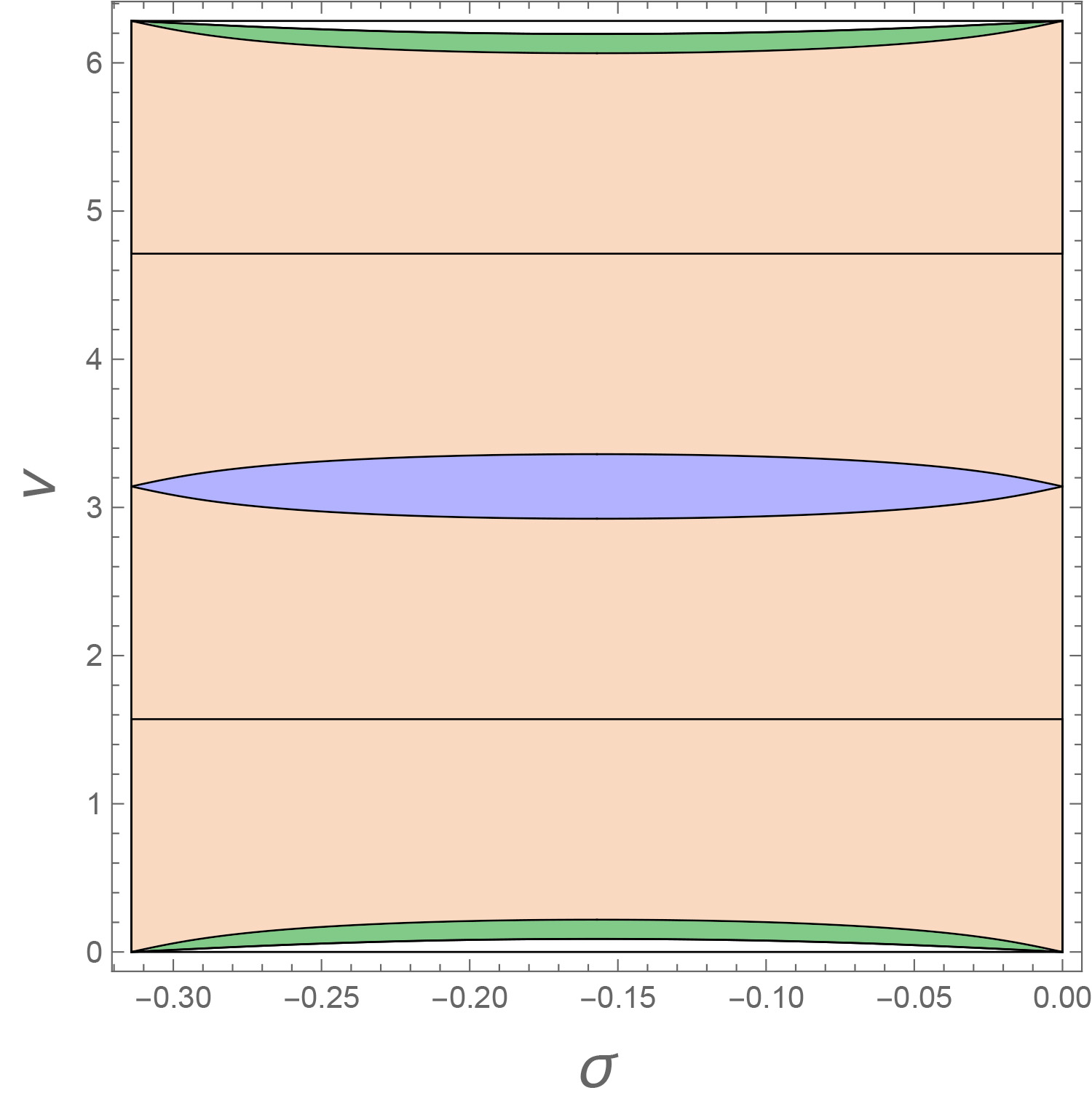}}\\\vspace{5mm}
\subfigure[]{
    \includegraphics[width=0.35\textwidth]{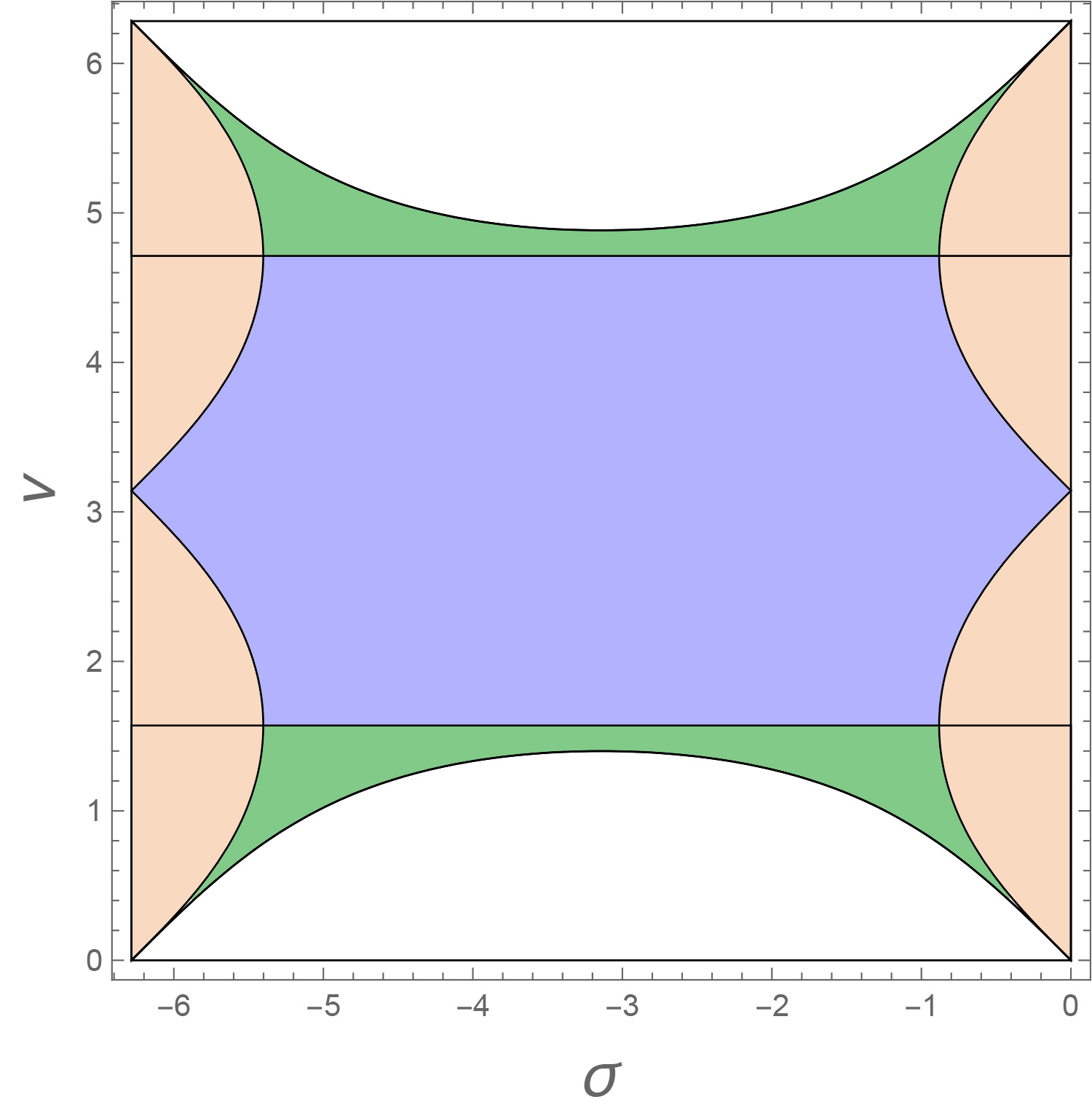}}
    \quad \quad \quad\quad \quad \quad
\subfigure[]{
    \includegraphics[width=0.35\textwidth]{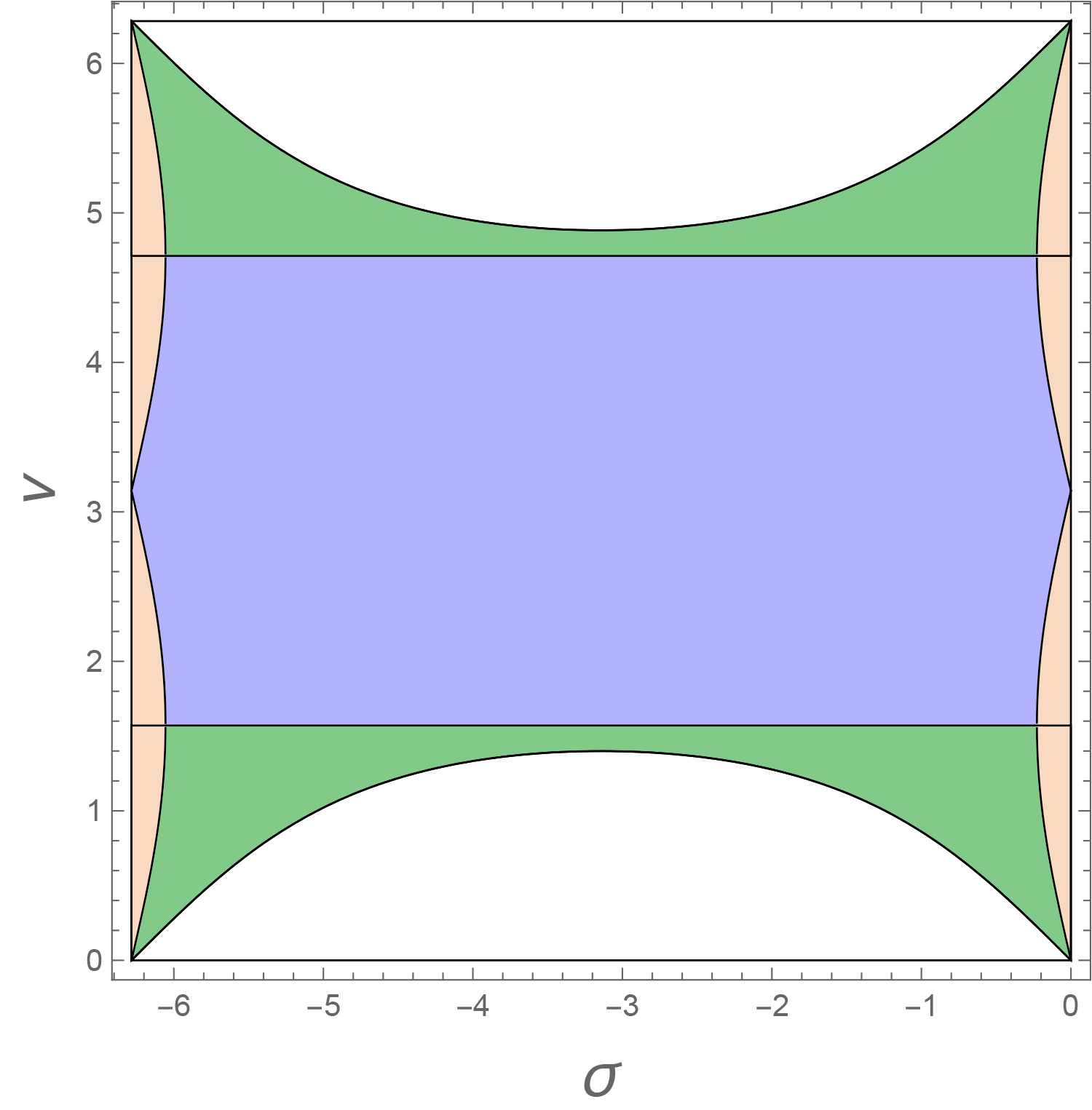}}
    \caption{Bulk teleportation and erasure in the one-sided infinite measurement setup. For heavy operator insertions in the white, blue, and green regions the connected geodesic is the shorter one, whereas for insertions in the orange regions the disconnected geodesic is shorter. The effect of heavy operator insertions in the blue (white) region can be reconstructed from the left (right) CFT independently of whether a measurement is performed or not. The effect of insertions in the green region can be reconstructed from the left CFT in the absence of measurement and from the right CFT in the presence of measurement, signaling bulk teleportation between the two sides. The effect of insertions in the orange regions cannot be reconstructed from either CFT in the presence of measurement: this information is being erased by the measurement. Note that teleportation occurs more extensively in the (connected) thermal phase and for large positive tensions, whereas erasure is more likely in the (disconnected) BTZ phase and for small (or negative) tensions. In all plots we set $R=1$, $\beta=2$ and choose the value of the IR cutoff to be $\epsilon=10^{-4}$. (a) BTZ phase with parameters $s=0.1$, $T=0.0001$. (b) BTZ phase with parameters $s=0.1$, $T=0.9$. (c) Thermal AdS phase with parameters $s=2$, $T=0.0001$. (d) Thermal AdS phase with parameters $s=2$, $T=0.9$.}
    \label{fig:teleportation1sided}
\end{figure}

The results of our analysis for the measurement of two semi-infinite intervals in the left CFT are reported in Figure \ref{fig:teleportation1sided}. We represent the boundary domain in BTZ/thermal AdS coordinates $(\sigma,\nu)$ and shade with different colors the regions in which operator insertions lead to teleportation, erasure, or neither. We find that bulk teleportation from the left to the right side occurs more extensively in the (connected) thermal phase and erasure is more likely in the (disconnected) BTZ phase. Within either of the two phases, teleportation is favored for large positive values of the tension, while erasure is favored for small (or negative) values of the tension. In particular, we focused here on the better-understood $T\geq 0$ case. However, applying our analysis to the $T<0$ case we find that even information associated with operator insertions with $y^*\in (\beta/4,\beta/2)$---which for $T\geq 0$ is accessible from the right CFT both in the absence and in the presence of measurement---can be erased.\footnote{This seems to suggest that by projecting part of the left CFT on a Cardy state with negative boundary entropy we are able to affect the entanglement wedge of the right CFT. This puzzling feature, which arose in a similar fashion in \cite{Antonini:2022sfm}, could be evidence of the non-physical nature of solutions involving negative tension branes.}

Note that this result is qualitatively different from the one obtained in lower dimensional models \cite{kourkoulou2017pure,Antonini:2022lmg}, where teleportation between the two sides occurs when most of the microscopic system living on one side is measured. We would also like to remark that, although the present analysis does not allow us to detect bulk teleportation from the measured part to the unmeasured part of the left CFT, the results of \cite{Antonini:2022sfm} suggest that this also generically takes place.

\begin{figure}
    \centering
    \subfigure[]{
    \includegraphics[width=0.35\textwidth]{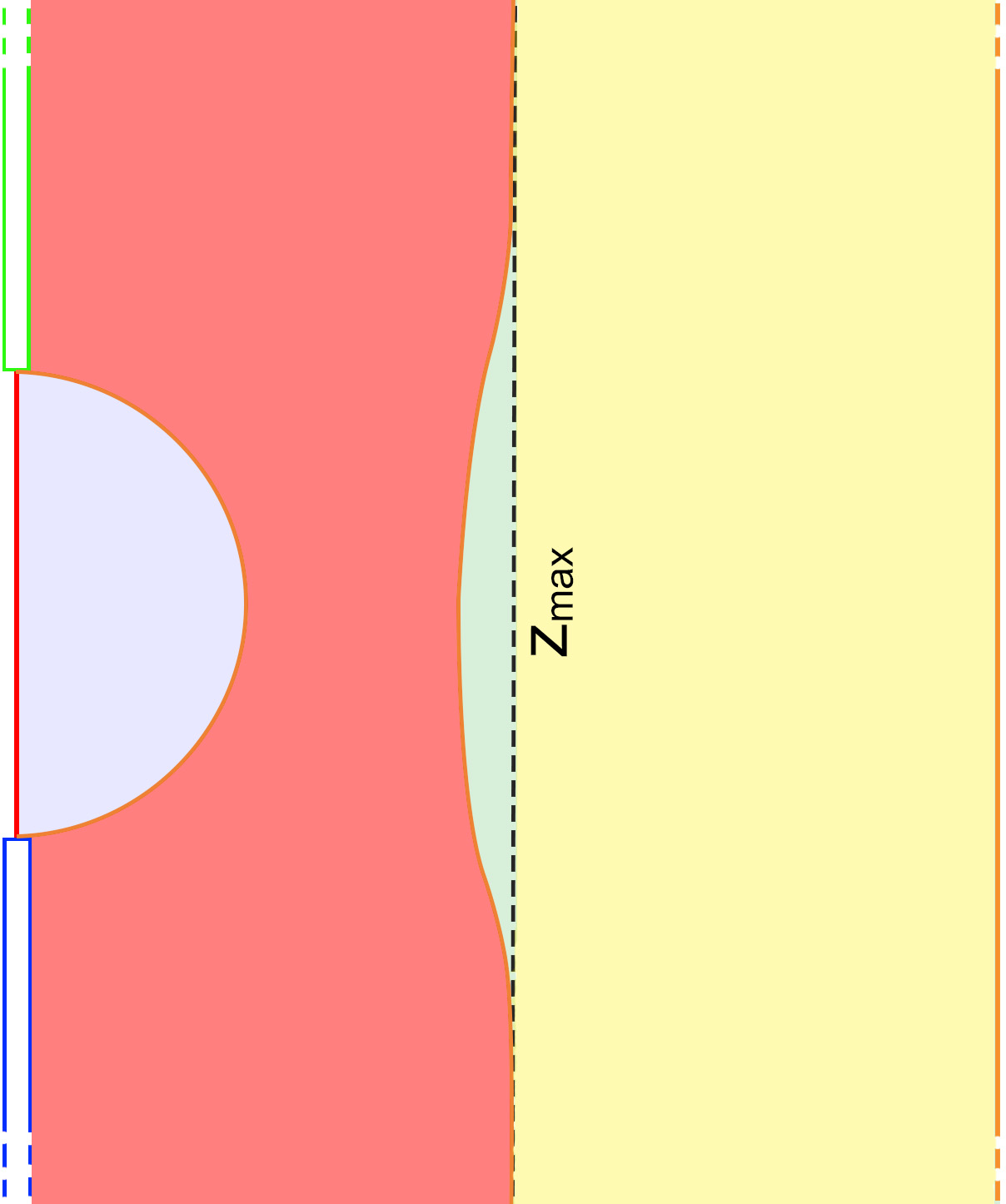}
    } \quad \quad \quad \quad \quad
\subfigure[]{
    \includegraphics[width=0.35\textwidth]{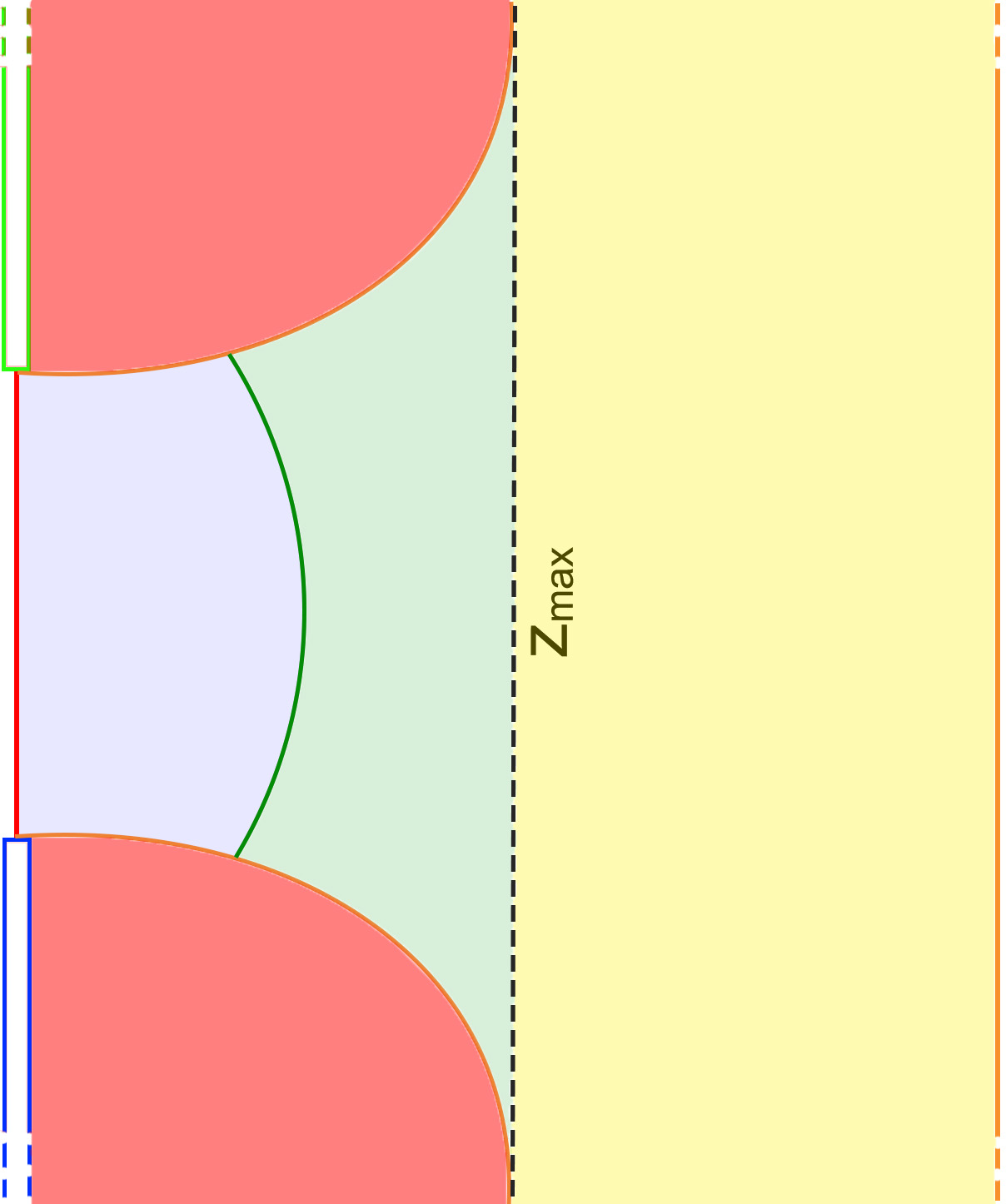}}
    \caption{A schematic illustration of the $y=0,\pm \beta/2$ bulk time reflection symmetric slice  for the one-sided infinite measurement setup with positive tension. The red (orange) line indicates the unmeasured part of the left CFT (the right CFT) and the two slits are represented in green and blue. The region shaded in red is bounded and cut off by the ETW brane. The blue region is the post-measurement entanglement wedge of the left CFT, whereas the union of the the yellow and the green regions is the post-measurement entanglement wedge of the right CFT. In particular, the yellow region is part of the entanglement wedge of the right CFT even without measurement, whereas the green region is ``teleported'' from the left to the right CFT by the measurement. (a) BTZ phase. The time-reflection symmetric slice is disconnected, implying that the RT surface for the right CFT (or, equivalently, the unmeasured part of the left CFT) is the empty set and the two CFTs are disentangled by the measurement. Only a small portion of the pre-measurement entanglement wedge of the left CFT is teleported into the right CFT. (b) Thermal AdS phase. The time reflection symmetric slice is connected and the RT surface, which has a finite length given by equation (\ref{eq:RTsurface}), is depicted as a dark green line. A large portion of the pre-measurement entanglement wedge of the left CFT is teleported into the right CFT.}
    \label{fig:1sidedcartoon}
\end{figure}

Finally, we report for completeness in Figure \ref{fig:1sidedcartoon} a schematic illustration of the $y=0,\pm \beta/2$ slice of the post-measurement geometry in the BTZ and thermal AdS phases. In the thermal AdS phase we also qualitatively depict the RT surface for the whole right CFT or, equivalently, for the unmeasured part of the left CFT (in the BTZ phase the RT surface is simply the empty set). The portions of the pre-measurement entanglement wedge of the left CFT ``teleported'' into the right CFT by the measurement in the two phases are also shown. We remark that this should be regarded only as a useful and intuitive cartoon of what the $y=0,\pm\beta/2$ slice would look like in a setup where the measurement is physically regularized; we remind that the measurement setup studied in the present paper is not regularized and the  $y=0,\pm\beta/2$ bulk time reflection symmetric slice is actually singular.

\subsection{Infinite intervals: two-sided measurement} \label{sec:infinite-two}

\subsubsection{Slit prescription} 
In the previous subsection we saw that the measurement of two semi-infinite intervals on one side of the TFD was able to disconnect the Einstein-Rosen bridge.  
In this subsection, we investigate an analogous question in the case where one semi-infinite interval is measured in each one of the two CFTs.

\begin{figure}
    \centering
\subfigure[$x$ coordinate]{
    \includegraphics[width=0.35\textwidth]{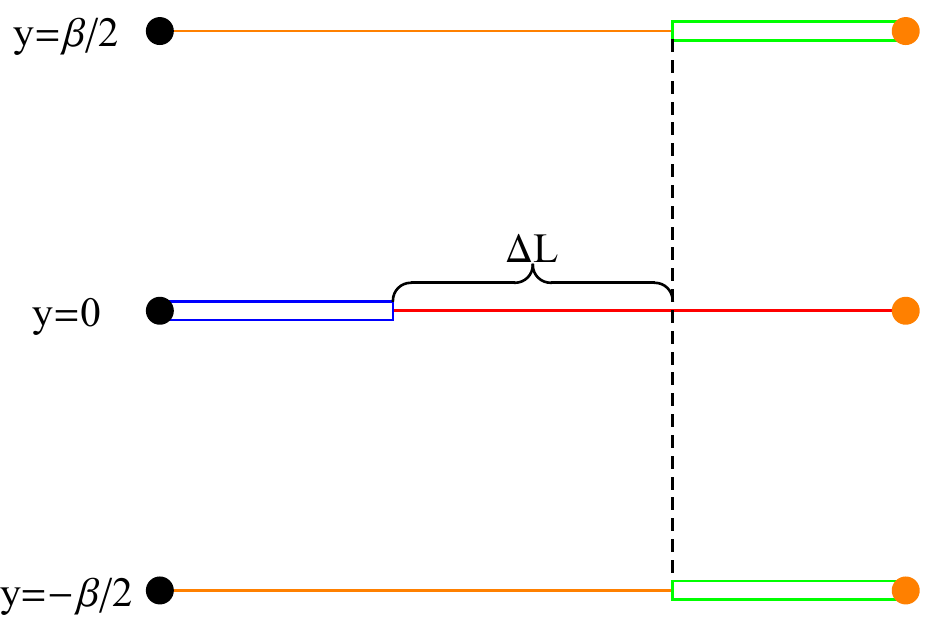}
    } \quad \quad \quad \quad \quad \quad
\subfigure[$X$ coordinate]{
    \includegraphics[width=0.35\textwidth]{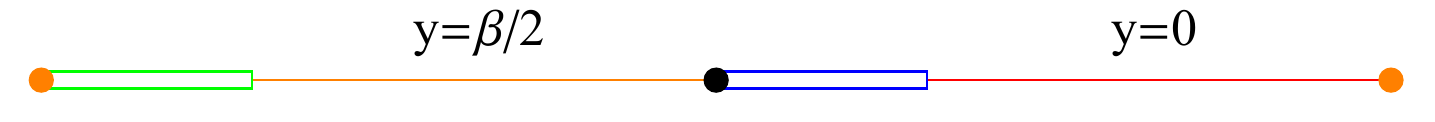}} \\\vspace{5mm}
\subfigure[$\zeta$ coordinate]{
    \includegraphics[width=0.35\textwidth]{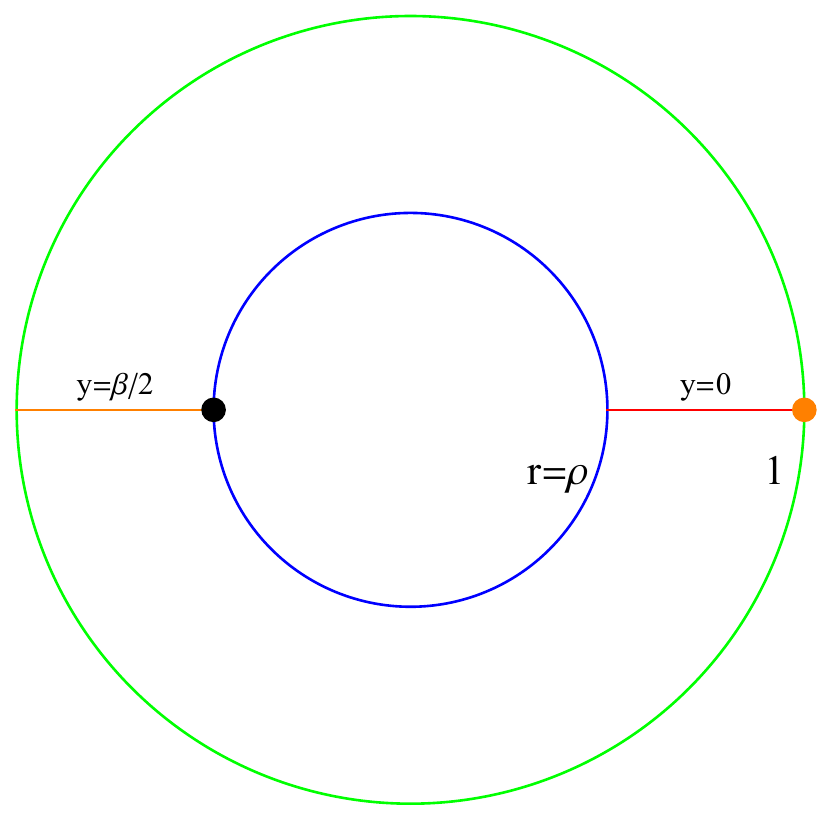}}
    \quad \quad \quad\quad \quad \quad
\subfigure[$w$ coordinate]{
    \includegraphics[width=0.35\textwidth]{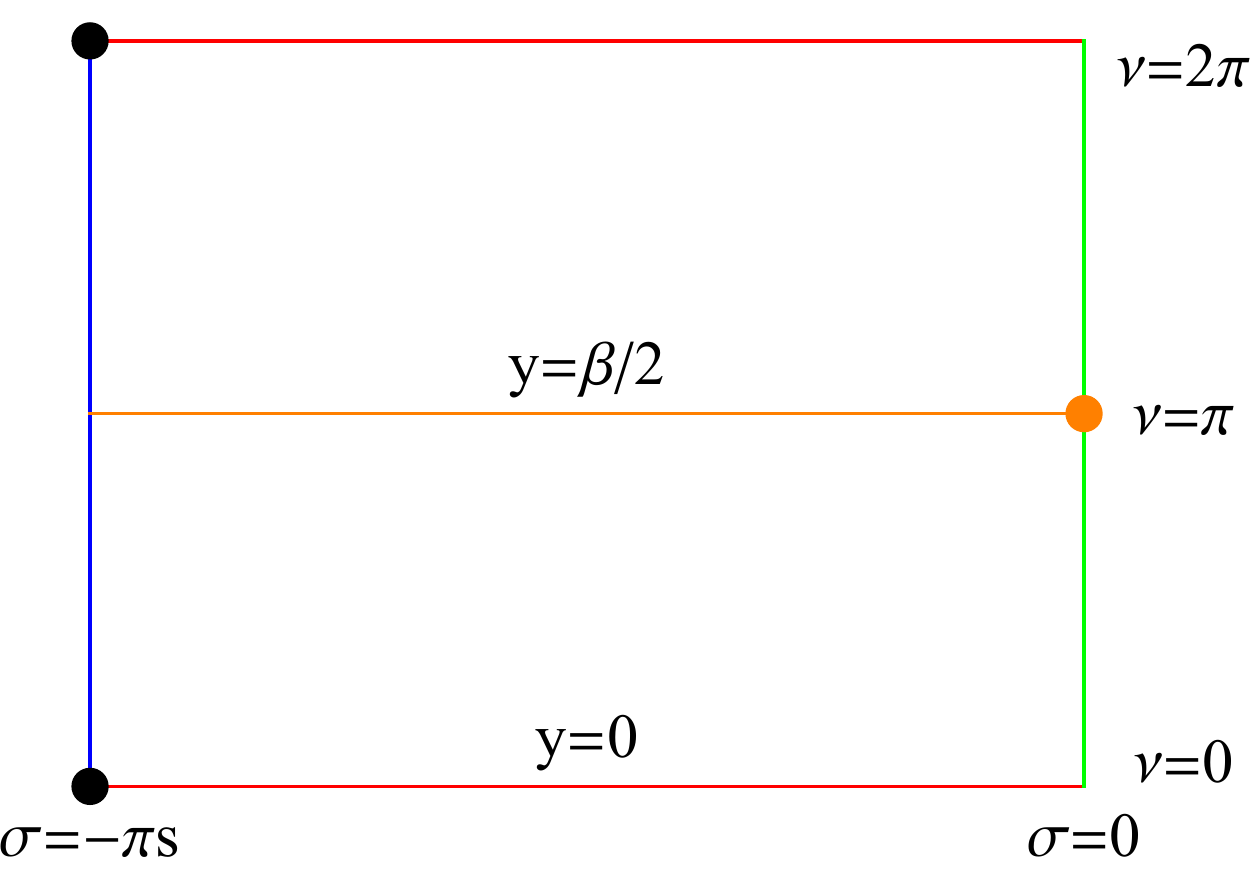}}
    \caption{The four different sets of coordinates used to describe the measurement of a semi-infinite interval in both of the two CFTs. The conformal maps between the different coordinates are given in Appendix \ref{append:infinite2sideconf}.
    The red (orange) line denotes the unmeasured part of the left CFT at $y=0$ (the right CFT at $y=\beta/2$). 
    They are time reflection symmetric lines. The green (blue) color denotes the first (second) slit. The black (orange) dots represent negative (positive) infinity in $x$ in both left and right CFTs.
    (a) We start with an infinitely long cylinder with one infinite slit in each of the two CFTs.
    The first (second) slit is located at $y=\beta/2$ ($y=0$), i.e. in the right (left) CFT. 
    $\Delta L$ denotes the region which is unmeasured in both CFTs.
    Recall that $y\sim y+\beta$, so that $y=-\beta/2=\beta/2$.
    The black (orange) dots represents negative (positive) infinity in $x$ for both left and right CFTs. 
    (b) We map this to the 2D plane with two radial slits on the real axis. 
    The first (second) slit is mapped to a semi-infinite (finite) segment.
    The unmeasured parts of the two CFTs are also mapped to the real axis, where the left (right) CFT is mapped to a semi-infinite (finite) segment. 
    Negative (positive) infinity is mapped to the origin (infinity).
    (c) We next map to the annulus. 
    The first slit is the outer edge $r= 1$ and the second slit is the inner edge $r= \rho$. 
    The unmeasured part of the left (right) CFT is mapped to a segment $r \in (\rho, 1)$ ($r \in (-1,-\rho)$).
    Negative (positive) infinity in $x$ is mapped to $r= \rho, \theta = \pi$ ($r=1, \theta = 0$). 
    Here we use $\zeta = r e^{i \theta}$, and $\rho$ is a constant determined by the measurement parameter $\Delta L$.
    (d) Finally, we map to a finite cylinder with $\nu \sim \nu + 2\pi$. 
    The fist (second) slit is mapped to the right (left) edge at $\sigma =0$ ($\sigma = -\pi s$). 
    The unmeasured part of the left (right) CFT is mapped to a segment $\nu = 0,2\pi$ ($\nu = \pi$). 
    Negative (positive) infinity in $x$ is mapped to $\sigma = -\pi s, \nu = 0,2\pi$ ($\sigma = 0, \nu = \pi$). }
   \label{fig:map-infinite-two}
\end{figure}

As usual, we start with a TFD state with temperature $\beta$ living in a 2D cylinder $(x,y)$ with $y \sim y + \beta$. 
The measurement is now described by two infinite slits located on the time reflection symmetric slice $y=0,\beta/2$ (see Fig. ~\ref{fig:map-infinite-two} (a))
\bea \label{eq:slit-2-2side}
    && \text{first slit:} \quad   x > \frac{\Delta L}2 , \quad y = \frac{\beta}2, \\
    && \text{second slit:} \quad x < - \frac{\Delta L}2  , \quad y = 0.
\eea
$\Delta L$ now controls the size of the region which is unmeasured in both CFTs. If $\Delta L>0$, there is an unmeasured region corresponding to the same range of $x$ coordinate in both CFTs, while if $\Delta L<0$ there is no such ``overlapping'' region.


As above, we implement a series of conformal transformations to map our domain to the finite cylinder so that we can construct the bulk dual.
The three conformal transformations, which are described in detail in Appendix \ref{append:infinite2sideconf}, are depicted in Fig.~\ref{fig:map-infinite-two}. 
The composed conformal transformation from the original $(x,y)$ coordinates (Fig.~\ref{fig:map-infinite-two} (a)) to the final $w$ coordinate (Fig.~\ref{fig:map-infinite-two} (d)) is given by
\bea
    x + i y = \frac{\beta}{2\pi} \log \left[ - \frac{\theta_3^2 \left(-\frac{i(\sigma + i \nu)}2 , e^{-\pi s} \right)}{\theta_1^2 \left(- \frac{i(\sigma+ i \nu)}2 , e^{-\pi s} \right)} \right].
    \label{eq:2sidedmap}
\eea
The parameter $s$ is related to the parameter $\Delta L$, see Fig. \ref{fig:s-infinite-two} and equation (\ref{eq:s-infinite-two}) in Appendix \ref{append:infinite2sideconf}. Recall that in the original $(x,y)$ coordinates, the left CFT corresponds to the $y=0$ line, and the right CFT to $y=\beta/2$. In the final $(\sigma,\nu)$ coordinates, the unmeasured part of the left CFT is mapped to the $\nu=0,2\pi$ segment and the unmeasured part of the right CFT is mapped to the $\nu=\pi$ segment. The two slits are again mapped to the circles $\sigma=0$ (first slit) and $\sigma=-\pi s$ (second slit), while negative (positive) infinity in $x$ is mapped to $\sigma=-\pi s$, $\nu=0,2\pi$ ($\sigma=0$, $\nu=\pi$).

Note that the final domain in $w$ coordinates is almost identical to the one obtained in Section \ref{sec:infinite-one} for the measurement of two semi-infinite intervals on only one of the two CFTs (see Fig. \ref{fig:map-infinite}). The only differences are where positive infinity in $x$ is mapped, by the relationship between $\Delta L$ and $s$, and by the fact that the location of the unmeasured parts of the left and right CFTs are switched. Therefore, all the discussion (see Sections \ref{sec:infinite-one-slit} and \ref{sec:infinite-one-HP}) about the phase structure and its implications for the connectivity of the Lorentzian geometry and the entanglement structure of the dual microscopic theory remains valid and unmodified in the two-sided measurement setup of the present subsection.


\begin{figure}
    \centering
    \includegraphics[width=0.5\textwidth]{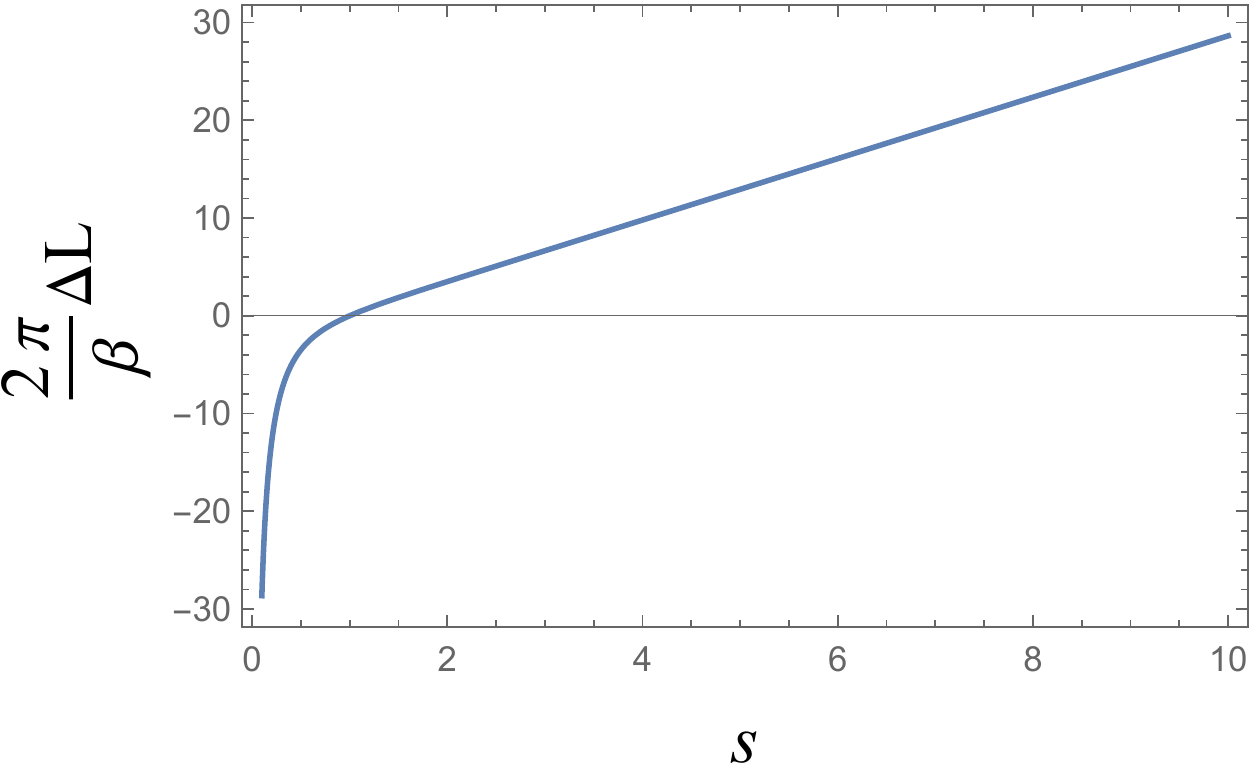}
    \caption{The relationship between the measurement parameter $\Delta L/\beta$ and the conformal transformation parameter $s$ for the case where semi-infinite intervals are measured on each side of the TFD. The parameter $s$ is a monotonically increasing function of $\Delta L$. In particular, when $\Delta L =0$, $s =1$. }
    \label{fig:s-infinite-two}
\end{figure}

\subsubsection{Bulk spacetime and Hawking-Page transition}

Given the similarities between the final domains, the spacetime dual to the final finite cylinder (in $w$ coordinates) will also be completely analogous to the one studied in Subsection \ref{sec:infinite-one-HP}, see Fig. \ref{fig:gravity-dual_infinite}. In particular, the phase boundary between the BTZ black hole phase and the thermal AdS phase is still determined by the value of the parameter $s$ and the brane tension $T$. The BTZ black hole phase again corresponds to a ``disconnected'' phase, in which the Einstein-Rosen bridge connecting the two CFTs in the Lorentzian spacetime dual to the TFD state is destroyed by the measurement. On the other hand, the thermal AdS phase is a ``connected'' phase, in which the Einstein-Rosen bridge is preserved after the measurement. The two phases correspond to a disentangled and an entangled phase in the dual CFTs, and the mutual information between the unmeasured regions in the two CFTs takes the same form (\ref{eq:mi}) as in the one-sided measurement case.

The only differences reside in the switched locations of the left and right CFTs (which is irrelevant for our discussion), in the relationship between $s$ and $\Delta L$---which is now given by equation (\ref{eq:s-infinite-two})---and in the meaning of the parameter $\Delta L$. As we have discussed, this quantifies the size of the region in $x$ coordinate which is unmeasured in both CFTs. Intuitively, if $\Delta L=0$ we are measuring exactly half of the system ($x>0$ for the right CFT and $x<0$ for the left CFT), if $\Delta L>0$ we are measuring less than half of the system, and vice versa for $\Delta L<0$. As we have seen, $\Delta L=0$ corresponds to $s=1$, which is the critical point for vanishing tension $T=0$. This leads to an intuitive result: in the $T=0$ case, i.e. if we postselect on a Cardy state with zero boundary entropy, measurement of more than half of the system will disentangle the two CFTs, destroy the Einstein-Rosen bridge, and disconnect the dual Lorentzian spacetime. 

\subsubsection{Heavy operator insertions, teleportation and information erasure}
\label{sec:2sidedteleportation}

The analysis of heavy operator insertions in the two-sided measurement setup is completely analogous to the one carried out for the one-sided measurement in Section \ref{sec:1sidedteleportation}. The only difference is that the map between the original $(x,y)$ coordinates and the BTZ/thermal AdS coordinates $(\sigma,\nu)$ is now given by equation (\ref{eq:2sidedmap}). Recall that with this map the left CFT is mapped to the $\nu=0,2\pi$ slice, whereas the right CFT is mapped to the $\nu=\pi$ slice. As a result, for given insertion points $P_1=(\sigma^*,\nu^*)$, $P_2=(\sigma^*,2\pi-\nu^*)$ for which the constant-$\sigma$ connected geodesic is dominant, if the geodesic crosses the reflection symmetric slice at $\nu=0,2\pi$ (i.e. if $\nu^*\in (0,\pi/2)$) the effect of the operator insertion can be detected from the left CFT; if the geodesic crosses the reflection symmetric slice at $\nu=\pi$ instead (i.e. if $\nu^*\in (\pi/2,\pi)$), the effect of the operator insertion can be detected from the right CFT. The lengths of the connected and disconnected geodesics in the two phases are still given by equations (\ref{eq:btzconn}), (\ref{eq:btzdisconn}), (\ref{eq:thermalconn}), and (\ref{eq:thermaldisconn}).

\begin{figure}
    \centering
    \subfigure[]{
    \includegraphics[width=0.35\textwidth]{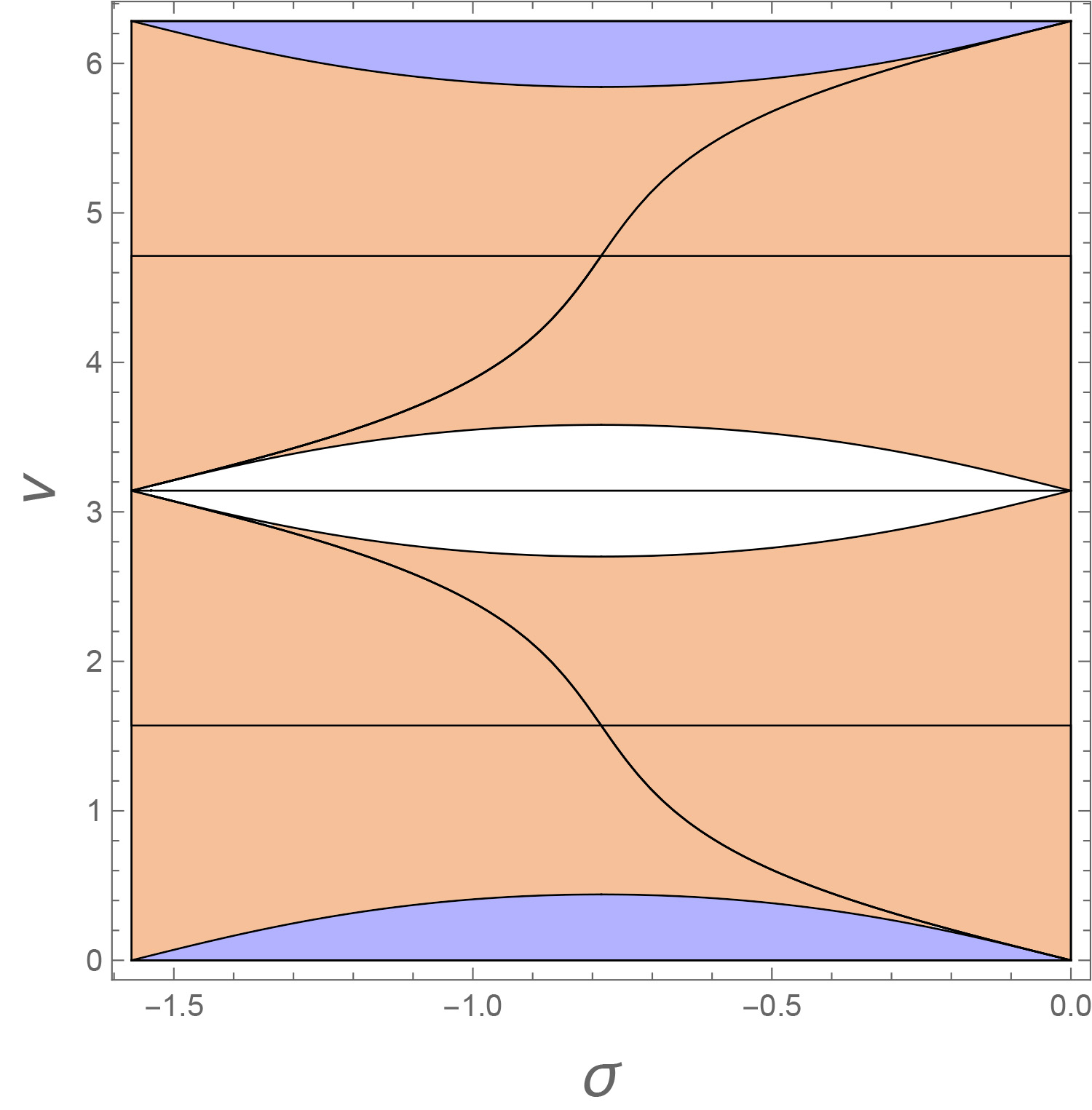}}
    \quad \quad \quad \quad \quad \quad
\subfigure[]{
    \includegraphics[width=0.35\textwidth]{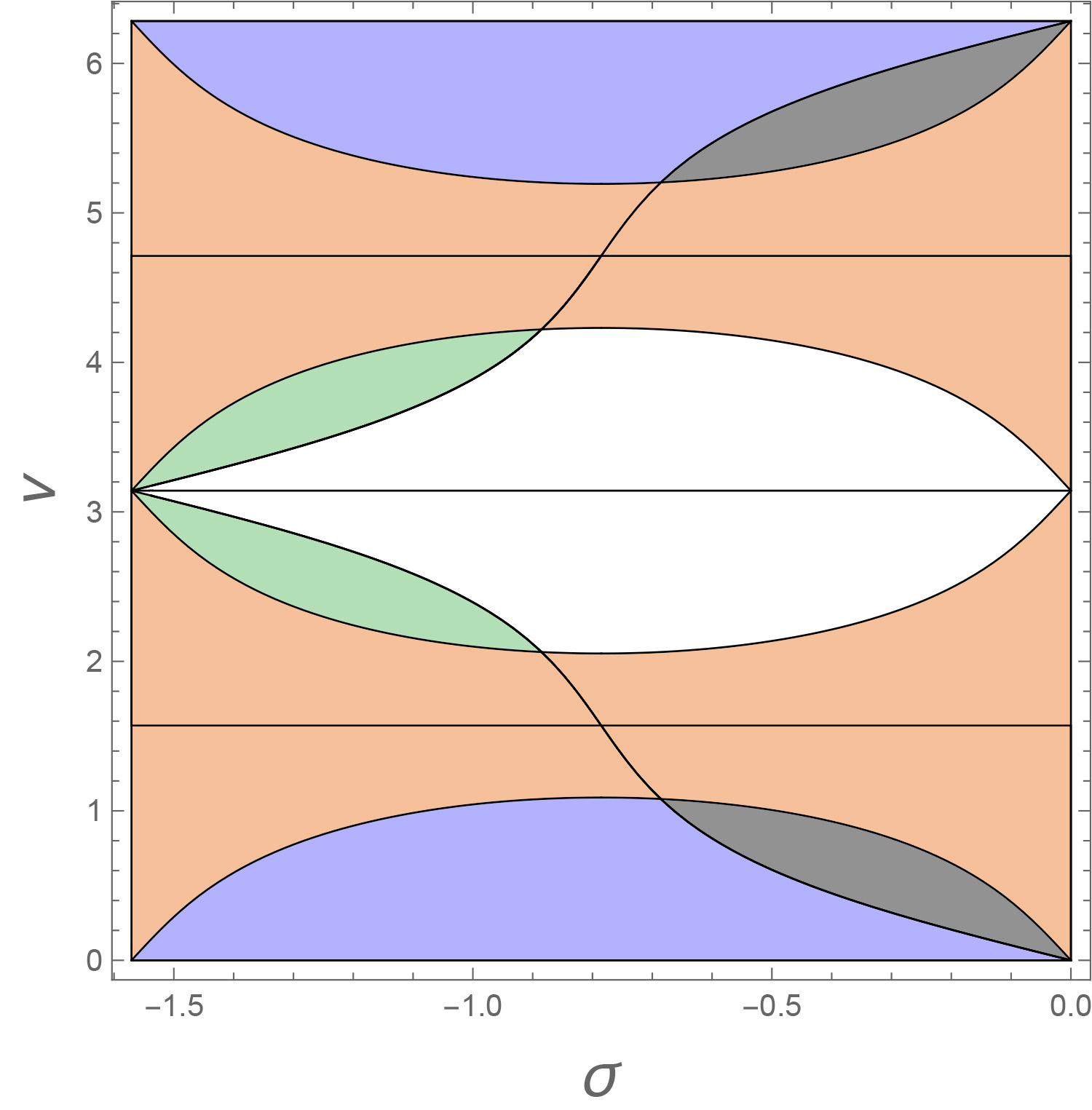}}\\\vspace{5mm}
\subfigure[]{
    \includegraphics[width=0.35\textwidth]{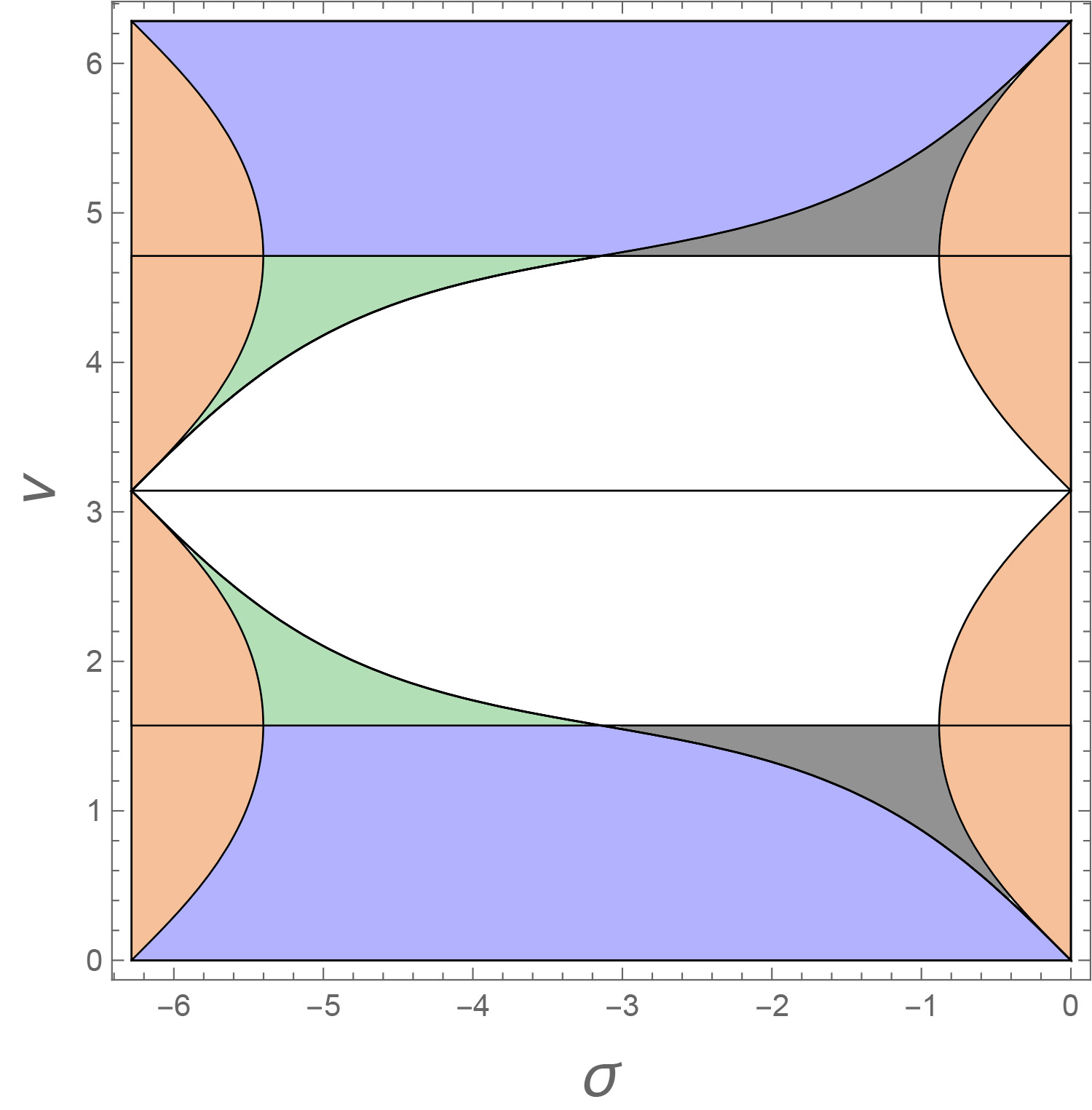}}
    \quad \quad \quad\quad \quad \quad
\subfigure[]{
    \includegraphics[width=0.35\textwidth]{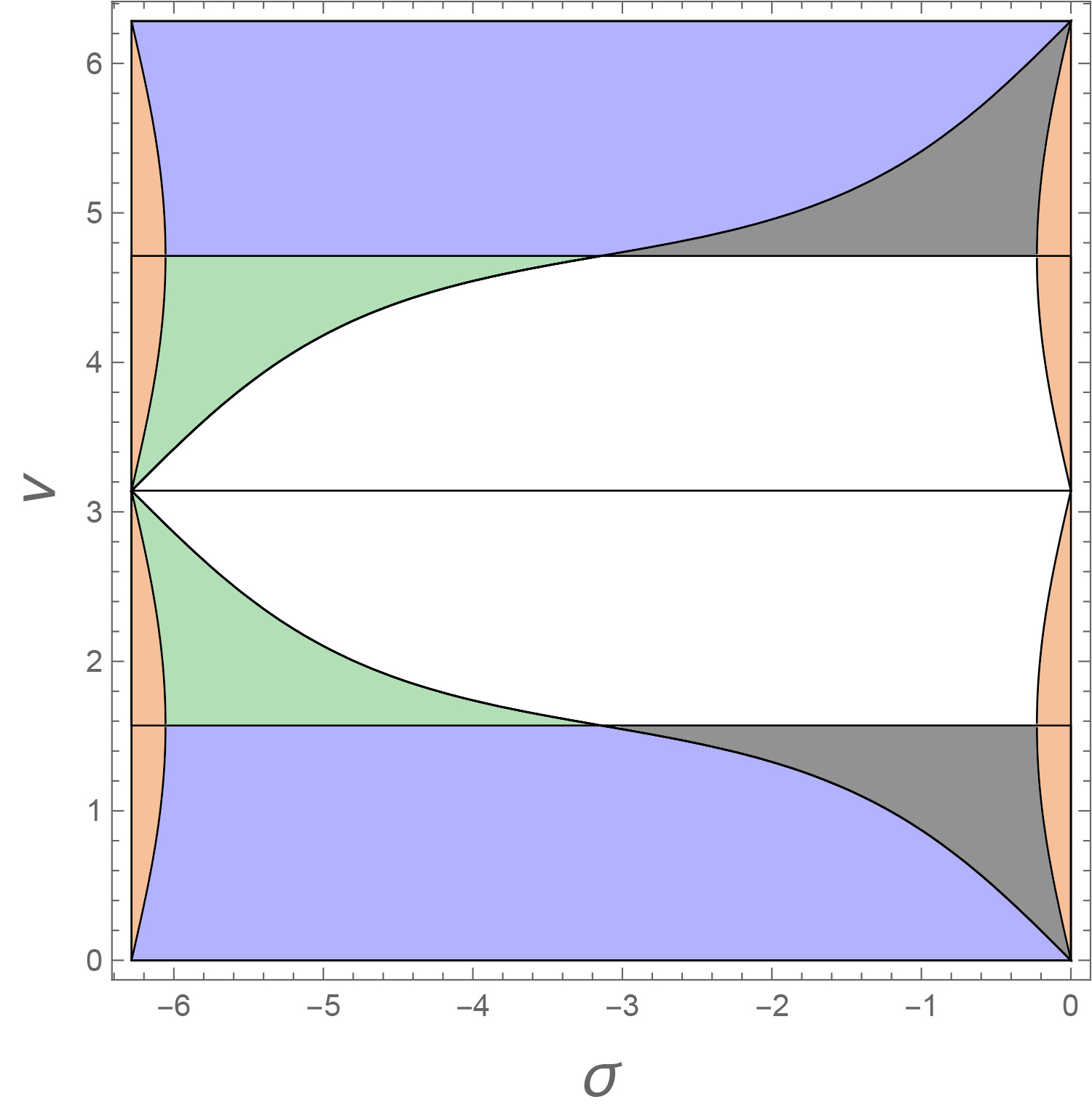}}
    \caption{For heavy operator insertions in the white, blue, green, and gray regions the connected geodesic is the shorter one, whereas for insertions in the orange regions the disconnected geodesic is shorter. The effect of heavy operator insertions in the blue (white) region can be reconstructed from the left (right) CFT independently of whether a measurement is performed or not. The effect of insertions in the green (gray) region can be reconstructed from the left (right) CFT in the absence of measurement and from the right (left) CFT in the presence of measurement, signaling symmetric bulk teleportation between the two sides. The effect of insertions in the orange regions cannot be reconstructed from either CFT in the presence of measurement: this information is being erased by the measurement. Teleportation occurs more extensively in the (connected) thermal phase and erasure is more likely in the (disconnected) BTZ phase, with teleportation favored for large positive values of the tension and erasure favored for small (or negative) values of the tension. In all plots we set $R=1$, $\beta=2$ and choose the value of the IR cutoff to be $\epsilon=10^{-4}$. (a) BTZ phase with parameters $s=0.5$, $T=0.0001$. (b) BTZ phase with parameters $s=0.5$, $T=0.9$. (c) Thermal AdS phase with parameters $s=2$, $T=0.0001$. (d) Thermal AdS phase with parameters $s=2$, $T=0.9$.}
    \label{fig:teleportation2sided}
\end{figure}

The results of our analysis for the two-sided measurement case are reported in Figure \ref{fig:teleportation2sided}. We again represent the boundary domain in the BTZ/thermal AdS coordinates $(\sigma,\nu)$ and shade with different colors the regions in which operator insertions lead to teleportation, erasure, or neither. Similar to the one-sided measurement analysis, we find that teleportation between the two sides occurs more extensively in the (connected) thermal phase and erasure is more likely in the (disconnected) BTZ phase, with teleportation favored for large positive values of the tension and erasure favored for small (or negative) values of the tension. We again focused our attention on the better-understood $T\geq 0$ case. Note that in the double-sided measurement case, part of the pre-measurement entanglement wedge of the left CFT is ``teleported'' into the right CFT by the measurement and, at the same time, part of the pre-measurement entanglement wedge of the right CFT is ``teleported'' into the left CFT.

\begin{figure}
    \centering
    \subfigure[]{
    \includegraphics[width=0.35\textwidth]{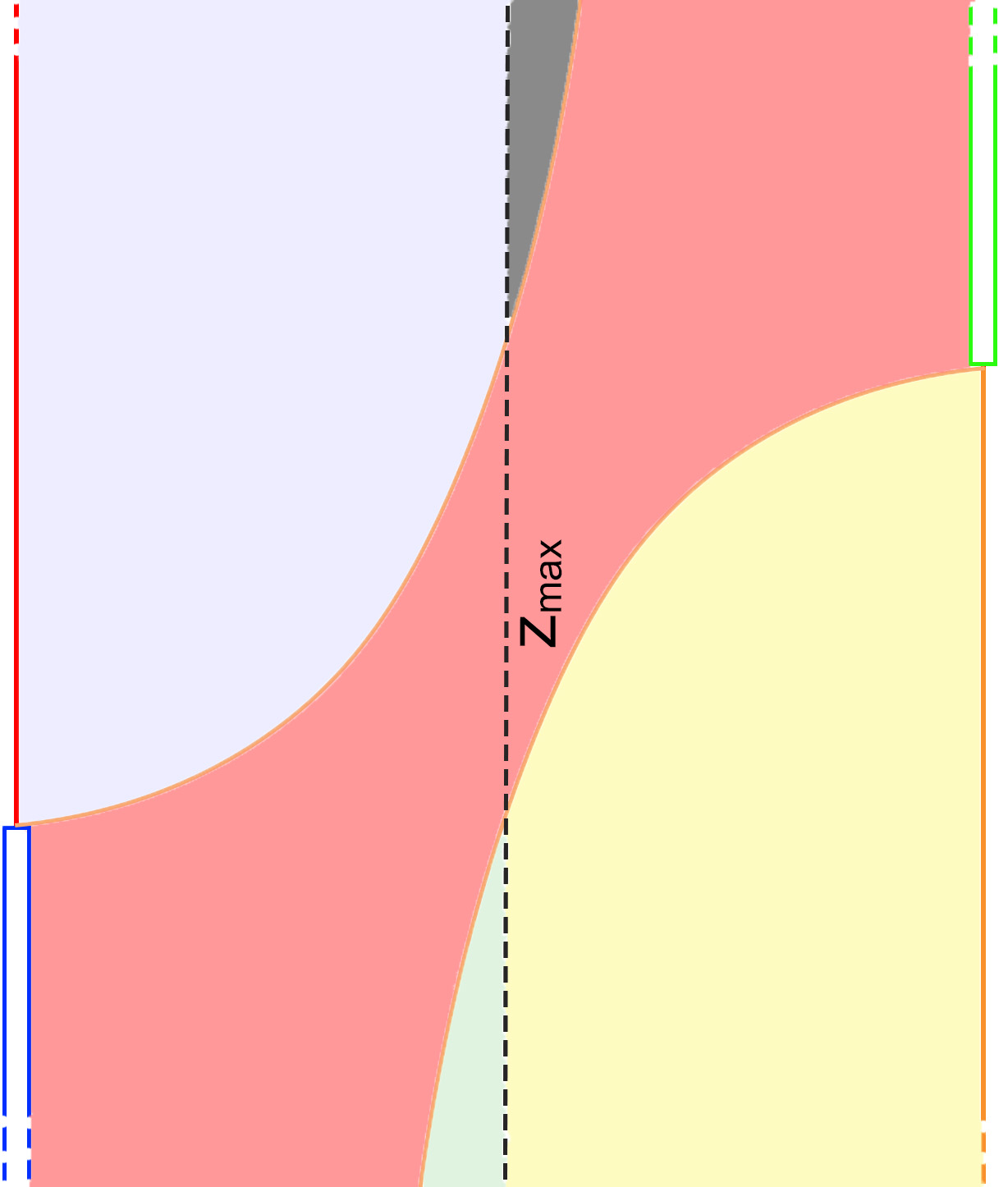}
    } \quad \quad \quad \quad \quad
\subfigure[]{
    \includegraphics[width=0.35\textwidth]{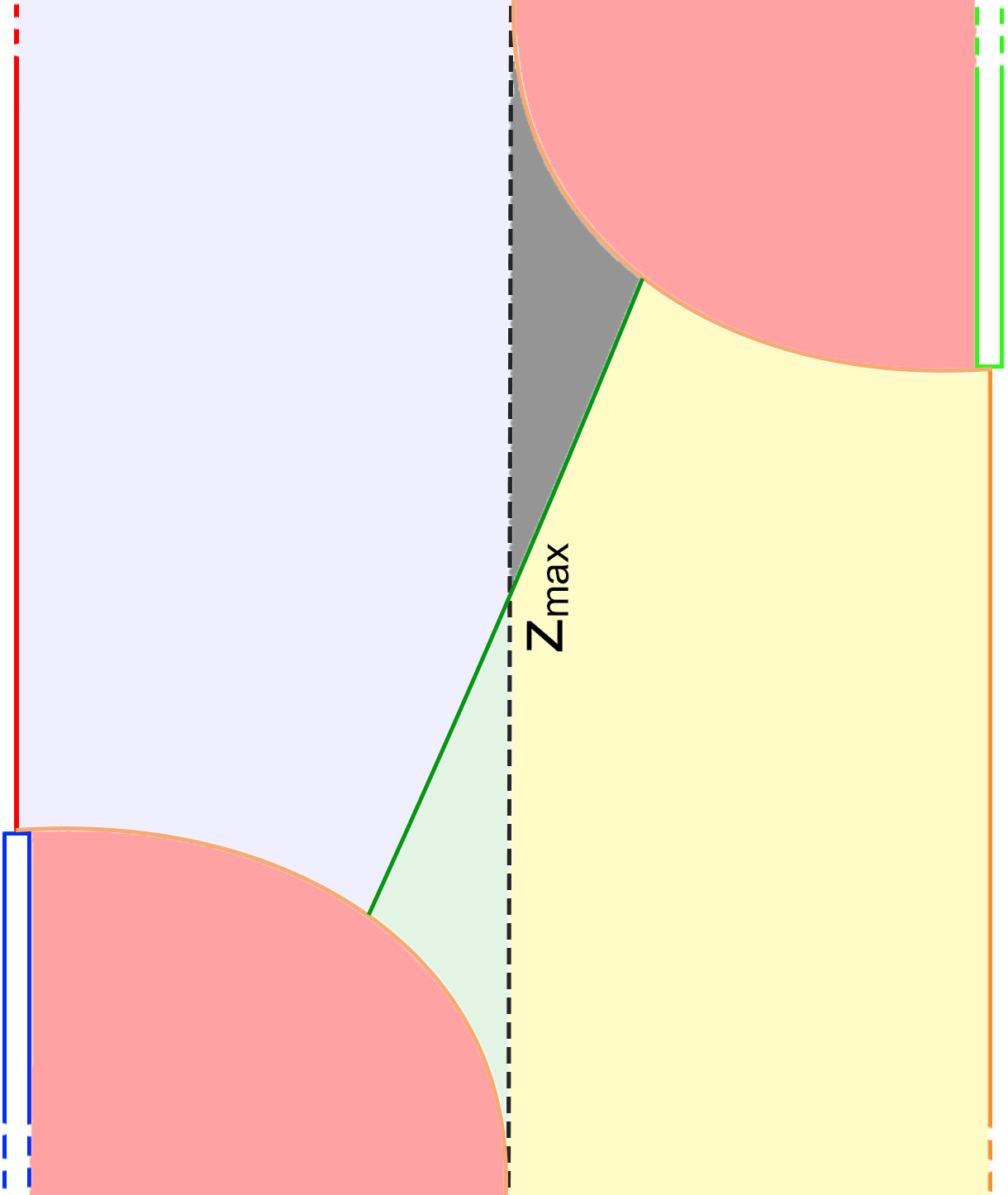}}
    \caption{Schematic illustration of the $y=0,\pm \beta/2$ bulk time reflection symmetric slice  for the two-sided infinite measurement setup with positive tension. The red (orange) line indicates the unmeasured part of the left (right) CFT and the two slits are represented in green and blue. The region shaded in red is bounded and cut off by the ETW brane. The union of the blue and the gray region is the post-measurement entanglement wedge of the left CFT, whereas the union of the yellow and green regions is the post-measurement entanglement wedge of the right CFT. In particular, the blue (yellow) region is part of the entanglement wedge of the left (right) CFT even without measurement, whereas the gray (green) region is ``teleported'' from the right (left) to the left (right) CFT by the measurement. (a) BTZ phase. The time-reflection symmetric slice is disconnected, implying that the RT surface for the unmeasured part of the right CFT (or, equivalently, the unmeasured part of the left CFT) is the empty set and the two CFTs are disentangled by the measurement. Only a small portion of the pre-measurement entanglement wedge of the left (right) CFT is teleported into the right (left) CFT. (b) Thermal AdS phase. The time reflection symmetric slice is connected and the RT surface, which has a finite length given by equation (\ref{eq:RTsurface}), is depicted as a dark green line. A large portion of the pre-measurement entanglement wedge of the left (right) CFT is teleported into the right (left) CFT.}
    \label{fig:2sidedcartoon}
\end{figure}

In Figure \ref{fig:2sidedcartoon} we depict a schematic illustration of the $y=0,\pm\beta/2$ slice of the post-measurement geometry in the two phases for the two-sided measurement setup, showing the bulk regions teleported from the left to the right and vice versa. In the thermal phase we also depict the RT surface for the unmeasured part of the left CFT (or, equivalently, the unmeasured part of the right CFT). In the BTZ phase the RT surface is again the empty set. As above, this illustration is just a cartoon of the $y=0,\pm\beta/2$ slice we would obtain in a physically regularized setup.

\section{Projective measurement in CFT thermofield doubles}\label{sec:cft}

In this section, we study the measurements considered in Section \ref{sec:bulk} purely from the boundary CFT point of view. In particular, we will focus on the cases considered in Sections \ref{sec:infinite-one} and \ref{sec:infinite-two} where infinite intervals are measured either in one or in both of the CFTs, and calculate the entanglement entropy between the two sides after the measurement is performed. This will allow us to characterize the measurement-induced Hawking-Page transition seen above in terms of an entangling phase transition in the microscopic system, and to reproduce (up to corrections due to $1/N$-suppressed effects) the results for the mutual information obtained by computing the holographic entanglement entropy in the bulk dual geometry. 
A general framework to carry out the calculations of interest has been developed in Ref.~\cite{rajabpour2016entanglement}. 
We will first outline the main steps of the calculation, which are common to the two setups, and then discuss the two cases in detail in Sections \ref{sec:CFT1side} and \ref{sec:CFT2sides}, respectively.

We start from the $X$ coordinates (see Fig.~\ref{fig:map-infinite} for the one-sided measurement and Fig.~\ref{fig:map-infinite-two} for the two-sided measurement). 
Because we are interested in computing the von Neumann entropy between the two sides, we can insert a twist operator on either side, i.e. at either $y=0$ or $y=\beta/2$ (since the full state is pure, the two choices lead to the same result).
Here we will insert a twist operator in the left CFT ($y=0$) for measurements on one side, and in the right CFT ($y= \beta/2$) for the measurements on both sides. 
Recall that the Renyi entropy is given by
\bea\label{eq:renyi}
    S_\alpha = \frac1{1-\alpha} \left( \log Z_\alpha - \alpha \log Z_1 \right),
\eea
where $Z_\alpha$ is the partition function of the CFT on a Riemann surface given by $\alpha$ copies of the original domain, and with a branch cut at $y=0$ or $y=\beta/2$ (depending on which side we are computing the Renyi entropy of) that connects the $i$-th and the $(i \pm 1)$-th Riemann sheet~\cite{calabrese2009entanglement}. This is equivalent to computing an $\alpha$-point function of appropriate twist fields in a single copy. The role of the twist operator is precisely to account for the correct boundary condition at the branch cut. 
The von Neumann entropy can then be obtained by analytically continuing $\alpha$ and taking the $\alpha \rightarrow 1$ limit.

\begin{figure}
    \centering
\subfigure[$\tilde X$ coordinate]{
    \includegraphics[width=0.4 \textwidth]{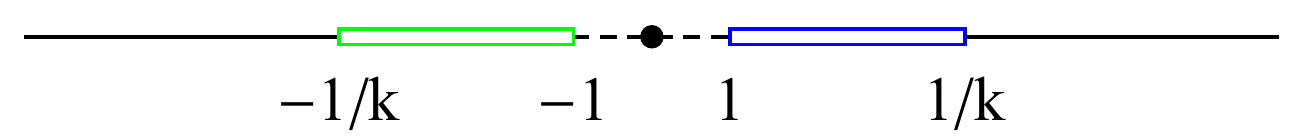}
    } \quad \quad \quad \quad \quad \quad
\subfigure[$\zeta$ coordinate]{
    \includegraphics[width=0.25\textwidth]{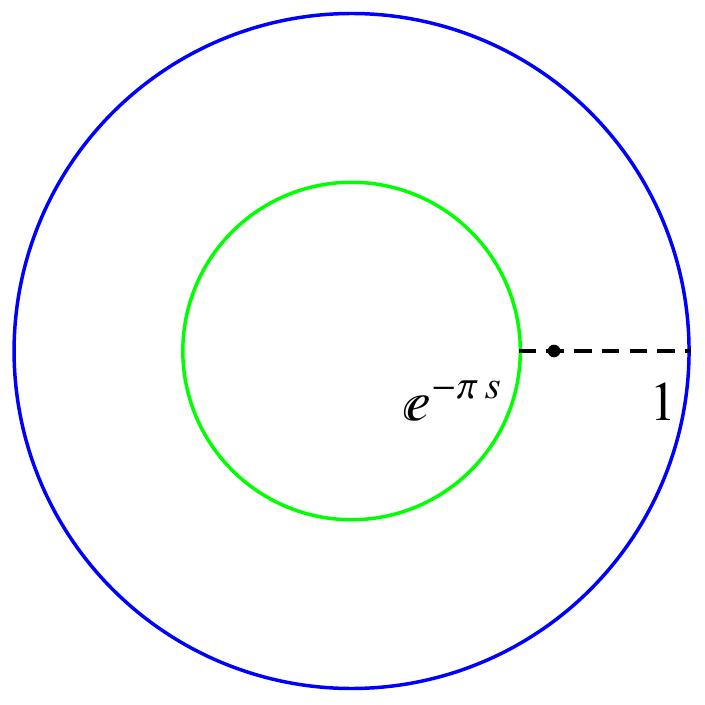}} \\\vspace{5mm}
    \caption{
    The green (blue) color denotes the first (second) slit.  
    The dashed line denotes a branch cut $(-1,1)$ induced by the twist operator.
    (a) 2D plane with two symmetric slits $(-1/k,-1)\cup (1,1/k)$, where $k < 1$ is a real number.  
    (b) Annulus. 
    The first slit is the outer edge $|\zeta| = 1$, the second slit is the inner edge $|\zeta|= e^{-\pi s}$, and the branch cut is indicated by the dashed line.
     }
   \label{fig:cft_symmetric_slit}
\end{figure}

In order to use the results of Ref.~\cite{rajabpour2016entanglement} directly, we consider a conformal transformation from the $X$ coordinate to a plane with two symmetric slits at $(- \frac1k , -1) \cup (1, \frac1k )$, which we denote by $\tilde X$.
The form of the conformal transformation in the two cases is given in Appendices \ref{append:infinite1sideconf} and \ref{append:infinite2sideconf}, respectively.
For both, the branch cut is mapped to $(-1,1)$, see Fig.~\ref{fig:cft_symmetric_slit}~(a) for an illustration. 

To evaluate the Renyi entropy (\ref{eq:renyi}), we follow \cite{rajabpour2016entanglement} and consider the free energy 
\bea\label{eq:free_energy}
F_\alpha = - \log Z_\alpha
\eea
first, which can be computed as follows by considering a small shift $\delta l$ of one of the slits. Let us denote by $l$ the distance between the two slits in $X$ and horizontally shift the second slit (the one mapped to $(1,1/k)$ in $\tilde{X}$) by an amount $\delta l$.
The change in the free energy of the system on the $\alpha$-sheeted Riemann surface due to the shift can then be written as~\cite{rajabpour2016entanglement}
\bea
    \delta F_\alpha(X) = - \frac{\delta l}{2\pi i} \oint dX \langle T(X) \rangle + c.c.
\eea
where $\langle T(X) \rangle$ denotes the vacuum expectation value of the stress tensor in $X$, and the integral's contour encloses the second slit. 
The next step is to map our domain in $\tilde{X}$ to an annulus in $\zeta$ coordinates using the conformal transformation
\bea
    \zeta(\tilde X) =  e^{-\frac{\pi s}2} \exp \left[ \pi s \frac{\sn^{-1}(\tilde X, k^2)}{2 K(k^2)} \right]
\eea
where $\sn$ is the elliptic sine function, $K$ is the complete elliptic integral of the first kind, and we have defined
\bea \label{eq:s_cft}
    s = \frac{2 K(k^2)}{K(1-k^2)}.
\eea
The slit at $(1,1/k)$ is mapped to the outer edge $|\zeta| = 1$, the slit at $(-1/k,-1)$ is mapped to the inner edge $|\zeta| = e^{-\pi s}$, and the branch cut at $\tilde{X}\in (-1,1)$ is mapped to $\zeta \in (e^{-\pi s},1)$, see Fig.~\ref{fig:cft_symmetric_slit}. We can now perform an additional conformal transformation $\zeta_\alpha =  \zeta^{1/\alpha}$ to ``unwind'' and obtain an annulus without branch cuts.

The purpose of this coordinate transformation is to simplify the evaluation of the free energy, of which we report here the final result (we refer the readers to Ref.~\cite{rajabpour2016entanglement} for a detailed derivation). 
The free energy consists of two parts:
\bea
    F_\alpha = F^\text{annu}_\alpha + F^\text{geom}_\alpha,
\eea
where $F^\text{annu}_\alpha$ is the free energy of the annulus and $F^\text{geom}_\alpha$ denotes the geometric contribution originating from the coordinate transformation between the plane in $X$ coordinates and the annulus.
The free energy in the annulus geometry is known and can be written in two equivalent forms~\cite{cardy2004boundary,rajabpour2016entanglement}: 
\begin{align}
    \label{eq:annulus_free_energy}
    &- F^\text{annu}_\alpha =  \log \left( q_\alpha ^{- \frac{c}{24}} (1 + \sum_j n_j q^{\Delta_j}_\alpha) \right) - c \frac{\pi s}{12\alpha}  \\
     &- F^\text{annu}_\alpha =\log \left( \tilde q_\alpha ^{- \frac{c}{24}} (b_0^2 + \sum_j b_j^2 \tilde q^{\Delta_j}_\alpha) \right) - c \frac{\pi s}{12\alpha} ,
    \label{eq:annulus_free_energy2}
\end{align}
where $c$ is the central charge, $n_j$ is the number of degenerate states with scaling dimension $\Delta_j$, $b_j$ are real numbers, and $q_\alpha = e^{- \frac{2\pi\alpha}s}$, $\tilde q_\alpha = e^{- \frac{2\pi s}{\alpha}}$. 
The first (second) expression is appropriate for the open (closed) string channel and is useful to compute the annulus partition function in the $s\ll 1$ ($s\gg 1$) limit. 
On the other hand, the geometric contribution is implicitly given by
\bea \label{eq:schwarzian}
    \frac{\delta F_\alpha^\text{geom}}{\delta l} = \frac{i c}{12 \pi} \oint dX \{ \zeta_\alpha(X), X \}, 
\eea
where again the integral's contour encloses the second slit, and $\{f(x),x\} = \frac{f'''(x)}{f'(x)} - \frac32 (\frac{f''(x)}{f'(x)})^2 $ is the Schwarzian derivative.
Note that $\zeta_\alpha(X)$ denotes the composed conformal transformation leading from the $X$ coordinates to the $\zeta_\alpha$ coordinates.
In terms of the free energy, the Renyi entropy is finally given by
\bea
    S_\alpha=S_\alpha^\text{annu}+ S_\alpha^\text{geom}= \frac1{\alpha - 1} \left[ F_\alpha^\text{annu} + F_\alpha^\text{geom} - \alpha (F_1^\text{annu} + F_1^\text{geom} \right)].
    \label{eq:renyi-free-energy}
\eea
With these results in hand, we are now ready to analyze the two cases in more detail and compute their respective entanglement entropies.

\subsection{Infinite intervals: one-sided measurement}
\label{sec:CFT1side}

The position of the two slits in $X$ coordinates in the one-sided measurement case is given in equation~(\ref{eq:slit1_infinite_one0}) and (\ref{eq:slit2_infinite_one0}). 
The conformal transformation from $X$ to $\tilde X$ is a simple Mobius transformation
\bea
    \tilde X( X ) = \frac{2a}{k} \frac{X}{a X + 1} - \frac1k
\eea
where we have defined 
\bea
    a =\frac1{\sqrt{h(h+l)}}, \quad 
    k = \frac{l+2h - 2\sqrt{h(l+h)}}{l},
\eea
for $h$ the length of the first slit and $l$ the distance between the two slits in $X$, and noting that the second slit has infinite length. Using equations (\ref{eq:slit1_infinite_one0}) and (\ref{eq:slit2_infinite_one0}), the explicit expressions for $h$ and $l$ are
\bea \label{eq:parameter_one_side}
    h= e^{- \frac{\pi \Delta L}{\beta}}, \quad l = e^{ \frac{\pi \Delta L}{\beta}} - e^{- \frac{\pi \Delta L}{\beta}} .
\eea

The geometric contribution can now be obtained in implicit form by evaluating equation~(\ref{eq:schwarzian}). 
Noting that there are poles at $\tilde X = \pm 1, \pm 1/k$ and that the contour integral picks up two poles at $\tilde X = 1, 1/k$, we arrive at~\cite{rajabpour2016entanglement}
\bea \label{eq:geom_one_side}
    \frac{\delta F_\alpha^\text{geom}}{\delta l} = \frac{c}{6} \frac{a(1- k)\left[-2\pi^2 + (1+ 6k+k^2)\alpha^2 K^2(1-k^2) \right]}{16 k (1+k) \alpha K^2(1-k^2)}.
\eea
This equation cannot be solved analytically for general values of $h$ and $l$, so we will focus on the $l \gg h$ ($\Delta L \gg \beta$, or $s\gg 1$) and $l \ll h$ ($\Delta L \ll \beta$, or $s\ll 1$) limits, which correspond to being safely in the thermal AdS and BTZ black hole phases, respectively.\footnote{This can be easily seen by inspection of Figs. \ref{fig:transition} and \ref{fig:s-infinite}: the thermal AdS (BTZ) phase is dominant for large (small) values of $s$, and $s$ is a monotonically increasing function of the ratio $\Delta L/\beta$.}

\subsubsection{$\Delta L \gg \beta$}

In the limit $l \gg h$, we can expand equation~(\ref{eq:geom_one_side}) at leading order in $h/l$ and then integrate over $l$. 
The result at leading order reads 
\bea
    F_\alpha^\text{geom} = \frac{c (\alpha^2 -1)}{{12}\alpha} \log \frac{l}h.
    \label{eq:geometriclargel}
\eea
Since in this limit we have $s\gg 1$, the last term in the annulus free energy (\ref{eq:annulus_free_energy2}) is cancelled by the $-c\log(\tilde{q}_\alpha)/24$ term, and the leading order annulus free energy is simply given by $F_\alpha^\text{annu}=-2\log b_0$. Summing such annulus contribution to the geometric free energy (\ref{eq:geometriclargel}), the Renyi entropy (\ref{eq:renyi-free-energy}) therefore gives 
%
\bea
\begin{split}
    S_\alpha &= \frac{(1+\alpha)c}{12\alpha}  \log \frac{l}h  + 2\log b_0 \\
    &\approx \frac{(1+\alpha)c}{12\alpha} \frac{2\pi \Delta L}{\beta} + 2\log b_0 \label{eq:renyilargel}
    \end{split}
\eea
where $\log b_0$ is the boundary entropy~\cite{affleck1991universal,rajabpour2016entanglement} and
in the last equality we used equation~(\ref{eq:parameter_one_side}) and kept only leading order terms in $\Delta L \gg \beta$.

\subsubsection{$\Delta L \ll \beta$}
In the limit $l \ll h$, we can compute the geometric contribution to the free energy by expanding equation~(\ref{eq:geom_one_side}) at leading order in $l/h$ and integrating over $l$. This leads to 
\bea \label{eq:geom_ond_side_2}
    S_{\alpha}^\text{geom}=\frac1{\alpha - 1} \left( F_\alpha^\text{geom} - \alpha F_1^\text{geom} \right) &=& \frac{(1+\alpha)\pi^2 c}{12 \alpha \log \frac{16 h}{l}} \approx \frac{(1+\alpha)\pi^2 c}{12 \alpha \log \frac{8\beta }{\pi \Delta L}}
\eea
where in the last equality we used equation~(\ref{eq:parameter_one_side}) and kept only leading order terms in $\Delta L \ll \beta$.
Next, we consider the annulus free energy.
In the limit $\Delta L \ll \beta$, according to~(\ref{eq:s_cft}), we have
\bea \label{eq:s-one-side}
    s = \frac{\pi}{\log \frac{8\beta}{\pi \Delta L}},
\eea
which implies $q_\alpha \ll 1$. 
Using equation (\ref{eq:s-one-side}), the contribution to the Renyi entropy that comes from the last term of the annulus free energy~(\ref{eq:annulus_free_energy}) is 
\bea
    - \frac{(1+\alpha)c \pi s}{12 \alpha}  = - \frac{(1+\alpha)\pi^2 c}{12 \alpha \log \frac{8\beta }{\pi \Delta L}}.
\eea 
This term exactly cancels the geometric contribution~(\ref{eq:geom_ond_side_2}). The $c\log(q_\alpha)/24$ term also gives a vanishing contribution to the Renyi entropy.
We are therefore left with only one term in~(\ref{eq:annulus_free_energy}) which we can write as
\bea
    \log \left(1 + \sum_j n_j q^{\Delta_j}_\alpha \right) \approx n_1 e^{-\Delta_1 \frac{2\pi \alpha}s} = n_1 \left( \frac{8\beta}{\pi \Delta L} \right)^{-2\alpha \Delta_1},
\eea
where in the first step we assumed that $\Delta_1$ is the smallest scaling dimension, and in the second step we used the relation~(\ref{eq:s-one-side}) for $s$.
This gives the Renyi entropy,
\bea
    S_\alpha = \begin{cases}
     \frac{\alpha}{\alpha - 1} n_1\left( \frac{8\beta}{\pi \Delta L}\right)^{-2\Delta_1} \quad & \alpha > 1 \\
     2\Delta_1 n_1\left( \frac{8\beta}{\pi \Delta L}\right)^{-2\Delta_1} \log \frac{8\beta}{\pi \Delta L}  \quad & \alpha = 1\\
     \frac1{1-\alpha} n_1\left( \frac{8\beta}{\pi \Delta L}\right)^{-2 \alpha \Delta_1} \quad & \alpha < 1
     \end{cases}
     \label{eq:renyifinalsmalll}
\eea

To compare these results with those from the bulk dual construction in Section \ref{sec:infinite-one}, we focus on the von Neumann entropy (e.g. the $\alpha =1$ case).
From the CFT calculations above we obtained the von Neumann entropy
\bea
    S^{lr} = \begin{cases} 
      \frac{c}{6} \frac{2\pi \Delta L}{\beta} + 2\log b_0 \quad & \Delta L \gg \beta\\
       2\Delta_1 n_1 \left( \frac{\pi \Delta L}{8\beta}\right)^{2\Delta_1} \log \frac{8\beta}{\pi \Delta L} \quad & \Delta L \ll \beta 
    \end{cases},
\eea
where the $lr$ superscript is a reminder that this is the entanglement entropy between left and right sides after measurement. 

We can compare this with the expression (\ref{eq:mi}) for the mutual information obtained computing the holographic entanglement entropy in the bulk dual setup. In the $\Delta L \gg \beta$ limit we have $s\approx \frac{2\Delta L}\beta$. By further identifying $c = \frac{3R}{2G_N}$ and the boundary entropy with the contribution dependent on the brane tension $T$ \cite{takayanagi2011holographic,fujita2011aspects}, we recover the result obtained for the thermal AdS phase of the gravity calculation of section \ref{sec:infinite-one}\footnote{We remind that  $I^{lr}=2S^{lr}$ because the full system is in a pure state.} (see the second line of equation (\ref{eq:mi})). 

In the BTZ phase, the classical bulk calculation gives vanishing mutual information between the two sides. 
This is what we should expect from a pure classical gravity calculation in the bulk: the $\Delta L \ll \beta$ result found here comes purely from the operator content in the CFT calculation and is independent of the central charge. It is therefore a subleading-in-$N$ contribution which we can expect to be captured by the holographic entanglement entropy only if we include bulk quantum fields (it would in fact be given by the bulk fields' entanglement entropy).

\subsection{Infinite intervals: two-sided  measurement}
\label{sec:CFT2sides}

The slits in the two-sided measurement case are given in equations~(\ref{eq:slit1_infinite_two0}) and (\ref{eq:slit2_infinite_two0}). 
The map from the $X$ to the $\tilde X$ coordinates is achieved via a Mobius transformation: 
\bea
    \tilde X = \frac{- X/k + a}{X + a},
\eea
with 
\bea
    a = - h - \sqrt{h(l + h)}, \quad k = \frac{l+2 h - 2\sqrt{h(l+h)} }l,
\eea
where $h$ is the length of the second slit and $l$ is the distance between the two slits in $X$ coordinates:
\bea \label{eq:parameter_two_side}
    h= e^{- \frac{\pi \Delta L}{\beta}}, \hspace{2cm} l = e^{ \frac{\pi \Delta L}{\beta}} .
\eea  

The geometric contribution can be obtained by evaluating~(\ref{eq:schwarzian}). 
Noting that there are poles at $\tilde X = \pm 1, \pm 1/k$ and that the contour integral picks up two poles at $\tilde X = 1, 1/k$, we arrive at
\bea \label{eq:geom_two_side}
    \frac{\delta F_\alpha^\text{geom}}{\delta l} = \frac{c \alpha}6 \left(  \frac{(1-k)\pi^2}{4a k(1+k)^2 \alpha^2 K^2(1-k^2)} + \frac{(k-1)(1+6k +k^2)}{8a k(1+k)^2} \right).
\eea
Once again, equation (\ref{eq:geom_two_side}) does not have an analytic solution for general $l$ and $h$.
We will then discuss again two limits, namely $\Delta L/\beta  \gg 1$ ($l \gg h$, or $s\gg 1$) and $\Delta L/\beta  \ll -1$ ($l \ll h$, or $s\ll 1$).

\subsubsection{$\Delta L/\beta \gg 1 $} 

In the limit $l \gg h$, we get the geometric contribution to the free energy by expanding equation~(\ref{eq:geom_two_side}) as a function of $h/l$ and then integrating over $l$. 
Then combining the contribution from the annulus free energy, we arrive at
\bea
    S_\alpha &=& \frac{(1+\alpha)c}{12\alpha}  \log \frac{l}h  + 2\log b_0, \\
    &\approx& \frac{(1+\alpha)c}{12\alpha} \frac{2\pi \Delta L}{\beta} + 2\log b_0.
\eea
This is similar to the one-sided measurement result, where $\log b_0$ is the boundary entropy originated from the annulus contribution~\cite{affleck1991universal,rajabpour2016entanglement}.
In the second line we used equation~(\ref{eq:parameter_two_side}) and took the limit $\Delta L /\beta \gg 1$.

\subsubsection{$\Delta L/\beta \ll -1 $}

In the limit $l \ll h$, we can expand equation~(\ref{eq:geom_two_side}) as a function of $l/h$ and integrate over $l$ to get the geometric free energy.  
This leads to 
\bea \label{eq:geom_two_side_2}
    S_\alpha^\text{geom}=\frac1{\alpha - 1} \left( F_\alpha^\text{geom} - \alpha F_1^\text{geom} \right) &=& \frac{(1+\alpha)\pi^2 c}{12 \alpha \log \frac{16 h}{l}} \approx -\frac{(1+\alpha)\pi c}{24 \alpha} \frac{\beta}{\Delta L},
\eea
where in the second equality we used equation~(\ref{eq:parameter_two_side}) and $\Delta L /\beta \ll  -1$.
Next, we consider the annulus free energy.
In this limit, from equation~(\ref{eq:s_cft}) we have
\bea \label{eq:s-two-side}
    s = - \frac{\beta}{2\Delta L},
\eea
which also indicates $q_\alpha \ll 1$. 
Using this relation, the contribution to the Renyi entropy that comes from the last term in~(\ref{eq:annulus_free_energy}) is 
\bea
    - \frac{(1+\alpha)c \pi s}{12 \alpha}  =  \frac{(1+\alpha)\pi c}{24 \alpha} \frac{\beta}{\Delta L}.
\eea 
This term exactly cancels the geometric contribution~(\ref{eq:geom_two_side_2}). 

What is left is then again the first term in~(\ref{eq:annulus_free_energy}), which we can write as (like in the one-sided case, the contribution from the $c\log (q_\alpha)/24$ term is vanishing)
\bea
    \log \left(1 + \sum_j n_j q^{\Delta_j}_\alpha \right) \approx n_1 e^{-\Delta_1 \frac{2\pi \alpha}s} = n_1 e^{2\alpha \Delta_1 \frac{2\pi \Delta L}\beta},
\eea
where in the second equality we used the relation~(\ref{eq:s-two-side}).
This gives the Renyi entropy
\bea
    S_\alpha = \begin{cases}
     \frac{\alpha}{\alpha - 1} n_1e^{2 \Delta_1 \frac{2\pi \Delta L}\beta} \quad & \alpha > 1 \\
     -2 \Delta_1 n_1 \frac{2\pi \Delta L}\beta e^{2 \Delta_1 \frac{2\pi \Delta L}\beta}   \quad & \alpha = 1\\
     \frac1{1-\alpha} n_1 e^{2\alpha \Delta_1 \frac{2\pi \Delta L}\beta} \quad & \alpha < 1
     \end{cases}
\eea
Note that in this limit $\Delta L < 0$. As we have discussed, this means that there is no unmeasured region corresponding to the same range of $x$ coordinate in both CFTs (intuitively, we have measured ``more than half" of both sides, see Fig.~\ref{fig:map-infinite-two} for an illustration). 
Notice that in the two-sided measurement case of this subsection the entanglement entropy between the two sides decays exponentially in $\Delta L/\beta$ as opposed to the power-law behavior (\ref{eq:renyifinalsmalll}) found in the one-sided measurement case of the previous subsection.

To compare these results with the gravity calculation, we can again focus on the von Neumann entropy.
From the CFT calculations we obtained the von Neumann entropy
\bea
    S^{lr} = \begin{cases} 
       \frac{c}{6} \frac{2\pi \Delta L}{\beta} + 2\log b_0 \quad & \Delta L/\beta \gg 1 \\
       -2 \Delta_1 n_1 \frac{2\pi \Delta L}\beta e^{2 \Delta_1 \frac{2\pi \Delta L}\beta} \quad & \Delta L/\beta \ll -1 
    \end{cases}
\eea

Noticing that $s\approx \frac{2\Delta L}\beta$ in the $\Delta L/\beta \gg 1$ limit and using the relation $c = \frac{3R}{2G_N}$, we obtain the same result obtained from the bulk holographic entanglement entropy calculation in the thermal AdS phase, see equation (\ref{eq:mi}).
Similar to the one-sided case, the bulk calculation gives vanishing mutual information between the two sides in the BTZ phase. As we have already explained, this is expected because we neglected the contribution of bulk quantum fields to the holographic entanglement entropy.

\section{Discussion}
\label{sec:discussion}

In this paper we considered local projective measurements performed on the TFD state of two copies of a 2D CFT on a line and studied the effects of the measurement on the entanglement structure of the microscopic system as well as on the bulk dual spacetime. We found that measuring finite intervals in one of the CFTs is not enough to completely disentangle the two CFTs, and the corresponding dual Lorentzian spacetime remains connected. On the other hand, measuring semi-infinite intervals in one or both the CFTs triggers an entangled/disentangled phase transition between the two CFTs, corresponding to a connected/disconnected phase transition in the bulk dual geometry. The disconnected phase is dominant when the measured regions are large enough and the boundary entropy of the Cardy state we postselect on (corresponding to the brane tension in the bulk description) is sufficiently small or negative. Therefore, we conclude that measurement of infinite subregions of the CFTs can destroy the Einstein-Rosen bridge connecting the two AdS asymptotic boundaries in the double-sided BTZ black hole geometry dual to the TFD state. We also quantified these statements by computing the post-measurement holographic mutual information between the two CFTs, and verified via a microscopic calculation performed in the dual CFT system, which yielded compatible results. 
Finally, we showed how inserting heavy operators in the boundary Euclidean path integrals preparing the microscopic state in the absence and presence of measurement allows us to detect whether, when a measurement is performed, the information associated with such operator insertions is ``teleported'' (in the sense of the ``bulk teleportation'' studied in \cite{Antonini:2022sfm}) from one CFT to the other, erased, or neither.
The results of the present paper extend those obtained in \cite{Antonini:2022sfm} and represent a higher-dimensional analogue of those presented in \cite{Antonini:2022lmg}.

Several open questions still remain, which could orient the direction of future research on this topic. 
First, it would be interesting to understand whether postselection procedures such as those investigated here facilitate the reconstruction of regions of bulk spacetime behind a black hole horizon. 
In fact, as we have seen, when one or two semi-infinite intervals are measured in, say, the left CFT, information that without measurement would be accessible from the left CFT becomes accessible from the right CFT. In other words, part of the pre-measurement entanglement wedge of the left CFT becomes part of the post-measurement entanglement wedge of the right CFT. 
Naively, one would expect the reconstruction of physics in such a ``teleported'' region to be extremely complex because it involves accessing information that, at least in the pre-measurement geometry, sits behind a horizon. 
However, provided that the observer on the right has knowledge of the measurement outcome (i.e. of the specific Cardy state we postselected on), it could be possible to reconstruct physics in this region in a much simpler way by evolving the right CFT in real time with a modified Hamiltonian conditioned on the measurement outcome, similar to the work of Kourkoulou and Maldacena \cite{kourkoulou2017pure}. 
This suggests that (partial) knowledge of the bulk state behind a horizon could largely simplify reconstruction of behind-the-horizon physics in various setups, possibly also in Python's lunch geometries. 
However, the implementation of this idea in our setup requires the definition of a non-singular Lorentzian geometry associated with the post-measurement state of interest, which in turn requires a well-defined regularization of the singularities arising in our setup and discussed in Sections \ref{sec:intro} and \ref{sec:bulk}. A second interesting direction would be to study tensor network models able to reproduce the results we obtained. This could shed light on the consequences of the measurement procedure for the complexity of bulk reconstruction, and in particular of the reconstruction of behind-the-horizon physics.

It would also be interesting to investigate whether the effects of heavy operator insertions which in our purely geometric approximation are ``erased'' by the measurement (in the sense we have explained) could still be accessible when bulk fields are included in our analysis. 
For example, 
a possible way to access the heavy operator insertion is to insert bulk operators in the left or right CFT on the time reversal invariant slice, which can be done, for instance, using HKLL reconstruction~\cite{hamilton2006holographic}, and calculate the matrix element between these two insertions: the heavy operator insertion in the Euclidean past and the bulk operator insertion on the time reversal invariant slice. 
It is natural to expect that the matrix element increases to an order one number in a proper normalization, when the geodesic from the heavy operator insertion is inside the entanglement wedge of either CFT.
Nevertheless, the matrix element can be small but nonvanishing even if the disconnected geodesic from the heavy operator insertion ends on the brane.
It would be interesting to investigate the behavior of the matrix element and its implication on the accessibility of the information of the heavy operator insertion.

Finally, a general and fundamental question is how generic the prescriptions used here and in \cite{numasawa2016epr,Antonini:2022sfm} are. 
In particular, the presence of ETW branes in the bulk description of the post-measurement state seems to be strictly related to the choice of postselecting on Cardy states. 
The dual description of different classes of measurement, however, are yet unknown. 
At the current stage, it is unclear whether the constructions used here could be generalized to describe more complicated setups by including additional elements (for instance branes that are not pure-tension, or additional modifications of the bulk spacetime besides the insertion of branes, such as the inclusion of shock waves or additional matter), or if a completely different framework is needed. 
Equally intriguing is the possibility that holographic tools similar to those employed here could be applied to describe the physics of decoherence \cite{Zou:2023rmw}.



\acknowledgments
We would like to thank Raphael Bousso, Alexey Milekhin, Vincent Su for helpful discussions. We acknowledge support from the Simons Foundation via It From Qubit (S-K.J.), from the U.S. Department of Energy grant DE-SC0009986 (S-K.J. and B.G.S.), from the AFOSR under FA9550-19-1-0360 (B.G-W.), and from the U.S. Department of Energy, Office of Science, Office of Advanced Scientific Computing Research, Accelerated Research for Quantum Computing program ``FAR-QC'' (S.A.). Part of S.-K. J.'s work is supported by a startup fund at Tulane University.

\appendix

\section{Details of the conformal transformations}
\label{app:a}

In this appendix we will delve into greater detail about the conformal transformations used in the main sections of the paper above. To better understand the effect of each conformal transformation, we will describe how the slits, the time reflection symmetric lines, and $x = \pm \infty$ are mapped through the different coordinates in the various measurement configurations considered above. 

\subsection{Finite intervals}
\label{sec:finite_conformal}
We start with the conformal transformations used to describe measurement of finite intervals considered in Section \ref{sec:finite} and illustrated in Fig. \ref{fig:map-finite}. 

Firstly, the infinitely long cylinder in $(x,y)$ coordinates (Fig.~\ref{fig:map-finite} (a)) is mapped to a 2D plane in the $X$ coordinate system\footnote{When, like in this case, we indicate a coordinate system by a single symbol, it is understood that we are considering complex coordinates, e.g. $X$ coordinate stands for the $(X,\bar{X})$ coordinate system.} (Fig.~\ref{fig:map-finite} (b)) by
\bea \label{eq:transform-1}
    X = e^{\frac{2\pi}\beta (x+i y)}, \quad \bar X = e^{\frac{2\pi}\beta (x- i y)}. 
\eea
Under this conformal map, the two parallel slits are mapped to two radial slits with angles $\pm \phi/2$, where we define 
\bea \label{eq:phi}
    \phi = 2\pi \Delta T /\beta.
\eea 
Also, notice that both of the slits span from $|X| = e^{-\pi \Delta L/\beta}$ to $|X| = e^{\pi \Delta L/\beta} $.
The time reflection symmetric lines are both mapped to the horizontal line $X= \bar X$, with negative (positive) infinity in $x$ coordinates mapped to the origin (infinity) in $X$ coordinates.

Secondly, the 2D plane in $X$ coordinates with two radial slits (Fig.~\ref{fig:map-finite} (b)) is mapped to an annulus in $\zeta$ coordinates (Fig.~\ref{fig:map-finite} (c)),
in which the two radial slits are mapped to the two circular edges of the annulus.
The conformal transformation mapping the annulus in $\zeta$ with external radius $|\zeta| = 1$ and internal radius $|\zeta| = \rho < 1$ to two radial slits in $X$ is given by \cite{crowdy2006conformal}:
\bea \label{eq:conformal_transform}
    X(\zeta) = C(\alpha,\beta) \frac{(\zeta - \alpha)(\zeta - \bar \alpha^{-1})}{(\zeta -\beta)(\zeta -\bar \beta^{-1})} \prod_{k=1}^\infty \frac{(1-\rho^{2k} \frac{\zeta}\alpha)(1-\rho^{2k} \frac\alpha\zeta)(1-\rho^{2k} \bar \alpha \zeta)(1- \rho^{2k} \frac1{\bar \alpha \zeta})}{(1-\rho^{2k} \frac{\zeta}\beta)(1-\rho^{2k} \frac\beta\zeta)(1-\rho^{2k} \bar \beta \zeta)(1- \rho^{2k} \frac1{\bar \beta \zeta})}
\eea
where $\zeta=\alpha$ is mapped to $X = 0$, and $\zeta=\beta$ to $|X| = \infty$. 
$C(\alpha,\beta)$ is a normalization prefactor. By choosing $\alpha = \sqrt{\rho} e^{i \phi}$, $\beta = \sqrt{\rho}$, and $C=e^{-i 3\phi/2}$, the two circles are mapped to two finite length radial slits, and equation~(\ref{eq:conformal_transform}) can be simplified to
\bea \label{eq:map-finite}
    X(\zeta) = e^{-i \frac\phi2 } \frac{\theta_4(\frac{-i \log \zeta -\phi}2, \sqrt\rho)}{\theta_4(\frac{-i \log \zeta}2, \sqrt\rho)} ,
\eea
where $\theta_{i}(\varphi, q) $ is the Jacobi theta function. In our conventions, the theta functions are defined as
\bea
    && \theta_3(\varphi,q) = \theta_4(\varphi + \frac\pi2, q) = \sum_{n= -\infty}^{\infty} q^{n^2} e^{i 2n \varphi} \nn \\
    && = \prod_{m=1}^\infty (1- q^{2m}) \left( 1 + q^{2m-1} e^{i 2\varphi}  \right) \left( 1 + q^{2m-1} e^{-i 2\varphi}  \right), 
    \\
    && \theta_2(\varphi,q) = \theta_1(\varphi+ \frac\pi2, q) = \sum_{n= -\infty}^{\infty} q^{(n+ \frac12)^2} e^{i(2n+1) \varphi} \nn \\
    && = 2 q^{1/4} \cos \varphi  \prod_{m=1}^\infty (1- q^{2m}) \left( 1 + q^{2m} e^{i 2\varphi}  \right) \left( 1 + q^{2m} e^{-i 2\varphi}  \right).
\eea

The conformal transformation~(\ref{eq:map-finite}) is determined by two parameters, $\rho$ and $\phi$.
$\phi $ is given by~(\ref{eq:phi}); in order to see how $\rho$ is related to the slit parameters, we note that the first (second) slit 
in $\zeta$ is given by $|\zeta| = 1$ ($|\zeta| = \rho$). 
Setting $\zeta = r e^{i \theta}$, the two slits are
\bea
    && \text{first slit: }
    X(e^{i\theta}) = e^{-i \frac\phi2 } \frac{\theta_4(\frac{\theta -\phi}2, \sqrt\rho)}{\theta_4(\frac{\theta }2, \sqrt\rho)}, \quad \theta \in (0,2\pi), \\
     &&\text{second slit: }
    X( \rho e^{i\theta}) = e^{i \frac\phi2 } \frac{\theta_4(\frac{\theta -\phi}2, \sqrt\rho)}{\theta_4(\frac{\theta }2, \sqrt\rho)}, \quad \theta \in (0,2\pi).
\eea
Since $\theta_4$ is a real function when its arguments are real, as $\theta$ moves from $\theta = 0$ to $\theta = 2\pi$, $X(e^{i\theta})$ ($X(\rho e^{-i\theta})$) is a radial slit with angle $-\phi/2$ ($\phi/2$). 
According to (\ref{eq:transform-1}), the slits span from $|X|= e^{-\pi\Delta L/\beta}$ to $|X|= e^{\pi \Delta L/\beta}$. 
Thus, when $\theta$ moves from $\theta = 0$ to $\theta = 2\pi$, the function $\frac{\theta_4(\frac{\theta -\phi}2, \sqrt\rho)}{\theta_4(\frac{\theta }2, \sqrt\rho)}$ will move between $e^{-\pi\Delta L/\beta}$ and $e^{\pi \Delta L/\beta}$, i.e., 
\bea \label{eq:rho}
    && \frac{\theta_4(\frac{\theta_\text{max} -\phi}2, \sqrt\rho)}{\theta_4(\frac{\theta_\text{max} }2, \sqrt\rho)} \equiv
    \max_{\theta \in (0,2\pi)}\frac{\theta_4(\frac{\theta -\phi}2, \sqrt\rho)}{\theta_4(\frac{\theta }2, \sqrt\rho)} =  e^{\pi \Delta L/\beta}, \\
    && \frac{\theta_4(\frac{\theta_\text{min} -\phi}2, \sqrt\rho)}{\theta_4(\frac{\theta_\text{min} }2, \sqrt\rho)} \equiv
    \min_{\theta \in (0,2\pi)}\frac{\theta_4(\frac{\theta -\phi}2, \sqrt\rho)}{\theta_4(\frac{\theta }2, \sqrt\rho)} =  e^{- \pi \Delta L/\beta},\label{eq:rho2}
\eea
where the first equality should be regarded as the definition of $\theta_\text{max}$ and $\theta_\text{min}$, and the second equality gives a relation between $\rho$ and the slit parameters $\Delta L$ and $\Delta T$.
Unfortunately, there is no simple expression for this relation, but 
as an illustration, we plot $\frac{\Delta L}{\Delta T}$ as a function of $\rho$ in Fig.~\ref{fig:rho-finite} for $\phi=0.2$. 
Note that when $\Delta T \ll \Delta L$, $\rho \rightarrow 1$. 
In particular, when $\Delta T \rightarrow 0$, the slit will reduce to the case in Fig.~\ref{fig:measure_imagine} (a), and the dual spacetime is described by an ETW brane anchored at the slit boundaries. 
In the following, we give a simple relation near $\rho = 1$.

The asymptotic form of $\theta_4(\varphi, q)$ near $q = 1$ is~\cite{banerjee2016lambert} 
\bea
    \theta_4(\varphi, q) \approx \sqrt{\frac{\pi}{\log q^{-1}}} e^{\frac{((\varphi)_\pi-\pi/2)^2}{\log q}}, \quad (\varphi)_\pi = \varphi \text{ mod }\pi,
\eea
from which, we arrive at
\bea
    \frac{\theta_4(\frac{\theta-\phi}2, \sqrt\rho)}{\theta_4(\frac{\theta}2, \sqrt{\rho})} \approx \exp\left[ \frac{((\theta-\phi)_{2\pi} - \pi)^2 - ((\theta)_{2\pi} - \pi)^2}{2\log \rho} \right].
\eea
In this case, it is not hard to see that $\theta_\text{min} = \phi$ and $\theta_\text{max} = 0$, which according to~(\ref{eq:rho}) leads to $\frac{2\pi}\beta \Delta L = -\frac{(2\pi- \phi)\phi}{\log \rho}$. 
Thus we have the following relation near $\rho = 1$,
\bea
     \rho = \exp \left[- 2\pi \frac{\Delta T }{\Delta L} \left(1- \frac{\Delta T}\beta \right) \right], \quad \phi = \frac{2\pi}\beta \Delta T.
     \label{eq:rhonear1}
\eea
Thus, $\log \rho \ge -\frac\pi2 \frac{\beta}{\Delta L}$.\footnote{The exponent on the right hand side of equation (\ref{eq:rhonear1}) is minimized for $\Delta T=\beta/2$, and this approximation is therefore valid in the limit $\frac{\Delta L}{\beta} \gg 1$.} 
We can then introduce another parameter $s$ defined as 
\bea \label{eq:s_finite}
\log \rho = - \pi s,
\eea
which near $\rho = 1$ takes the form
\bea
    s = \frac{2\Delta T}{\Delta L}\left(1-\frac{\Delta T}\beta \right), 
\eea
and so is roughly the ratio between the imaginary time evolution and the length of the measured interval. 
As we have seen in Section \ref{sec:transition}, a phase transition in the brane configuration in the bulk---corresponding to an entanglement phase transition in the dual CFT---can be characterized in terms of a critical value $s_c(T)$ of the parameter $s$ we just introduced. Note that $s_c(T)$ depends on the brane tension $T$, see~(\ref{eq:transition}), i.e. on the boundary entropy of the specific Cardy state we project on.

On the other hand, the asymptotic form of $\theta_4(\varphi,q)$ near $q=0$ is
\bea
    \frac{\theta_4(\frac{\theta-\phi}2, \sqrt\rho)}{\theta_4(\frac{\theta}2, \sqrt{\rho})} \approx \exp\left[2\sqrt \rho( \cos \theta - \cos(\theta - \phi)) \right].
\eea
Therefore, near $\rho = 0$ (corresponding to $\Delta T \gg \Delta L$) we have the relation
\bea
    \rho=e^{-\pi s} = \left( \frac{\pi \Delta L}{4\beta \sin \frac{\phi}2} \right)^2,
\label{eq:thermal_rho}
\eea
which implies that the approximation is valid when $\frac{\Delta L}{\beta} \ll \sin \left( \frac{\pi}\beta \Delta T \right)$.

The time reflection symmetric lines are mapped to $r = \sqrt \rho$, i.e.,
\bea    
    X(\sqrt \rho e^{i\theta} ) = \frac{\theta_1(\frac{\theta -\phi}2, \sqrt \rho)}{\theta_1(\frac{\theta}2, \sqrt \rho)}, \quad \theta \in (0,2\pi).
\eea
Since $\lim_{\theta \rightarrow 0^+} \theta_1(\theta,q) = -\infty $ and $\lim_{\theta \rightarrow 2\pi^-} \theta_1(\theta,q) = \infty $, the origin $X = \bar X = 0$ is mapped to $r = \sqrt \rho, \theta = \phi$, and $X = \bar X = \infty $ ($X = \bar X = -\infty$) is mapped to $r = \sqrt \rho, \theta = 0^-$ ($r = \sqrt \rho, \theta = 0^+$). 
See Fig.~\ref{fig:map-finite} (c) for an illustration.

Finally, the annulus in $\zeta$ coordinates [Fig.~\ref{fig:map-finite} (c)] is mapped to a finite cylinder in $w$ coordinates [Fig.~\ref{fig:map-finite} (d)] 
by 
\bea \label{eq:transform-3}
    \zeta = e^{ \sqrt2 w} = e^{\sigma + i \nu}, \quad \bar \zeta = e^{ \sqrt2 \bar w} = e^{\sigma - i\nu}, \quad 
    w = \frac{\sigma + i \nu}{\sqrt2}, \quad w = \frac{\sigma - i \nu}{\sqrt2}.
\eea
Now the two slits are located at
\bea
\label{eq:first}    && \text{first slit:}   \quad \sigma = 0, \\
\label{eq:second}    && \text{second slit:}  \quad \sigma =  -\pi s,
\eea
and it is clear from the transformation that $\nu \sim \nu +  2\pi$. 
The time reflection symmetric lines are mapped to a circle at $\sigma = -\pi s/2$, where the left (right) CFT is mapped to the segment $\nu \in (\phi, 2\pi)$ [$\nu \in (0,\phi)$]. 
Negative (positive) infinity in $x$ is mapped to $\sigma=-\pi s/2, \nu = \phi$ ($\sigma=-\pi s/2, \nu = 0,2\pi$).
See Fig.~\ref{fig:map-finite} (d) for an illustration.

To summarize, the conformal transformation from original $x$ coordinate [Fig.~\ref{fig:map-finite} (a)] to the final $w$ coordinate [Fig.~\ref{fig:map-finite} (d)] is 
\bea
    x + i y = \frac{\beta}{2\pi} \log \left[ e^{-i \phi} \frac{\theta_4\left(\frac{-i (\sigma +i\nu) - \phi}2, e^{-\pi s/2}\right)}{\theta_4\left(\frac{-i (\sigma + i\nu)}2, e^{-\pi s/2}\right)} \right].
\eea

\subsection{Infinite intervals: one-sided  measurement}
\label{append:infinite1sideconf}

In this section, we describe the conformal transformations illustrated in Fig. \ref{fig:map-infinite} and used for infinite intervals measured on one sided, as studied in Section \ref{sec:infinite-one}.

In the first conformal transformation, the infinitely long cylinder in $(x,y)$ coordinates (Fig.~\ref{fig:map-infinite} (a)) is mapped to a 2D plane in $X$ coordinates (Fig.~\ref{fig:map-infinite} (b)) using~(\ref{eq:transform-1}). 
Under this conformal map, the two measured slits (at $x\in (-\infty,\Delta L/2)$ and $x\in(\Delta L/2,\infty)$, with $y=0$)  are mapped to $X= \bar X$, and 
\bea
    && \text{first slit:} \quad   0 < X < e^{- \frac{\pi \Delta L}\beta} , \quad  \label{eq:slit1_infinite_one0} \\
    && \text{second slit:} \quad e^{ \frac{\pi \Delta L}\beta} < X.
    \label{eq:slit2_infinite_one0}
\eea
The right CFT and the unmeasured part of the left CFT are likewise mapped to the horizontal line $X= \bar X$, with the left CFT mapped to $e^{- \frac{\pi \Delta L}\beta} < X < e^{ \frac{\pi \Delta L}\beta}$, and the right CFT is mapped to $X<0$.
$x=-\infty$ is mapped to the origin, and $x=+\infty$ is mapped to $X=\pm \infty$.

In the second conformal transformation, the 2D plane with slits on the real axis (Fig.~\ref{fig:map-infinite} (b)) is mapped to an annulus (Fig.~\ref{fig:map-infinite} (c)), with the two slits mapped to the two circular edges of the annulus.
The map between these circles and the initial,  infinitely long slits can be found using the conformal transformation (\ref{eq:conformal_transform}) and taking $\alpha$ and $\beta$ to be at the circles. 
For the one-sided measurement case of interest here, setting $\alpha = \rho$, $\beta = 1$, and $C = \sqrt{\rho}$ in (\ref{eq:conformal_transform}) leads to
\bea
    X(\zeta) 
    = \frac{\theta_4^2(-\frac{i\log \zeta}2 , \rho)}{\theta_1^2(- \frac{i\log \zeta}2 , \rho)}. 
\eea
The conformal transformation is determined by a single parameter $\rho = e^{-\pi s}$. 
In order to see how it is related to $\Delta L$ (the size of the unmeasured region in the left CFT), we note that the first and second slits are mapped to $|\zeta| = \rho$ and $|\zeta| = 1$, respectively. 
Setting $\zeta = r e^{i \theta}$, the two slits are given by
\bea
    && \text{first slit: }
    X( \rho e^{i\theta}) =  \frac{\theta_1^2(\frac{\theta}2, \rho)}{\theta_4^2(\frac{\theta}2, \rho)}, \quad \theta \in (0,2\pi),
     \label{eq:slit1_infinite_one1}\\
     &&\text{second slit: }
     X(e^{i\theta}) =  \frac{\theta_4^2(\frac{\theta}2, \rho)}{\theta_1^2(\frac{\theta }2, \rho)}, \quad \theta \in (0,2\pi),
     \label{eq:slit2_infinite_one1}.
\eea
Both functions are non-negative. Further, it is not hard to see that the maximum of $X$ on the first slit (\ref{eq:slit1_infinite_one1}) and the minimum of $X$ on the second slit (\ref{eq:slit2_infinite_one1}) both occur at $\theta = \pi$. 
Matching these to (\ref{eq:slit1_infinite_one0}) and (\ref{eq:slit2_infinite_one0}) implies that we have the following relation
\bea \label{eq:s-infinite-one}
    e^{\frac{\pi \Delta L }\beta } =  \frac{\theta_4^2(\frac\pi2, \rho)}{\theta_1^2(\frac\pi2, \rho)} = \frac{\theta_4^2(\frac\pi2, e^{-\pi s})}{\theta_1^2(\frac\pi2, e^{-\pi s})}.
\eea
Note that $s$ is a monotonic, increasing function of $\Delta L$ (see Fig.~\ref{fig:s-infinite}). Two helpful special values are given by $s=0$ (corresponding to $\Delta L= 0$) and $s=1$ (corresponding to to $\Delta L = \log 2 \frac{\beta}{2\pi}$). Unlike the finite intervals case studied in Section \ref{sec:finite}, the time reflection symmetric lines are now mapped to two segments: the unmeasured part of the left CFT is mapped to $\zeta\in(-1,-\rho)$, whereas the right CFT is mapped to $\zeta \in (\rho , 1) $. 
Positive infinity in $x$ is mapped to $\zeta = 1$, and negative infinity to $\zeta = \rho$.

The final coordinate transformation maps the annulus in $\zeta$ coordinates (Fig.~\ref{fig:map-infinite} (c)) to a cylinder with finite length in $w$ coordinates (Fig.~\ref{fig:map-infinite} (d)) by~(\ref{eq:transform-3}). As in Appendix \ref{sec:finite_conformal}, the two slits are located at
\bea
    && \text{first slit:}   \quad \sigma = -\pi s, \\
    && \text{second slit:}  \quad \sigma = 0.
\eea
However, the unmeasured part of the left CFT is now mapped to the segment $\nu = \pi$, while the right CFT is mapped to the segment $\nu = 0$.
Negative spatial infinity in $x$ is mapped to $\sigma = -\pi s, \nu = 0$ (while positive infinity is mapped to $\sigma = 0, \nu = 0$).

To summarize, the conformal transformation from the original $(x,y)$ coordinates (Fig.~\ref{fig:map-infinite} (a)) to the final $w$ coordinate (Fig.~\ref{fig:map-infinite} (d)) is given by 
\bea
    x + i y = \frac{\beta}{2\pi} \log \left[ \frac{\theta_4^2 \left(-\frac{i(\sigma + i \nu)}2 , e^{-\pi s} \right)}{\theta_1^2 \left(- \frac{i(\sigma+ i \nu)}2 , e^{-\pi s} \right)} \right].
\eea

\subsection{Infinite intervals: two-sided  measurement}
\label{append:infinite2sideconf}

Finally, we describe here the conformal transformations illustrated in Fig. \ref{fig:map-infinite-two} used for the case where infinite intervals are measured in both CFTs, as considered in Section \ref{sec:infinite-two}.

The first conformal transformation maps the original, infinite cylinder in $(x,y)$ coordinates (Fig.~\ref{fig:map-infinite-two} (a))  to the 2D $X$ plane (Fig.~\ref{fig:map-infinite-two} (b)) using~(\ref{eq:transform-1}). 
The two infinite slits are mapped to $X= \bar X$ and 
\bea
    && \text{first slit:} \quad    X < -e^{ \frac{\pi \Delta L}\beta} , \quad  \label{eq:slit1_infinite_two0} \\
    && \text{second slit:} \quad 0 < X < e^{ -\frac{\pi \Delta L}\beta} .
    \label{eq:slit2_infinite_two0}
\eea
The unmeasured regions of the two CFTs are also both mapped to the horizontal line $X= \bar X$, where the left CFT ($y=0$) is mapped to $X> e^{ \frac{-\pi \Delta L}\beta}$ and the right CFT ($y=\beta/2)$ is mapped to $-e^{ \frac{\pi \Delta L}\beta} < X < 0$ .
Finally, $x=-\infty$ is mapped to the origin, while $x=\infty$ is mapped to $X = \infty$.

The second conformal transformation maps the 2D plane with two slits (Fig.~\ref{fig:map-infinite-two} (b)) to an annulus (Fig.~\ref{fig:map-infinite-two} (c)), in which the slits are mapped to the two circular edges of the annulus.
The transformation is given by equation (\ref{eq:conformal_transform}) with $\alpha=-\rho$, $\beta=1$, and $C=\sqrt{\rho}$ and takes the form
\bea \label{eq:map-infinite-two}
   X(\zeta) =  - \frac{\theta_3^2(- \frac{i\log \zeta}2, \rho)}{\theta_1^2(- \frac{i\log \zeta}2, \rho)}.
\eea
This map is determined by a single parameter $\rho = e^{-\pi s}$. 
In order to see how it is related to the slit parameter, we note that the first (second) slit is given by $|\zeta| = 1$ ($|\zeta| = \rho$). 
Setting $\zeta = r e^{i \theta}$, the two slits are
\bea
    && \text{first slit: }
    X(e^{i\theta}) =  - \frac{\theta_3^2( \frac{\theta}2, \rho)}{\theta_1^2( \frac{\theta}2, \rho)}, \quad \theta \in (0,2\pi) \label{eq:slit1_infinite_two1}\\
     &&\text{second slit: }
    X( \rho e^{i\theta}) =  \frac{\theta_2^2( \frac{\theta}2, \rho)}{\theta_4^2( \frac{\theta}2, \rho)}, \quad \theta \in (0,2\pi). \label{eq:slit2_infinite_two1}
\eea
Note that the maximum of the expression for the second slit (\ref{eq:slit2_infinite_two1}) occurs at $\theta = 0$. 
By matching with equation (\ref{eq:slit2_infinite_two0}), we obtain the relation\footnote{Equivalently, we can match the maxima in equations (\ref{eq:slit1_infinite_two0}) and (\ref{eq:slit1_infinite_two1}).}
\bea \label{eq:s-infinite-two}
   e^{-\frac{\pi \Delta L}\beta} =  \frac{\theta_2^2(0, \rho)}{\theta_4^2(0, \rho)} = \frac{\theta_2^2(0, e^{-\pi s})}{\theta_4^2(0, e^{-\pi s})}.
\eea
$s$ is again a monotonically increasing function of $\Delta L$ (see Fig.~\ref{fig:s-infinite-two}), and when $s=1$, $ \Delta L = 0$. The unmeasured regions of the two CFTs are mapped to two segments $r\in(\rho,1)$ (for the left CFT) and $r \in (-1,-\rho) $ (for the right CFT). 
$x = \pm \infty$ are mapped to $\zeta = 1$ and $\zeta = -\rho$, respectively.


For the final conformal transformation, the $\zeta$ annulus (Fig.~\ref{fig:map-infinite-two} (c)) is mapped to a cylinder with finite length in $w$ coordinates (Fig.~\ref{fig:map-infinite-two} (d)) by equation~(\ref{eq:transform-3}). 
With this map, the two slits are located at
\bea
    && \text{first slit:}   \quad \sigma = 0, \\
    && \text{second slit:}  \quad \sigma = -\pi s.
\eea
Further, the unmeasured region of the left (right) CFT is mapped to the segment $\nu =0$ ($\nu = \pi$). Spatial negative (positive) infinity in $x$ is mapped to $\sigma = -\pi s, \nu = 0,2\pi$ ($\sigma = 0, \nu = \pi$).

To summarize, the conformal transformation from original $(x,y)$ coordinates (Fig.~\ref{fig:map-infinite-two} (a)) to final $w$ coordinates (Fig.~\ref{fig:map-infinite-two} (d)) is given by
\bea
    x + i y = \frac{\beta}{2\pi} \log \left[ - \frac{\theta_3^2 \left(-\frac{i(\sigma + i \nu)}2 , e^{-\pi s} \right)}{\theta_1^2 \left(- \frac{i(\sigma+ i \nu)}2 , e^{-\pi s} \right)} \right].
\eea


\section{Brane trajectories}  
\label{app:b}

From the Euclidean action~(\ref{eq:action}), the equation of motion for the brane is 
\bea \label{eq:brane_eom}
    K_{\mu\nu} = (K-T) h_{\mu\nu},
\eea
where $K_{\mu\nu}$ is the extrinsic curvature, $K$ its trace, $T$ is the tension of the brane, and $h_{\mu\nu}$ the metric induced on the brane.
In this appendix, we will solve equation~(\ref{eq:brane_eom}) to obtain the brane trajectories in the BTZ black hole and thermal AdS backgrounds.

\subsection{Brane in BTZ black hole}
\label{append:brane-BTZ}


The BTZ black hole metric reads
\bea \label{eq:metric_BTZapp}
    ds^2 = \frac{R^2}{z^2} \left( \frac{dz^2}{h(z)} + h(z) d\sigma^2  +  d\nu^2 \right), \quad h(z) = 1- \frac{z^2}{z_H^2},
\eea
where $z=z_H$ indicates the black hole horizon. 
The periodicity of $\sigma$ is fixed by smoothness to be $2\pi z_H$. 
However, in the BCFT $\sigma$ is cut off at two boundaries $\sigma=-\pi s,0$ corresponding to the two slits introduced by the measurement, as shown in Fig.~\ref{fig:map-finite} (d). According to the AdS/BCFT prescription \cite{takayanagi2011holographic,fujita2011aspects}, the brane must then be anchored at $\sigma=-\pi s,0$ at the boundary $z=0$.
We will determine $z_H$ in the following by imposing such a boundary condition on the brane trajectory.

Because the metric does not depend on $\nu$, we consider a brane given by $( \nu, z, \sigma(z))$ with coordinate $( \nu,z)$. 
Two basis vectors tangent to the brane are 
\bea
    e_\nu^\mu = (1,0,0), \quad e_z^\mu = (0,1,\sigma').
\eea
where $\sigma' = \sigma'(z) = \partial_z \sigma(z)$. The unit normal can be obtained as
\bea
    n_\mu = (0,-\sigma',1) \frac{R}{z} \frac1{\sqrt{h \sigma^{\prime 2}+ \frac1h}}, \quad n^\mu = (0, -h^2 \sigma', 1) \frac{z}R \frac1{\sqrt{h(h^2 \sigma^{\prime 2}+1)}}.
\eea
The induced metric on the brane is given by
\bea
    h_{\nu\nu} = \frac{R^2}{z^2}, \quad h_{zz} = \frac{R^2}{z^2} \left( h \sigma^{\prime 2} + \frac1h \right). 
\eea
Further, from these quantities, the extrinsic curvature reads
\bea
    K_{\nu\nu} = \partial_0 n_0 - \Gamma_{00}^\mu n_\mu = \frac{Rh \sigma'}{z^2 \sqrt{h \sigma^{\prime 2} + \frac1h} }.
\eea

To solve (\ref{eq:brane_eom}), we can first contract the indices to get $K = 2T$, then use this to eliminate $K$ to get
\bea \label{eq:brane_emo2}
    K_{\mu\nu} = T h_{\mu\nu}. 
\eea
A simple observation is that the $K_{\nu\nu}$ component involves only first order derivatives, so we get
\bea
    \sigma'(z) = \pm \frac{RT}{h\sqrt{h-R^2 T^2}}, \quad 
\eea
which has solution
\bea
    \sigma(z) = \pm z_H \tan^{-1} \left[ \frac{RTz}{z_H \sqrt{h - R^2 T^2}} \right],
\eea
up to an additive integration constant. 
Because the brane should be anchored at $\sigma =0$ and $\sigma = \pi s$\footnote{We remind that $\sigma=\pm \pi z_H=\pm \pi s$ are identified.} for $z=0$, we can fix the condition $z_H = s$, and arrive at the following brane trajectory for $T\in (0, 1/R)$:
\bea  \label{aeq:brane_BTZ1}
    \sigma(z) = \begin{cases} s \tan^{-1} \left[ \frac{RTz}{s \sqrt{h - R^2 T^2}} \right], & \quad  0 < \sigma < \pi s/2 \\ \\
     s \left( \pi - \tan^{-1} \left[ \frac{RTz}{s \sqrt{h - R^2 T^2}} \right] \right), & \quad  \pi s/2 < \sigma < \pi s.
    \end{cases}
\eea
whereas for $T\in (-1/R, 0)$ we obtain
\bea  \label{aeq:brane_BTZ2}
    \sigma(z) = \begin{cases} s \tan^{-1} \left[ \frac{RTz}{s \sqrt{h - R^2 T^2}} \right], & \quad  -\pi s/2 < \sigma < 0 \\ \\
     s \left( -\pi - \tan^{-1} \left[ \frac{RTz}{s \sqrt{h - R^2 T^2}} \right] \right), & \quad  -\pi s < \sigma < -\pi s/2.
    \end{cases}
\eea

Note that in the BTZ phase we have a single connected brane (see e.g. Fig. \ref{fig:gravity-dual}).

\subsection{Brane in thermal AdS} \label{append:brane_AdS}

The metric of thermal AdS reads
\bea \label{eq:metric_AdS}
    ds^2 = \frac{R^2}{z^2} \left( \frac{dz^2}{f(z)} + f(z) d\nu^2  +  d\sigma^2 \right), \quad f(z) = 1- z^2. 
\eea
where the maximum value $z=1$ is determined by the periodicity of $\nu = \nu + 2\pi$. 
Again, the brane ends at $\sigma = 0$ and $\sigma = - \pi s$.

Consider a brane given by $(\nu, z, \sigma(z))$ with coordinate $(\nu, z)$. 
Two basis vectors tangent to the brane are 
\bea
    e_\nu^\mu = (1,0,0), \quad e_z^\mu = (0,1,\sigma').
\eea
The unit normal is
\bea
    n_\mu = (0,-\sigma',1) \frac{R}{z} \frac1{\sqrt{h \sigma^{\prime 2}+ \frac1h}}, \quad n^\mu = (0, -h^2 \sigma', 1) \frac{z}R \frac1{\sqrt{h(h^2 \sigma^{\prime 2}+1)}}.
\eea

The induced metric on the brane is
\bea
    h_{\nu\nu} = \frac{R^2}{z^2} f, \quad h_{zz} = \frac{R^2}{z^2} \left(\sigma^{\prime 2} + \frac1f \right). 
\eea
And the extrinsic curvature reads
\bea
    K_{\nu\nu} = \left( f - \frac{z}2 f'\right) \frac{R}{z^2} \frac{f \sigma'}{\sqrt{f \sigma^{\prime2} + 1}}. 
\eea

The $K_{\nu\nu}$ component of the equation of motion (\ref{eq:brane_emo2}) involves only first order derivatives, so we get
\bea
    \sigma'(z) = \pm \frac{RT}{\sqrt{\left(f- \frac{z}2 f'\right)^2 - R^2 T^2 f}}, \quad 
\eea
which leads to the solution
\bea
    \sigma(z) = \pm  \sinh^{-1}  \left( \frac{RTz}{\sqrt{1-R^2T^2}} \right)
\eea
up to an additive integration constant.

By imposing that the brane anchors at $\sigma = 0$ and $\sigma = - \pi s$, we finally obtain the solution
\bea \label{aeq:brane_AdS}
    \sigma(z) = \begin{cases}
        \sinh^{-1}  \left( \frac{RTz}{\sqrt{1-R^2T^2}} \right), & \quad \sigma > -\pi s/2  \\ \\
        -\pi s - \sinh^{-1}  \left( \frac{RTz}{\sqrt{1-R^2T^2}},  \right). & \quad \quad \sigma < -\pi s/2
    \end{cases}
\eea
which is valid for $T\in (-T_*,1/R)$, with $T_*$ defined in equation (\ref{eq:limitension}).
Note that in the thermal AdS phase we have two disconnected branes (see e.g. Fig. \ref{fig:gravity-dual}), and the first (second) line of equation (\ref{aeq:brane_AdS}) is for the brane trajectory anchoring at $\sigma = 0$ ($\sigma = -\pi s$). 
The range of the $\sigma$ coordinate in the bulk is given by $\sigma \in \left(-\pi s - \sinh^{-1}  \left( \frac{RTz}{\sqrt{1-R^2T^2}},  \right) ,\sinh^{-1}  \left( \frac{RTz}{\sqrt{1-R^2T^2}} \right) \right) $.



\section{On-shell action and Hawking-Page transition} \label{append:transition}


In this appendix we evaluate the on-shell action for both the BTZ black hole phase and the thermal AdS phase.

For the BTZ black hole, the metric is given by equation~(\ref{eq:metric_BTZ}) and the brane trajectory by equations~(\ref{aeq:brane_BTZ1}), (\ref{aeq:brane_BTZ2}) for positive and negative tension, respectively. 
The on-shell brane action is given by
\bea
    - \frac{1}{8\pi G_N} \int \sqrt{h} (K-T) 
    &=& - 2\cdot \frac{T}{4 G_N} \int_{\epsilon}^{z_*}  dz \frac{R^2}{z^2} \frac1{\sqrt{ h - R^2 T^2}} 
\eea
where 
the factor of 2 comes from dividing the brane trajectory into two segments, both spanning from $z=\epsilon$ to the turning point $z=z_*$, and $\epsilon$ is the cutoff at the boundary.
The bulk term is (using $\mathcal R = - \frac6{R^2}$ and $\Lambda = \frac1{R^2}$), 
\bea
    - \frac{1}{16\pi G_N} \int \sqrt{g} (R-2\Lambda ) 
    &=& \frac{R}{2G_N} \left(\int_{\epsilon}^{z_*} \frac{dz}{z^3} (2 \sigma(z) + \pi s) + \int_{z_*}^{s}  \frac{dz}{z^3}   2 \pi s \right)
\eea
where in the first integrand we used 
$\sigma(z) = s \tan^{-1} \left[ \frac{RTz}{s \sqrt{h - R^2 T^2}} \right]$.
Combining the boundary and bulk terms, integrating by parts, and adding appropriate counterterms (see \cite{numasawa2016epr,Antonini:2022sfm} for a detailed derivation), the on-shell action is
\bea
    I_\text{BTZ} = 
    - \frac{\pi R}{4 G_N s}. 
\eea

For the thermal AdS phase, the metric is given by equation~(\ref{eq:metric_AdS}) and the brane trajectory by equation~(\ref{aeq:brane_AdS}).
The evaluation of the on-shell action is similar to the BTZ black hole phase
(see \cite{numasawa2016epr,Antonini:2022sfm} for details). 
The final result is
\bea
    I_\text{AdS} = - \frac{R \pi s}{4 G_N} - \frac{R}{2G_N} \tanh^{-1} RT. 
\eea

By equating the on-shell actions for the BTZ black hole and thermal AdS phases, we find that the Hawking-Page transition is located at
\bea
    RT = \tanh \left[ \frac{\pi}2 \left( \frac{1}{s_c} - s_c \right) \right]
\eea
where $s_c$ is the critical value of $s$. 
For $s < s_c$ the BTZ black hole is the dominant saddle in the gravitational Euclidean path integral and when $s > s_c$ the thermal AdS spacetime is the dominant saddle.

\section{Embedding coordinate and geodesic length} \label{append:geodesic}

In this section, we compute length of geodesics in BTZ via the coordinates in an embedding space, as needed to calculate the entropy via the RT formula for the setup studied in Section \ref{sec:hologeodesic}. 
The Euclidean $AdS_3$ spacetime can be embedded as a 3D submanifold of a 4D flat spacetime with coordinates $(Y^{-1}, Y^0, Y^1, Y^2)$, given by
\bea 
    -(Y^{-1})^2 + (Y^0)^2 + (Y^1)^2 + (Y^2)^2 = -R^2,
\eea
where $R$ is the AdS radius. 
The 4D metric in the embedding coordinates is $ds^2 = -(dY^{-1})^2 + (dY^{0})^2 + (dY^{1})^2  + (dY^{2})^2 $. The BTZ black hole coordinates used  above in \ref{eq:metric_BTZ} are related to these embedding coordinates by
\bea
Y^{-1} &=& R \frac{z_H}{z} \cosh \frac{\nu}{z_H}, \\
Y^0 &=& R \sqrt{\frac{z_H^2}{z^2}-1} \sin \frac{\sigma}{z_H}, \\
Y^1 &=& R \frac{z_H}{z} \sinh \frac{\nu}{z_H}, \\
Y^2 &=& R \sqrt{\frac{z_H^2}{z^2}-1} \cos \frac{\sigma}{z_H}.
\eea
Using these embedding coordinates, the geodesic length between two points  $Y_i = Y_i(\sigma_i, \nu_i, z_i)$, $i=1,2$ is 
\bea
   R^2 \cosh \left(\frac{D(Y_1, Y_2)}{R}\right) =  -Y_1 \cdot Y_2,
\eea
(see e.g. \cite{Shenker:2013pqa}) where $D(Y_1,Y_2)$ denotes the geodesic length and the right-hand side denotes the inner product in the embedding space, $Y_1 \cdot Y_2 = -Y^{-1}_1Y^{-1}_2 + Y^0_1Y^0_2 + Y^1_1Y^1_2 + Y^2_1 Y^2_2$ \footnote{Equivalently, the geodesic length can be related to the embedding space coordinates via $4 R^2 \sinh[\frac{D(Y_1,Y_2)}{R}] = C$, for $C$ the ``chordal distance'' in the embedding space, given by $C = -(\Delta Y^{-1})^2 + (\Delta Y^0)^2 + (\Delta Y^1)^2 + (\Delta Y^2)^2$, as in \cite{Louko:2000tp}.}.
Written in the BTZ black hole coordinates, the geodesic length is then
\bea
    D(Y_1, Y_2) = R \cosh^{-1}\left(\frac{z_H^2}{z_1 z_2}\left[ \cosh \frac{\nu_1-\nu_2}{z_H} - \sqrt{\left(1- \frac{z_1^2}{z_H^2}\right)\left(1- \frac{z_2^2}{z_H^2}\right)}\cos\frac{\sigma_1 - \sigma_2}{z_H} \right] \right). \nn \\
\eea

\bibliographystyle{jhep}
\bibliography{reference}

\providecommand{\href}[2]{#2}\begingroup\raggedright\begin{thebibliography}{10}

\bibitem{Maldacena:1997re}
J.~M. Maldacena, \emph{{The Large N limit of superconformal field theories and
  supergravity}}, \href{http://dx.doi.org/10.1023/A:1026654312961}{\emph{Adv.
  Theor. Math. Phys.} {\bf 2} (1998) 231--252},
  [\href{https://arxiv.org/abs/hep-th/9711200}{{\tt hep-th/9711200}}].

\bibitem{Witten:1998qj}
E.~Witten, \emph{{Anti-de Sitter space and holography}},
  \href{http://dx.doi.org/10.4310/ATMP.1998.v2.n2.a2}{\emph{Adv. Theor. Math.
  Phys.} {\bf 2} (1998) 253--291},
  [\href{https://arxiv.org/abs/hep-th/9802150}{{\tt hep-th/9802150}}].

\bibitem{Gubser:1998bc}
S.~S. Gubser, I.~R. Klebanov and A.~M. Polyakov, \emph{{Gauge theory
  correlators from noncritical string theory}},
  \href{http://dx.doi.org/10.1016/S0370-2693(98)00377-3}{\emph{Phys. Lett. B}
  {\bf 428} (1998) 105--114}, [\href{https://arxiv.org/abs/hep-th/9802109}{{\tt
  hep-th/9802109}}].

\bibitem{Aharony:1999ti}
O.~Aharony, S.~S. Gubser, J.~M. Maldacena, H.~Ooguri and Y.~Oz, \emph{{Large N
  field theories, string theory and gravity}},
  \href{http://dx.doi.org/10.1016/S0370-1573(99)00083-6}{\emph{Phys. Rept.}
  {\bf 323} (2000) 183--386}, [\href{https://arxiv.org/abs/hep-th/9905111}{{\tt
  hep-th/9905111}}].

\bibitem{Ryu2006a}
S.~Ryu and T.~Takayanagi, \emph{{Aspects of Holographic Entanglement Entropy}},
  \href{http://dx.doi.org/10.1088/1126-6708/2006/08/045}{\emph{JHEP} {\bf 08}
  (2006) 045}, [\href{https://arxiv.org/abs/hep-th/0605073}{{\tt
  hep-th/0605073}}].

\bibitem{Ryu2006b}
S.~Ryu and T.~Takayanagi, \emph{Holographic derivation of entanglement entropy
  from the anti--de sitter space/conformal field theory correspondence},
  {\emph{Physical review letters} {\bf 96} (2006) 181602}.

\bibitem{Hubeny:2007xt}
V.~E. Hubeny, M.~Rangamani and T.~Takayanagi, \emph{{A Covariant holographic
  entanglement entropy proposal}},
  \href{http://dx.doi.org/10.1088/1126-6708/2007/07/062}{\emph{JHEP} {\bf 07}
  (2007) 062}, [\href{https://arxiv.org/abs/0705.0016}{{\tt 0705.0016}}].

\bibitem{Swingle:2009bg}
B.~Swingle, \emph{{Entanglement Renormalization and Holography}},
  \href{http://dx.doi.org/10.1103/PhysRevD.86.065007}{\emph{Phys. Rev. D} {\bf
  86} (2012) 065007}, [\href{https://arxiv.org/abs/0905.1317}{{\tt
  0905.1317}}].

\bibitem{VanRaamsdonk:2010pw}
M.~Van~Raamsdonk, \emph{{Building up spacetime with quantum entanglement}},
  \href{http://dx.doi.org/10.1142/S0218271810018529}{\emph{Gen. Rel. Grav.}
  {\bf 42} (2010) 2323--2329}, [\href{https://arxiv.org/abs/1005.3035}{{\tt
  1005.3035}}].

\bibitem{Maldacena:2013xja}
J.~Maldacena and L.~Susskind, \emph{{Cool horizons for entangled black holes}},
  \href{http://dx.doi.org/10.1002/prop.201300020}{\emph{Fortsch. Phys.} {\bf
  61} (2013) 781--811}, [\href{https://arxiv.org/abs/1306.0533}{{\tt
  1306.0533}}].

\bibitem{Engelhardt:2014gca}
N.~Engelhardt and A.~C. Wall, \emph{{Quantum Extremal Surfaces: Holographic
  Entanglement Entropy beyond the Classical Regime}},
  \href{http://dx.doi.org/10.1007/JHEP01(2015)073}{\emph{JHEP} {\bf 01} (2015)
  073}, [\href{https://arxiv.org/abs/1408.3203}{{\tt 1408.3203}}].

\bibitem{Dong:2016eik}
X.~Dong, D.~Harlow and A.~C. Wall, \emph{{Reconstruction of Bulk Operators
  within the Entanglement Wedge in Gauge-Gravity Duality}},
  \href{http://dx.doi.org/10.1103/PhysRevLett.117.021601}{\emph{Phys. Rev.
  Lett.} {\bf 117} (2016) 021601},
  [\href{https://arxiv.org/abs/1601.05416}{{\tt 1601.05416}}].

\bibitem{Harlow:2016vwg}
D.~Harlow, \emph{{The Ryu\textendash{}Takayanagi Formula from Quantum Error
  Correction}},
  \href{http://dx.doi.org/10.1007/s00220-017-2904-z}{\emph{Commun. Math. Phys.}
  {\bf 354} (2017) 865--912}, [\href{https://arxiv.org/abs/1607.03901}{{\tt
  1607.03901}}].

\bibitem{numasawa2016epr}
T.~Numasawa, N.~Shiba, T.~Takayanagi and K.~Watanabe, \emph{{EPR Pairs, Local
  Projections and Quantum Teleportation in Holography}},
  \href{http://dx.doi.org/10.1007/JHEP08(2016)077}{\emph{JHEP} {\bf 08} (2016)
  077}, [\href{https://arxiv.org/abs/1604.01772}{{\tt 1604.01772}}].

\bibitem{Antonini:2022sfm}
S.~Antonini, G.~Bentsen, C.~Cao, J.~Harper, S.-K. Jian and B.~Swingle,
  \emph{{Holographic measurement and bulk teleportation}},
  \href{http://dx.doi.org/10.1007/JHEP12(2022)124}{\emph{JHEP} {\bf 12} (2022)
  124}, [\href{https://arxiv.org/abs/2209.12903}{{\tt 2209.12903}}].

\bibitem{Maldacena:2001kr}
J.~M. Maldacena, \emph{{Eternal black holes in anti-de Sitter}},
  \href{http://dx.doi.org/10.1088/1126-6708/2003/04/021}{\emph{JHEP} {\bf 04}
  (2003) 021}, [\href{https://arxiv.org/abs/hep-th/0106112}{{\tt
  hep-th/0106112}}].

\bibitem{kourkoulou2017pure}
I.~Kourkoulou and J.~Maldacena, \emph{Pure states in the syk model and nearly-$
  ads\_2 $ gravity}, {\emph{arXiv preprint arXiv:1707.02325} (2017) }.

\bibitem{Cooper:2018cmb}
S.~Cooper, M.~Rozali, B.~Swingle, M.~Van~Raamsdonk, C.~Waddell and D.~Wakeham,
  \emph{{Black hole microstate cosmology}},
  \href{http://dx.doi.org/10.1007/JHEP07(2019)065}{\emph{JHEP} {\bf 07} (2019)
  065}, [\href{https://arxiv.org/abs/1810.10601}{{\tt 1810.10601}}].

\bibitem{Antonini:2019qkt}
S.~Antonini and B.~Swingle, \emph{{Cosmology at the end of the world}},
  \href{http://dx.doi.org/10.1038/s41567-020-0909-6}{\emph{Nature Phys.} {\bf
  16} (2020) 881--886}, [\href{https://arxiv.org/abs/1907.06667}{{\tt
  1907.06667}}].

\bibitem{Antonini:2021xar}
S.~Antonini and B.~Swingle, \emph{{Holographic boundary states and
  dimensionally reduced braneworld spacetimes}},
  \href{http://dx.doi.org/10.1103/PhysRevD.104.046023}{\emph{Phys. Rev. D} {\bf
  104} (2021) 046023}, [\href{https://arxiv.org/abs/2105.02912}{{\tt
  2105.02912}}].

\bibitem{Milekhin:2022bzx}
A.~Milekhin and F.~K. Popov, \emph{{Measurement-induced phase transition in
  teleportation and wormholes}},  \href{https://arxiv.org/abs/2210.03083}{{\tt
  2210.03083}}.

\bibitem{Antonini:2022lmg}
S.~Antonini, B.~Grado-White, S.-K. Jian and B.~Swingle, \emph{{Holographic
  measurement and quantum teleportation in the SYK thermofield double}},
  \href{http://dx.doi.org/10.1007/JHEP02(2023)095}{\emph{JHEP} {\bf 02} (2023)
  095}, [\href{https://arxiv.org/abs/2211.07658}{{\tt 2211.07658}}].

\bibitem{rajabpour2015post}
M.~Rajabpour, \emph{Post-measurement bipartite entanglement entropy in
  conformal field theories}, {\emph{Physical Review B} {\bf 92} (2015) 075108}.

\bibitem{rajabpour2016entanglement}
M.~A. Rajabpour, \emph{{Entanglement entropy after a partial projective
  measurement in $1+1$ dimensional conformal field theories: exact results}},
  \href{http://dx.doi.org/10.1088/1742-5468/2016/06/063109}{\emph{J. Stat.
  Mech.} {\bf 1606} (2016) 063109},
  [\href{https://arxiv.org/abs/1512.03940}{{\tt 1512.03940}}].

\bibitem{cardy1989boundary}
J.~L. Cardy, \emph{Boundary conditions, fusion rules and the verlinde formula},
  {\emph{Nuclear Physics B} {\bf 324} (1989) 581--596}.

\bibitem{miyaji2014boundary}
M.~{Miyaji}, S.~{Ryu}, T.~{Takayanagi} and X.~{Wen}, \emph{{Boundary states as
  holographic duals of trivial spacetimes}},
  \href{http://dx.doi.org/10.1007/JHEP05(2015)152}{\emph{Journal of High Energy
  Physics} {\bf 2015} (May, 2015) 152},
  [\href{https://arxiv.org/abs/1412.6226}{{\tt 1412.6226}}].

\bibitem{takayanagi2011holographic}
T.~Takayanagi, \emph{Holographic dual of a boundary conformal field theory},
  {\emph{Physical review letters} {\bf 107} (2011) 101602}.

\bibitem{fujita2011aspects}
M.~Fujita, T.~Takayanagi and E.~Tonni, \emph{Aspects of ads/bcft},
  {\emph{Journal of High Energy Physics} {\bf 2011} (2011) 1--40}.

\bibitem{affleck1991universal}
I.~Affleck and A.~W. Ludwig, \emph{Universal noninteger ‘‘ground-state
  degeneracy’’in critical quantum systems}, {\emph{Physical Review Letters}
  {\bf 67} (1991) 161}.

\bibitem{cardy2004boundary}
J.~Cardy, \emph{Boundary conformal field theory}, {\emph{arXiv preprint
  hep-th/0411189} (2004) }.

\bibitem{Miyaji:2022dna}
M.~Miyaji and C.~Murdia, \emph{{Holographic BCFT with a Defect on the
  End-of-the-World brane}},
  \href{http://dx.doi.org/10.1007/JHEP11(2022)123}{\emph{JHEP} {\bf 11} (2022)
  123}, [\href{https://arxiv.org/abs/2208.13783}{{\tt 2208.13783}}].

\bibitem{Barcelo:2000ta}
C.~Barcelo and M.~Visser, \emph{{Brane surgery: Energy conditions, traversable
  wormholes, and voids}},
  \href{http://dx.doi.org/10.1016/S0550-3213(00)00379-5}{\emph{Nucl. Phys. B}
  {\bf 584} (2000) 415--435}, [\href{https://arxiv.org/abs/hep-th/0004022}{{\tt
  hep-th/0004022}}].

\bibitem{Marolf:2002np}
D.~Marolf and S.~F. Ross, \emph{{Stringy negative tension branes and the second
  law of thermodynamics}},
  \href{http://dx.doi.org/10.1088/1126-6708/2002/04/008}{\emph{JHEP} {\bf 04}
  (2002) 008}, [\href{https://arxiv.org/abs/hep-th/0202091}{{\tt
  hep-th/0202091}}].

\bibitem{Burgess:2002vu}
C.~P. Burgess, F.~Quevedo, S.~J. Rey, G.~Tasinato and I.~Zavala,
  \emph{{Cosmological space-times from negative tension brane backgrounds}},
  \href{http://dx.doi.org/10.1088/1126-6708/2002/10/028}{\emph{JHEP} {\bf 10}
  (2002) 028}, [\href{https://arxiv.org/abs/hep-th/0207104}{{\tt
  hep-th/0207104}}].

\bibitem{roberts2012time}
M.~M. Roberts, \emph{Time evolution of entanglement entropy from a pulse},
  {\emph{Journal of High Energy Physics} {\bf 2012} (2012) 1--14}.

\bibitem{Faulkner:2018faa}
T.~Faulkner, M.~Li and H.~Wang, \emph{{A modular toolkit for bulk
  reconstruction}},
  \href{http://dx.doi.org/10.1007/JHEP04(2019)119}{\emph{JHEP} {\bf 04} (2019)
  119}, [\href{https://arxiv.org/abs/1806.10560}{{\tt 1806.10560}}].

\bibitem{calabrese2009entanglement}
P.~Calabrese and J.~Cardy, \emph{Entanglement entropy and conformal field
  theory}, {\emph{Journal of physics a: mathematical and theoretical} {\bf 42}
  (2009) 504005}.

\bibitem{hamilton2006holographic}
A.~Hamilton, D.~Kabat, G.~Lifschytz and D.~A. Lowe, \emph{Holographic
  representation of local bulk operators}, {\emph{Physical Review D} {\bf 74}
  (2006) 066009}.

\bibitem{Zou:2023rmw}
Y.~Zou, S.~Sang and T.~H. Hsieh, \emph{{Channeling quantum criticality}},
  \href{https://arxiv.org/abs/2301.07141}{{\tt 2301.07141}}.

\bibitem{crowdy2006conformal}
D.~Crowdy and J.~Marshall, \emph{Conformal mappings between canonical multiply
  connected domains}, {\emph{Computational Methods and Function Theory} {\bf 6}
  (2006) 59--76}.

\bibitem{banerjee2016lambert}
S.~Banerjee and B.~Wilkerson, \emph{Lambert series and q-functions near q= 1},
  {\emph{arXiv preprint arXiv:1602.01085} (2016) }.

\bibitem{Shenker:2013pqa}
S.~H. Shenker and D.~Stanford, \emph{{Black holes and the butterfly effect}},
  \href{http://dx.doi.org/10.1007/JHEP03(2014)067}{\emph{JHEP} {\bf 03} (2014)
  067}, [\href{https://arxiv.org/abs/1306.0622}{{\tt 1306.0622}}].

\bibitem{Louko:2000tp}
J.~Louko, D.~Marolf and S.~F. Ross, \emph{{On geodesic propagators and black
  hole holography}},
  \href{http://dx.doi.org/10.1103/PhysRevD.62.044041}{\emph{Phys. Rev. D} {\bf
  62} (2000) 044041}, [\href{https://arxiv.org/abs/hep-th/0002111}{{\tt
  hep-th/0002111}}].

\end{thebibliography}\endgroup

\end{document}